%% file: main.tex
\documentclass[12pt]{ociamthesis}  

\usepackage[utf8]{inputenc}
\usepackage{amssymb}
\usepackage[english]{babel}
\usepackage[colorinlistoftodos]{todonotes}
\usepackage{times}
\usepackage{latexsym}
\usepackage{graphicx}
\usepackage{subfigure}
\usepackage{comment}
\usepackage{amsmath}
\usepackage{url}
\usepackage{hyperref}
\usepackage{tikz}
\usepackage{float}
\usepackage{enumitem}
\usepackage[T1]{fontenc}
\usepackage{fixmath}
\usepackage{booktabs}
\usepackage{multirow}
\usepackage{siunitx}
\usepackage{amsfonts}
\usepackage{algorithm}
\usepackage{algpseudocode}
\usepackage{color}
\usepackage{etoolbox}
\usepackage{tcolorbox}
\usepackage{listings}
\usepackage{makecell}
\usepackage{upquote}
\usepackage{wrapfig}
\usepackage{fancybox}
\usepackage{pifont}
\usepackage{xcolor}
\usepackage{setspace}
\usepackage{longtable}
\usepackage{multirow}
\usepackage{breakcites}
\newtheorem{test}{\textbf{Answers to RQ}}
\newenvironment{boxedtest}
{\noindent\begin{Sbox}\begin{minipage}{\linewidth-7.5\fboxrule-4\fboxsep-1pt}\begin{test}}
{\end{test}\end{minipage}\end{Sbox}\doublebox{\TheSbox}}

\usepackage{titlesec}
\titleformat{\chapter}[display]   
{\normalfont\huge\bfseries}{\chaptertitlename\ \thechapter}{20pt}{\LARGE}  
\titlespacing*{\chapter}{0pt}{0pt}{40pt}

\lstdefinestyle{JAVA}{
  language=JAVA,
  moredelim=[is][\underbar]{_}{_},
}

\definecolor{KWColor}{rgb}{0.37,0.08,0.25}
\definecolor{CommentColor}{rgb}{0.133,0.545,0.133}
\definecolor{StringColor}{rgb}{0,0.126,0.941}
\title{Demystifying security and compatibility issues in Android Apps}   
\author{Xiaoyu Sun}             
\college{Faculty of Information Technology}  
\degree{Doctor of Philosophy}     
\degreedate{January 2023}         

\begin{document}
\emergencystretch 1em
\baselineskip=25pt

\setcounter{secnumdepth}{3}
\setcounter{tocdepth}{3}

\maketitle                  

\begin{romanpages}          

\include{i_copyright}       
\include{ii_abstract}        
\include{iii_declaration}   
\include{iv_publication}    
\include{vi_acknowledgements} 

\tableofcontents            
\listoffigures              
\listoftables

\end{romanpages}            

\include{chapter1}

\include{chapter2}

\include{chapter3}
\include{chapter4}

\include{chapter5}

\include{chapter6}


\addcontentsline{toc}{chapter}{Bibliography}
\renewcommand{\bibname}{References}
\bibliography{refs}        
\bibliographystyle{apalike}  

\end{document}

%% file: i_copyright.tex
\chapter*{Copyright}

\textcopyright  \textbf{2022 - Xiaoyu Sun}


\noindent \textbf{Copyright Notice} 
\textit{
I certify that I have made all reasonable efforts to secure copyright permissions for third-party content included in this thesis and have not knowingly added copyright content to my work without the owner's permission.
}

%% file: ii_abstract.tex
\chapter*{Abstract}
Never before has any OS been so popular as Android. Existing mobile phones are not simply devices for making phone calls and receiving SMS messages, but powerful communication and entertainment platforms for web surfing, social networking, etc. Even though the Android OS offers powerful communication and application execution capabilities, it is riddled with defects (e.g., security risks, and compatibility issues), new vulnerabilities come to light daily, and bugs cost the economy tens of billions of dollars annually. For example, malicious apps (e.g., back-doors, fraud apps, ransomware, spyware, etc.) are reported~\cite{googlereport} to exhibit malicious behaviours, including privacy stealing, unwanted programs installed, etc. To counteract these threats, many  works have been proposed that rely on static analysis techniques to detect such issues. However,  static techniques are not sufficient on their own to detect such defects precisely. This  will likely yield false positive results as static analysis has to make some trade-offs when handling complicated cases (e.g., object-sensitive vs. object-insensitive). In addition, static analysis techniques will also likely suffer from soundness issues because some complicated features (e.g., reflection, obfuscation, and hardening) are difficult to be handled~\cite{sun2021taming,samhi2022jucify}.

My research during my Ph.D. is among the first to tackle these problems directly: I develop several tools that attempt to automatically integrate static program analysis techniques with dynamic testing techniques for detecting Android defects in real-world applications. I have demonstrated that my approaches  are more effective and precise in detecting more security issues than state-of-the-art program analysis tools. Specifically, my works highlight several security/compatibility aspects of the Android framework and unveil the discovery of various security/compatibility threats that were never been investigated before. 
I present three different approaches to solve these issues:
\begin{itemize}
    \item {\sc \textbf{SEEKER} -- }
    While extremely valuable to achieve many advanced functions, mobile phone sensors can be abused by attackers to implement malicious activities in Android apps, as experimentally demonstrated by many state-of-the-art studies. There is hence a strong need to regulate the usage of mobile sensors so as to keep them from being exploited by malicious attackers. However, despite the fact that various efforts have been put in achieving this, i.e., detecting privacy leaks in Android apps, I have not yet found approaches to automatically detect sensor leaks in Android apps. To fill the gap, I designed and implemented a novel prototype tool, SEEKER, that extends the famous FlowDroid~\cite{arzt2014flowdroid} tool to detect sensor-based data leaks in Android apps. SEEKER conducts sensor-focused static taint analyses directly on the Android apps’ bytecode and reports not only sensor-triggered privacy leaks but also the sensor types involved in the leaks. Experimental results using over 40,000 real-world Android apps show that SEEKER is effective in detecting 1,964 sensor leaks in Android apps, and malicious apps are more interested in leaking sensor data than benign apps. Moreover, our improvement for supporting field sources detection has been merged to the original version FlowDroid via pull \#385 on GitHub\cite{FlowDroidMerge}), which I believe could be adapted to analyze other sensitive field-triggered leaks.
    \item {\sc \textbf{HiSenDroid -- }}
    Security of Android devices is now paramount, given their wide adoption among consumers for many sensitive tasks and data management. As researchers develop tools for statically or dynamically detecting suspicious apps, malware writers regularly update their attack mechanisms to hide malicious behavior implementation. This poses two problems to current research techniques: static analysis approaches, given their over-approximations, can report an overwhelming number of false alarms, while dynamic approaches will miss those behaviors that are hidden through evasion techniques. I propose in this work a static approach specifically targeted at highlighting hidden sensitive operations, mainly sensitive data flows. The prototype version of HiSenDroid has been evaluated on a large-scale dataset of thousands of malware and goodware samples on which it successfully revealed anti-analysis code snippets aiming at evading detection by dynamic analysis. I further experimentally show that, with FlowDroid, some of the hidden sensitive behaviors would eventually lead to private data leaks. Those leaks would have been hard to spot either manually among the large number of false positives reported by the state of the art static analyzers, or by dynamic tools. Overall, by putting the light on hidden sensitive operations, HiSenDroid helps security analysts in validating potential sensitive data operations, which would be previously unnoticed.
   
    \item {\sc \textbf{JUnitTestGen} -- } 
    Despite being one of the largest and most popular projects, the official Android framework has only provided test cases for less than 30\% of its APIs. Such a poor test case coverage rate has led to many compatibility issues that can cause apps to crash at runtime on specific Android devices, resulting in poor user experiences for both apps and the Android ecosystem. To mitigate this impact, various approaches have been proposed to automatically detect such compatibility issues. Unfortunately, these approaches have only focused on detecting signature-induced compatibility issues (i.e., a certain API does not exist in certain Android versions), leaving other equally important types of compatibility issues unresolved. In this work, I propose a novel prototype tool, JUnitTestGen, to fill this gap by mining existing Android API usage to generate unit test cases. After locating Android API usage in given real-world Android apps, JUnitTestGen performs inter-procedural backward data-flow analysis to generate a minimal executable code snippet (i.e., test case). Experimental results on thousands of real-world Android apps show that JUnitTestGen is effective in generating valid unit test cases for Android APIs. I show that these generated test cases are indeed helpful for pinpointing compatibility issues, including ones involving semantic code changes.
    
\end{itemize}

%% file: iii_declaration.tex
\chapter*{Thesis including published works declaration}
I hereby declare that this thesis contains no material which has been accepted for the award of any other degree or diploma at any university or equivalent institution and that, to the best of my knowledge and belief, this thesis contains no material previously published or written by another person, except where due reference is made in the text of the thesis. 

This thesis includes 3 original papers published in peer-reviewed journals/conferences. The core theme of the thesis is demystifying security and compatibility issues in Android Apps. The ideas, development and writing up of all the papers in the thesis were the principal responsibility of myself, the student, working within the Faculty of Information Technology of Monash University under the supervision of John Grundy and Li Li. 

The inclusion of co-authors reflects the fact that the work came from active collaboration between researchers and acknowledges input into team-based research. In the case of Chapter 3,4,5, my contribution to the work involved the following:

\begin{figure}[H]
    \centering 
    \includegraphics[width=0.85\textwidth]{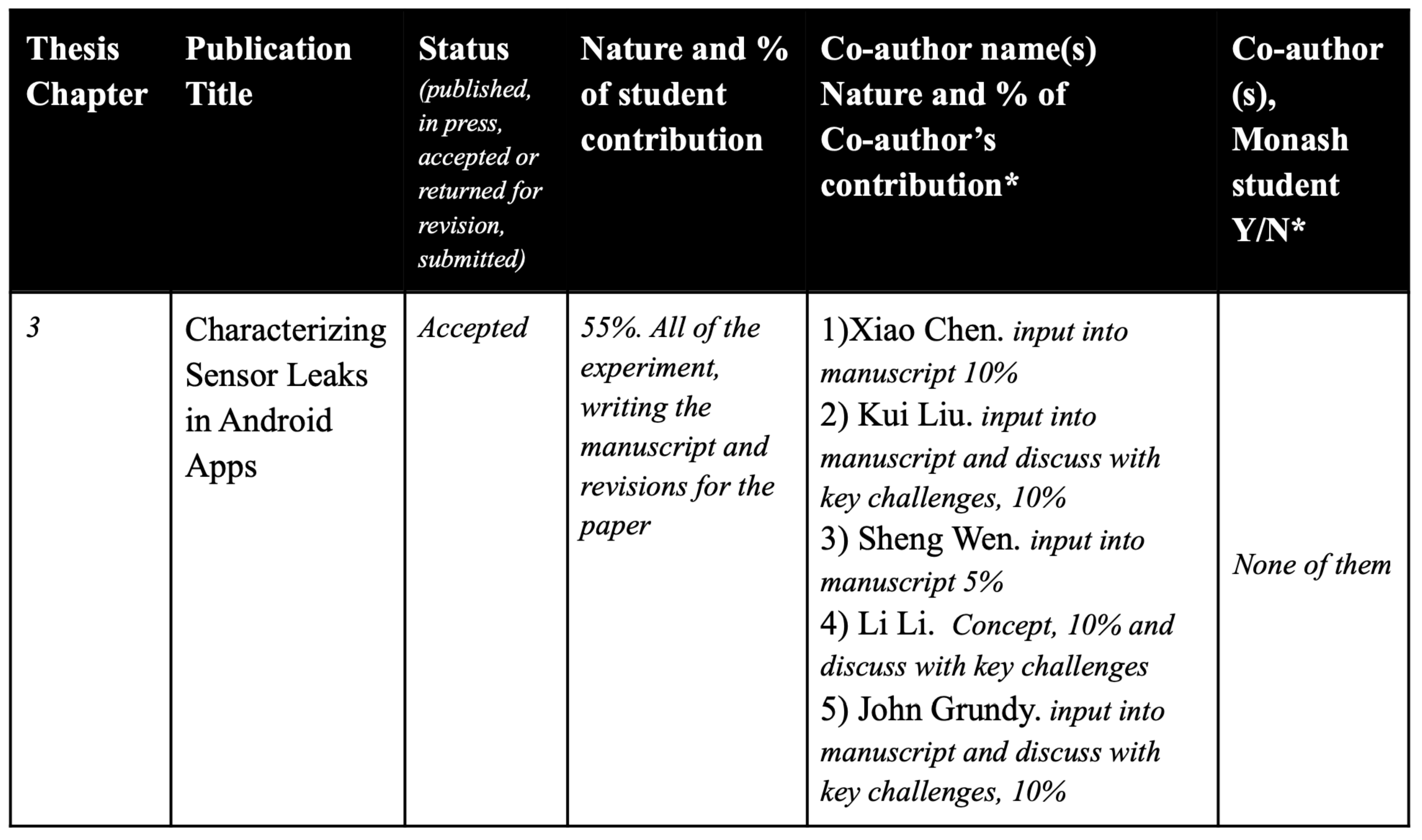}
    \label{fig:clarification}
\end{figure}

\begin{figure}[H]
    \centering 
    \includegraphics[width=0.85\textwidth]{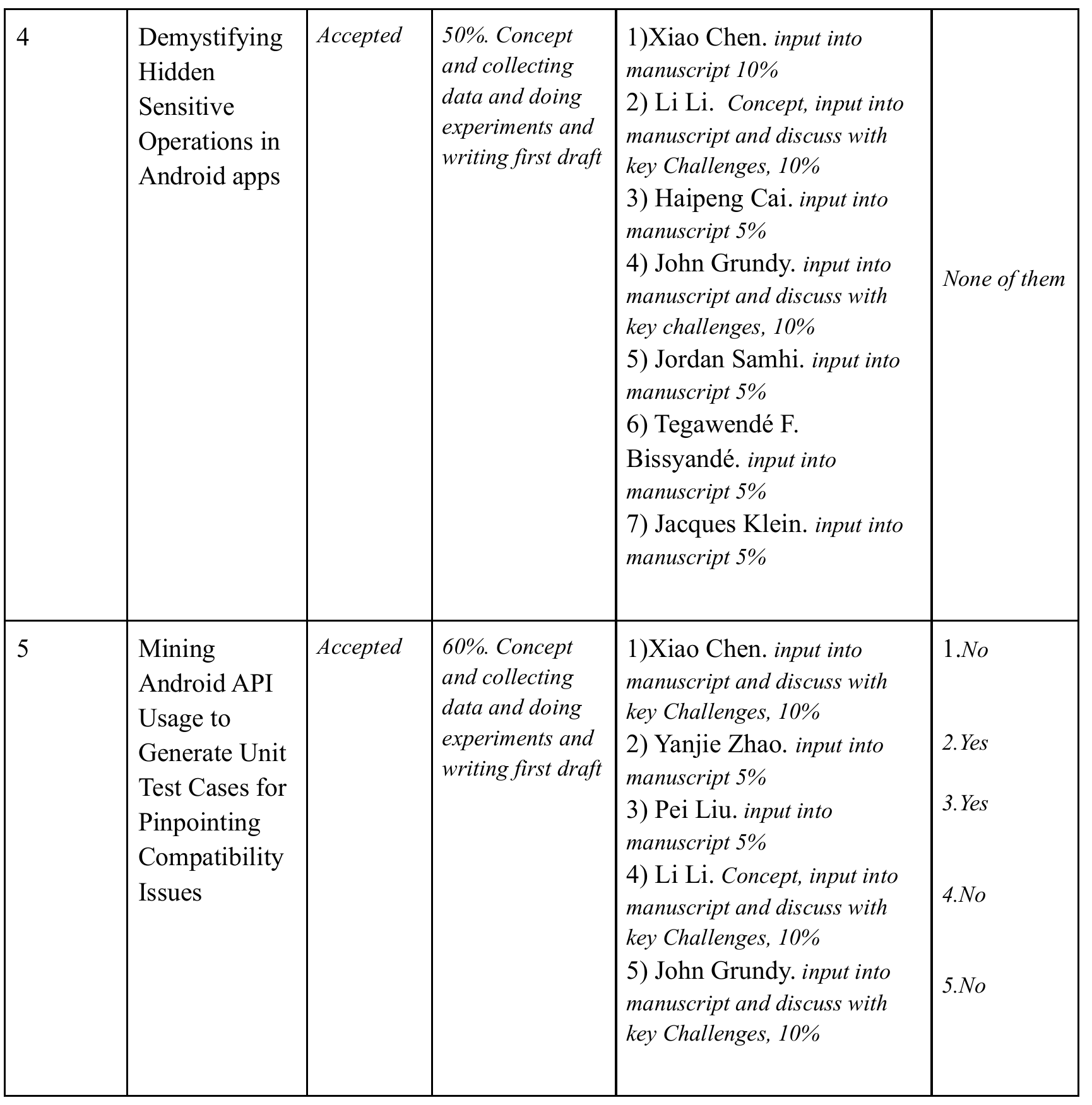}
    \label{fig:clarification}
\end{figure}

I have renumbered sections of published papers in order to generate a consistent presentation within the thesis. \par

\textbf{Student name:} Xiaoyu Sun \par
\textbf{Student signature: } Xiaoyu Sun  \qquad  \qquad \qquad \qquad \qquad \qquad \textbf{Date: 27-Jan-2023}

The undersigned hereby certify that the above declaration correctly reflects the nature and extent of
the student’s and co-authors’ contributions to this work. In instances where I am not the responsible
author I have consulted with the responsible author to agree on the respective contributions of the
authors.

\textbf{Student name:} John Grundy \par 
\textbf{Student signature: } John Grundy   \qquad  \qquad \qquad \qquad \qquad \qquad \textbf{Date: 27-Jan-2023}

%% file: iv_publication.tex
\chapter*{Publications}
\noindent \textbf{Publications included in this thesis}
\begin{enumerate}
    \item \textbf{Xiaoyu Sun}, Xiao Chen, Kui Liu, Sheng Wen, Li Li, and John Grundy. 2021. Characterizing Sensor Leaks in Android Apps. The IEEE 32nd International Symposium on Software Reliability Engineering (ISSRE) 2021, 12 pages. (Chapter 3)
    
    \item \textbf{Xiaoyu Sun}, Xiao Chen, Li Li, Haipeng Cai, John Grundy, Jordan Samhi, Tegawend\'e F. Bissyand\'e and Jacques Klein, 2022, Demystifying Hidden Sensitive Operations in Android apps. 
    ACM Transactions on Software Engineering and Methodology (TOSEM) 2022, 30 pages. (Chapter 4)
    
    \item \textbf{Xiaoyu Sun}, Xiao Chen, Yanjie Zhao, Pei Liu, John Grundy and Li Li. 2022. Mining Android API Usage to Generate Unit Test Cases for Pinpointing Compatibility Issues.  The 37th IEEE/ACM International Conference on Automated Software Engineering (ASE) 2022, 13 pages. (Chapter 5)

\end{enumerate}


\noindent \textbf{Other publications during candidature}
\begin{enumerate}
    \item \textbf{Xiaoyu Sun}, Li Li, Tegawend\'e F. Bissyand\'e, Jacques Klein, Damien Octeau, and John Grundy, 2021, Taming reflection: An essential step toward whole-program analysis of android apps. 
    ACM Transactions on Software Engineering and Methodology (TOSEM) 2021, 35 pages.

     \item \textbf{Xiaoyu Sun}, Xiao Chen, Yonghui Liu, John Grundy, Li Li. 2022. Taming Android Fragmentation through Lightweight Crowdsourced Testing. IEEE Transactions on Software Engineering (TSE) 2022 -- Under Major Revision.
    
\end{enumerate}

%% file: vi_acknowledgements.tex
\chapter*{Acknowledgements}
I am incredibly fortunate to have many people to thank.

First and foremost, I would like to thank my supervisors Li Li and John Grundy, who guided me into the land of program analysis. Their immense knowledge and plentiful experience have encouraged me in all the time of my academic research and daily life. Every meeting I had with Li and John was fruitful – they could consistently provide insightful suggestions and was also willing to dive deeply into technical discussions. It is their mentorship and encouragement which made this thesis possible.

I would like to thank Michael Sheng, who hosted me during my visit to Macquarie University. Michael is insightful and has the fantastic ability to explain complicated things in an abstract and easy-to-follow manner. I am also very grateful for his long-lasting support with my research career.

Great thanks to Jin-Song Dong, who hosted me during my visit at  Griffith University. Dong has lots of exciting ideas, and his research vision has always inspired me. I expect much more collaboration to come. I also thank Zhe Hou and the whole research team for their generous support of this visit.

I would like to thank all the people who contributed in some way to the work described in this thesis. Every result described in this thesis was accomplished with the help and support of fellow labmates and collaborators. Jacques Klein and I worked together on several projects, and without his efforts my job would have undoubtedly been more difficult. I greatly benefited from his keen scientific insight, his knack for solving seemingly intractable practical difficulties, and his ability to put complex ideas into simple terms. Yanjie Zhao joined our research group when I was beginning my first year as a graduate student, and, with her knowledge of program analysis and relentless work ethic, she was able to successfully carry out the fixable patch for incompatibility in Android. I was fortunate to have the chance to work with Dr. Pei Liu, who worked closely
with me during my PhD. Moreover, I would like to thank the various members from the University of Luxembourg with whom I had the opportunity to work and have not already mentioned: Jordan Samhi, Jun Gao and Tegawend\'e F Bissyand\'e.

Last but not least, I would like to thank my parents for their unconditional love and support. I made it through thanks to all of you. 

%% file: chapter1.tex
\chapter{Introduction}

\section{Background}
\subsection{Android Security and Compatibility Issues}
Android is the most adopted mobile operating system in terms of users, applications and developers~\cite{IDCReport}. However, its popularity means that good Android applications must co-exist with apps with security defects. Reports on many different kinds of Android defects are presented in the technology and lay media. For example, Trojans are the most prominent mobile threats as they constitute over 95\% of mobile malware~\cite{android_security_issues}. Over 98\% of mobile banking attacks target Android devices, which also comes as no surprise~\cite{android_security_issues}. Also, cyber scams (e.g., phishing and social media scams) are blamed to obtain sensitive personal information (e.g., financial information or account passwords). In fact, during the pandemic, phishing attacks grew by 600\% and became the top infection method in 2021~\cite{phishing_attacks}. The increasing number of Android security issues in the recent years put both Android users and app developers in danger. So the need for security is greater than ever for revealing these severe security issues. 

On the other hand, Android is also known suffer from fragmentation problems (a.k.a, compatibility issues). At a high level, Android Fragmentation refers to the fact that there are a massive number of different Android OS versions available on the market. Not every Android user would update their particular OS with the pace of Android framework evolution. As reported, Android API evolution is a critical issue in software maintenance ~\cite{nagappan2016future,li2018moonlightbox,martin2016survey,li2016accessing,oliveira2018android,lamothe2020a3,li2018characterising,zhou2016api,dig2005role,kapur2010refactoring,liu2022deep,sun2021characterizing}. McDonnell et al. ~\cite{mcdonnell2013empirical} have shown that the Android
system updates 115 APIs per month on average, while app developers usually adopt the new APIs much more slower. The slow adoption of API updates may raise various issues, such as security and compatibility. An empirical study on StackOverflow conducted by Linares-Vasquez et al. ~\cite{linares2014api} suggests that API updates would trigger more discussions, especially if APIs are removed from the Android system. They also revealed that users are in more favour of apps that use less fault and change-prone APIs \cite{linares2013api, bavota2014impact}, as these apps would likely introduce fewer failures, crashes and other bugs.

In this thesis, I wanted to focus on addressing specific types of security and compatibility issues commonly found in Android applications. 

\subsection{Sensor Leak Detection} I found that mobile phone sensors can be abused by attackers to implement malicious activities, as experimentally demonstrated by many state-of-the-art studies. Hence there is a strong need to regulate the usage of mobile sensors so as to keep them from being exploited by malicious attackers. However, despite the fact that various efforts have been put in achieving this, i.e., detecting privacy leaks in Android apps, I have not yet found approaches to automatically detect sensor leaks in Android apps. To fill the gap, I designed and implemented a novel prototype tool, SEEKER, that extends the famous FlowDroid tool to detect sensor-based data leaks in Android apps. SEEKER conducts sensor-focused static taint analyses directly on the Android apps' bytecode and reports not only sensor-triggered privacy leaks but also the sensor types involved in the leaks.

\subsection{Hidden Sensitive Operation Detection} Other malicious behaviours I wanted to investigate relate to `sensitive' operations in apps. I found that malware writers regularly update their attack mechanisms to hide malicious behaviour implementation, in order to bypass static code analysis (e.g., via obfuscation) and even dynamic detection (e.g., sensing of sandbox execution). In practice, sophisticated code obfuscation techniques \cite{moser2007limits} are being leveraged by attackers to hide their malicious program behavior, leading to false negatives in most static analyses thus resulting in imprecise and unsound results.
Camouflage techniques have been frequently leveraged by attackers to evade dynamic testing approaches~\cite{rasthofer2017making,egele2008survey}.
Attackers often introduce a so-called logic bomb or time bomb to set off malicious functions only after certain conditions are met.
For instance, after knowing that Google  applies a dynamic analysis tool called \emph{bouncer} to scan every app submitted to Google Play for five minutes, as revealed by Oberheide et al.~\cite{oberheide2012dissecting}, a bunch of malicious apps has been created and been demonstrated to be capable of penetrating Google's bouncer vetting system by simply waiting five minutes before triggering their malicious behavior.

To cope with such hidden malicious behaviors, researchers have devised new detection approaches.
For example, Fratantonio et al.~\cite{fratantonio2016triggerscope} have proposed an approach called TriggerScope to detect hidden behaviors triggered by predefined circumstances such as events related to location, time, and SMS. 
However, TriggerScope is not capable of detecting such malicious activities hidden behind other trigger types, such as the existence of other services (i.e., other than location, time and SMS).
In line with this research, Pan et al.~\cite{pan2017dark} have proposed a machine learning-based approach aiming to discover unknown trigger types.
Their approach, however, needs to manually label a dataset for training, which is known to be resource-intensive and error-prone.

Static analyzers suffer less than dynamic approaches from evasion techniques such as logic bomb or time bomb. 
In particular, regarding sensitive flow detection (also called privacy leak detection), numerous static analysis tools have been proposed such as FlowDroid~\cite{arzt2014flowdroid} (and its extension \textsc{IccTA}~\cite{li2015iccta}), \textsc{Amandroid}~\cite{wei2014amandroid},  or \textsc{DroidSafe}~\cite{gordon2015information}. Although these tools are able to track sensitive flows (which are often hidden) by bringing  key new contributions to the research community, they still face some well-known limitations~\cite{AnalyzingAnalyzers-ISSTA2018}: their inherent over-approximations inevitably lead to false alarms, which, for some analyzers, occur at a high rate, making them impractical. Consequently, when building on static analysis, manual investigation is often required.
Unfortunately, such efforts cannot scale. Dynamic validation then appears as an alternative. Unfortunately, runtime execution often misses hidden sensitive flows due to the implementation of evasion techniques by attackers.
While some effort (e.g.,~\cite{fratantonio2016triggerscope,pan2017dark}) has been put to \emph{characterize} Hidden Sensitive Operations (HSOs) in Android apps, our community has not yet proposed dedicated approaches to \emph{detect and explain} such operations, allowing attackers to achieve malicious behaviors while bypassing certain security vetting mechanisms.

I filled this research gap by proposing a new prototype tool, HiSenDroid, which deploys an automated static app analyzer tailored for detecting \textit{hidden} sensitive operations. 

\subsection{Compatibility Issues Detection} My focus of research has also included compatibility issues that might also lead to poor user experience. The Android framework provides thousands of public APIs that are heavily leveraged by app developers to facilitate their development of Android apps. Ideally, each such public API should be provided with a set of unit tests to ensure that the API is correctly implemented and the continuous evolution of the framework will not change its semantics.
Unfortunately, based on my preliminary investigation, less than 30\% of APIs, are provided with unit test cases, leaving the majority of APIs uncovered.
This is unacceptable considering that the Android framework nowadays has become one of the most popular projects (with millions of devices running it).

This poor test coverage of Android APIs has led to serious compatibility issues in the Android ecosystem, as recently shown~\cite{li2018cid,wei2016taming,mutchler2016target,zhang2015compatibility,ham2011mobile,huang2018understanding}.
For example, Li et al. ~\cite{li2018cid} demonstrate that various Android APIs suffer from compatibility issues as the evolution of the Android framework will regularly remove APIs from or add APIs into the framework. Such API removal or addition can result in no such class or method runtime exceptions when the corresponding app is running on certain framework versions.
Liu et al.~\cite{liu2021identifying} further present an approach to detect silently-evolved Android APIs, which could cause another type of compatibility issue as their semantics are altered (while not explicitly documented) due to the evolution of the Android framework.
Moreover, Wei et al.~\cite{wei2019pivot} experimentally show that some Android APIs could even be customized by smartphone manufacturers, leading to another type of compatibility issue that causes Android apps to crash on certain devices while behaving normally on others.
The authors further propose a prototype tool called PIVOT to automatically learn device-specific compatibility issues from existing Android apps. Their experiments on a set of top-ranked Google Play apps have discovered 17 device-specific compatibility issues.

To the best of our knowledge, the state-of-the-art works targeting compatibility issue detection leverage static analysis techniques to achieve their purpose.
However, as known to the community, the static analysis will likely yield false positive results as it has to make some trade-offs when handling complicated cases (e.g., object-sensitive vs. object-insensitive).
In addition, the static analysis will also likely suffer from soundness issues because some complicated features (e.g., reflection, obfuscation, and hardening) are difficult to be handled~\cite{sun2021taming,samhi2022jucify}.
Furthermore, except for syntactic changes, compatibility issues could also be triggered by semantic changes, which are non-trivial to be handled statically.
Indeed, as demonstrated by Liu et al.~\cite{liu2021identifying}, there are various semantic change-induced compatibility issues in the Android ecosystem that remain undetected after various static compatibility issue detection approaches are proposed to the community.

Moreover, static app analysis can only be leveraged to perform post-momentum analysis (i.e., after the compatibility issues are introduced to the community). 
They cannot stop the problems from being distributed into the community -- many Android apps, including very popular ones, still suffer from compatibility issues.
To mitigate this problem, I argued that incompatible Android APIs should be addressed as early as possible, i.e., ideally, at the time when they are introduced to the framework.
This could be achieved by providing unit tests for every API introduced to the framework and regressively testing the APIs against Android devices with different manufacturers and different framework versions.
However, it is time-consuming to manually write and maintain unit tests for each Android API (which probably explains why there is only a small set of APIs covered by unit tests at the moment).
There is hence a need to automatically generate compatibility unit tests for Android APIs.

To adddress this, I developed a prototype tool, JUnitTestGen, that attempts to automatically generate test cases for Android APIs based on their practical usage in real-world apps.

\section{Motivations and Objectives}
\subsection{Motivations}
To overcome the security and compatibility issues, other researchers have proposed various approaches to mitigate the usage of security/compatibility issues in Android apps~\cite{wei2016taming, wei2019pivot, xia2020android, crandall2006temporal,brumley2008automatically,zheng2012smartdroid}.
These approaches mainly leverage static analysis to achieve their purpose.
Unfortunately, static analysis is known to likely generate false-positive and false-negative results and is yet hard to handle many issues that involve sophisticated semantic changes~\cite{liu2021identifying}. In addition, the static analysis will also likely suffer from soundness issues because some complicated features (e.g., hidden sensitive operations) are difficult to be handled~\cite{sun2021taming,samhi2022jucify}.
Therefore, there is a imperative need to provide more precise and comprehensive approaches to complement existing works in handling sophisticated security/compatibility issues. Here, I broke down the motivations regarding to the three research projects as follows:

\begin{itemize}
   \item \textbf{Sensor Leaks in Android Apps: }
   
   As a concrete example, Lu et al. \cite{lu2018snoopy} revealed that sensitive intercepting password could be accessed through motion data on the smartwatch's onboard sensors. They proposed \emph{Snoopy}, a password extraction and inference approach via sensor data for PIN attack, which could affect smartwatch users in a non-invasive way.
\emph{Snoopy} extracts the segments of motion data when users entered passwords and then applies deep learning techniques to infer the actual passwords. Figure~\ref{fig:snoopy_example} gives two examples of the differences of the motion sensor data changes when the user swipes or taps a password on a smartwatch.
\emph{Snoopy} demonstrates the feasibility of sensor data leaks by intercepting password information entered on smartwatches. 
Such real-world sensor-enabled attacks motivated us to provide automatic tools for characterizing universal sensor leaks in Android Apps that have been long overlooked.

\begin{figure}[!t]
    \centering 
    \includegraphics[width=0.85\textwidth]{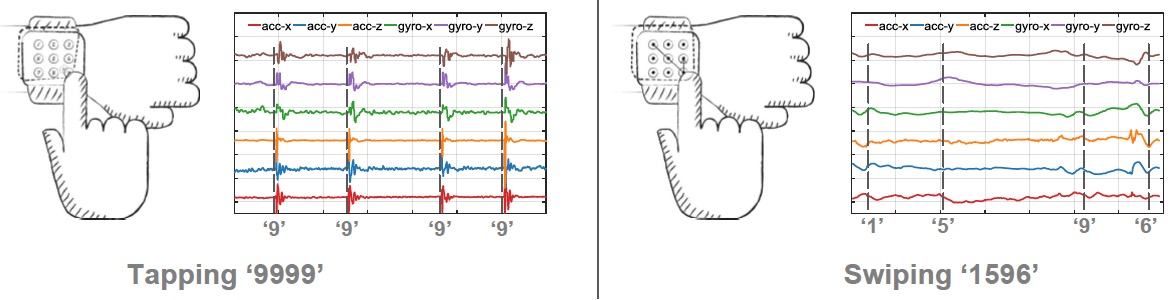}
    \caption{The snoopy example of sniffing smartwatch passwords via censoring motion sensor data \cite{lu2018snoopy}.}
    \label{fig:snoopy_example}
\end{figure}

    \item \textbf{Hidden Sensitive Operations in Android apps:}
    
\begin{lstlisting}[
caption={An example of a real-world hidden sensitive data flow.},
label=code:emu_HSO,
float=t,
firstnumber=1]
public class MainActivity extends AppCompatActivity {
 protected void onCreate(Bundle savedInstanceState) {
  SmsManager smsManager = SmsManager.getDefault();
  ED ed = new ED(this);
  StringBuilder message = new StringBuilder();

  if(ed.checkPackageName()) {
   TelephonyManager tm = (TelephonyManager)     getSystemService(Context.TELEPHONY_SERVICE);
   String imei = tm.getDeviceId();
   String phoneNumber = tm.getLine1Number();
   String subscriberId = tm.getSubscriberId();
   message.append(imei);
   message.append(phoneNumber);
   message.append(subscriberId);
   smsManager.sendDataMessage("+115800763861", null, (short)1001, message.toString().getBytes(), null, null);
  } else {
   //benign string operations
}}

public class ED {
 public ED(Context pContext) {
  mContext = pContext;
  mListPackageName.add("com.google.android...genymotion");
  mListPackageName.add("com.bluestacks");
  mListPackageName.add("com.bignox.app");
 }
 public boolean checkPackageName() {
  if (!isCheckPackage || mListPackageName.isEmpty()) {
   return false;
  }
  final PackageManager packageManager = mContext.getPackageManager();
  for (final String pkgName : mListPackageName) {
   final Intent tryIntent = packageManager.getLaunchIntentForPackage(pkgName);
   if (tryIntent != null) {
    final List<ResolveInfo> resolveInfos = packageManager.queryIntentActivities(tryIntent, PackageManager.MATCH_DEFAULT_ONLY);
    if (!resolveInfos.isEmpty()) {
     return true;
    }
   }
  }
  return false;
}
\end{lstlisting}

Listing~\ref{code:emu_HSO} exemplifies a simplified code snippet illustrating these definitions in practice. Note that Listing 1 presents the typical characteristics of an HSO in many real-world apps that I have manually analyzed. 
At line 7, the app firstly checks if it is running on one of the popular Android emulators (i.e., \emph{genymotion, bluestacks, and bignox}). If not, the app reads the device information and sends it to a hard-coded phone number through an SMS. Otherwise, if an emulator environment is detected, it will only perform some unharmful string operations (ignored).
In this example, three private data -- namely the device's IMEI, IMSI, and phone number -- are retrieved in lines 9-11 and sent to a hard-coded phone number via SMS (line 15).
All of these three leaks are hidden behind the trigger condition \emph{ed.checkPackageName()} (line 7). The trigger condition checks the return value of a self-defined method \emph{checkPackageName()} (line 30), which is determined by several other \emph{if-conditions} defined in the invoking method (lines 31,37,39). Finally, the trigger condition in the HSO is traced back to a system API \emph{PackageManager.queryIntentActivities()} (line 38).
This trigger condition examines whether popular Android emulator packages (lines 26-28) are available in the device, i.e., checking if the app is running on these emulators. If the running environment is not one of the hard-coded emulators, the HSO will be performed. Otherwise, benign string operations are executed (lines 17-19).

    \item \textbf{Compatibility Issues in Android apps:}
    
I conducted a preliminary study investigating the test case coverage in the Android framework. Specifically, I downloaded the source code of AOSP from API level 21 to 30 and then calculated the number of public APIs\footnote{The APIs in \texttt{platform/frameworks/base} path.} and their corresponding unit test cases provided by Google. Our results revealed that on average less than 30\% of Android framework APIs have provided test cases in each API level, indicating the Android framework has a poor test case coverage.
When more APIs are provided with unit test cases, more compatibility issues of APIs will likely be identified during regression testing. This will enable them to be fixed at an earlier stage to avoid the introduction of compatibility issues in the first place. 
To this end, I proposed to effectively and efficiently detect compatibility issues through a dynamic testing approach that fulfills its objective by automatically generating valid test cases by mining API usages from real-world Android apps.

\end{itemize}

\subsection{Research objectives}

There still widely exist many undetected security and compatibility issues in Android applications. To tackle these, I developed several automatic tools in detecting these issues to the following several aspects:
\begin{enumerate}
\item Sensor leaks in Android apps can be abused by attackers to implement malicious activities, i.e., inferring actual password based on motion sensor data. Also, there is no approaches to automatically detect sensor leaks in Android apps.
\item Malware writers regularly update their attack mechanisms to hide malicious behavior implementation. Such behaviors will only be triggered under special circumstances such as at a specific location or in a certain time period. This poses two problems to current research techniques: static analysis approaches, given their over-approximations, can report an overwhelming number of false alarms, while dynamic approaches will miss those behaviors that are hidden through evasion techniques. 
\item Despite being one of the largest and most popular projects, the official Android framework has only provided test cases for less than 30\% of its APIs.
Such a poor test case coverage rate has led to many compatibility issues that can cause apps to crash at runtime on specific Android devices, resulting in poor user experiences for both apps and the Android ecosystem.
\end{enumerate}
Three key research objectives are raised regarding the above three reasons as follows: 
\begin{itemize}
    \item \textbf{Research Objective 1:} \textit{Characterizing Sensor Leaks in Android Apps. An automatic approach should be proposed to detect sensor leaks in Android apps before publishing them onto app markets. Specifically, I needed to pinpoint privacy leaks that are triggered by sensor-related fields and methods. In addition, the sensor type information should be identified to further help security analysts to understand sensor leaks.}
    
    \item \textbf{Research Objective 2:}
    \textit{Demystifying Hidden Sensitive Operations in Android apps. Static code analysis should be applied to uncover hidden sensitive operations that will only be triggered under special circumstances such as at a specific location or in a certain time period. Additionally, I needed to provide details aiming at helping security analysts understand why a given hidden sensitive operation is flagged as such.}
    
    \item \textbf{Research Objective 3:}
    \textit{Mining Android API Usage to Generate Unit Test Cases for Pinpointing Compatibility Issues. Firstly, static code analysis techniques should be applied to mine the API usages from real-world applications. Then, after constructing test cases for Android APIs, the dynamic execution of such tests on Android devices can help to pinpoint compatibility issues, including not only signature-based but also semantics-based compatibility issues.}
\end{itemize}

\section{Thesis Scope}

This thesis focuses on providing  more precise and comprehensive results in detecting security/compatibility issues in Android applications. 
I aim to develop automatic tools that can tame these issues in an automatic way.
Considering the state-of-the-art tools are not sufficient to detect all sorts of security/compatibility issues, it is therefore imperative to propose novel techniques to address the aforementioned research objectives respectively: 

\begin{itemize}
    \item {\sc \textbf{SEEKER} --} Leverages static analysis to automatically detect privacy leaks originated from Android sensors -- Research Objective 1
    
    \item {\sc \textbf{HiSenDroid} --} I leveraged control flow and data flow analyses to identify the unique code level characteristics of hidden sensitive operations  -- Research Objective 2
    
    \item {\sc \textbf{JUnitTestGen} --} I performed field-aware, inter-procedural backward data-flow analysis to automatically generate unit test cases for APIs based on their existing usages  -- Research Objective 3
\end{itemize}

\noindent{\sc \textbf{SEEKER}}\textbf{ (Chapter3): }
Despite being needed to support the implementation of many diverse Android apps, mobile phone sensors can also be abused to achieve malicious behaviors. There have been many reports of  apps that exploit sensors in Android devices to conduct malicious activities. For example, sensor data are known to be leaked out because they are not protected by any permissions in Android. There is hence a strong need to regulate the usage of mobile sensors so as to keep them from being exploited by malicious attackers. However, despite the fact that various efforts have been put in achieving this, i.e., detecting privacy leaks in Android apps, I have not yet found approaches to automatically detect sensor leaks in Android apps.

To fill the gap, I have presented a novel tool, SEEKER~\cite{sun2021characterizing}, for characterizing sensor leaks in Android apps. Specifically, I extended the famous FlowDroid~\cite{arzt2014flowdroid} tool to detect sensor-based data leaks in Android apps. SEEKER conducts sensor-focused static taint analyses directly on the Android apps’ bytecode and reports not only sensor-triggered privacy leaks but also the sensor types involved in the leaks. 

Our experimental results on a large scale of real-world Android apps indicate that our tool is effective in identifying all types of potential sensor leaks in Android apps. Our tool is not only capable of detecting sensor leaks, but also pinpointing general privacy leaks that are triggered by class fields. 
Although there are related works on sensor usage analysis, to the best of our knowledge, there is no other work that thoroughly analyses Android sensor leakage. Unlike previous works, our tool is the first one  to characterize all kinds of sensor leaks in Android apps. I extended FlowDroid for supporting field sources detection (i.e., merged to FlowDroid via pull \#385 on Github\cite{FlowDroidMerge}), which I believed could be adapted to analyze other sensitive field-triggered leaks.
To benefit our fellow researchers and practitioners towards achieving this, I have made our approach open source at the following Github site.

\begin{itemize}
    \item \textit{This work has led to a research paper -- \textbf{Characterizing Sensor Leaks in Android Apps}, which has been published in the IEEE 32nd International Symposium on Software Reliability Engineering (ISSRE) in 2021.} 
\end{itemize}

\noindent{\sc \textbf{HiSenDroid}}\textbf{ (Chapter4): }
The emergence of many different malware detection techniques has stimulated malware attackers into being more innovative to increasingly better hide malicious behaviour, in order to bypass static code analysis (e.g., via obfuscation) and even dynamic detection (e.g., sensing of sandbox execution). 
In practice, sophisticated code obfuscation techniques \cite{moser2007limits} are being leveraged by attackers to hide their malicious program behavior, leading to false negatives in most static analyses thus resulting in imprecise and unsound results.

To cope with such hidden malicious behaviors, I presented to the community a prototype tool called HiSenDroid, which performs a static code analysis to uncover hidden sensitive operations that will only be triggered under special circumstances such as at a specific location or in a certain time period. Additionally, HiSenDroid goes one step deeper to provide details aiming at helping security analysts understand why a given hidden sensitive operation is flagged as such.
Experimental results over 20,000 apps, including both malicious and benign apps, show that hidden sensitive operations are indeed quite frequently presented in Android apps and HiSenDroid is effective in automatically discovering them.
I further experimentally showed that, with FlowDroid, some of the hidden sensitive behaviours would eventually lead to private data leaks. Those leaks would have been hard to spot either manually among the large number of false positives reported by the state-of-the-art static analyzers, or by dynamic tools. Overall, by putting the light on hidden sensitive operations, HiSenDroid helps security analysts in validating potential sensitive data operations, which would be previously unnoticed.

\begin{itemize}
    \item \textit{This work has led to a research paper -- \textbf{Demystifying Hidden Sensitive Operations in Android apps}, which has been published in the ACM Transactions on Software Engineering and Methodology (TOSEM) in 2022.}
\end{itemize}

\noindent{\sc \textbf{JUnitTestGen}}\textbf{ (Chapter5): }
Despite being one of the largest and most popular projects, the official Android framework has only provided test cases for less than 30\% of its APIs, leaving the majority of APIs uncovered. This poor test coverage of Android APIs has led to serious compatibility issues in the Android ecosystem, as recently shown~\cite{li2018cid,wei2016taming,mutchler2016target,zhang2015compatibility,ham2011mobile,huang2018understanding}.

To the best of our knowledge, the state-of-the-art works targeting compatibility issue detection leverage static analysis techniques to achieve their purpose.
However, as known to the community, the static analysis will likely yield false positive results as it has to make some trade-offs when handling complicated cases (e.g., object-sensitive vs. object-insensitive).
In addition, the static analysis will also likely suffer from soundness issues because some complicated features (e.g., reflection, obfuscation, and hardening) are difficult to be handled~\cite{sun2021taming,samhi2022jucify}. Moreover, static app analysis can only be leveraged to perform post-momentum analysis (i.e., after the compatibility issues are introduced to the community). 
They cannot stop the problems from being distributed into the community -- many Android apps, including very popular ones, still suffer from compatibility issues.

To mitigate this problem, I presented a prototype tool, JUnitTestGen, that attempts to automatically generate test cases for Android APIs based on their practical usage in real-world apps. Specifically, after locating existing API usages in real-world Android apps, JUnitTestGen performs field-aware, inter-procedural backward data-flow analysis to infer the API caller instance and its parameter values.
JUnitTestGen then leverages the inferred information to reconstruct a minimal executable code snippet for the API under testing.
Experimental results on thousands of Android apps show that JUnitTestGen is effective in generating test cases for Android APIs. It achieves an 80.4\% of success rate in generating valid test cases. 
These test cases subsequently allow our approach to pinpoint various types of compatibility issues, outperforming a state-of-the-art generic test generation tool named EvoSuite, which can only generate test cases to reveal a small subset of compatibility issues.
Furthermore, I demonstrated the usefulness of JUnitTestGen by comparing it against a state-of-the-art static analysis-based compatibility issue detector called CiD. JUnitTestGen is able to mitigate CiD's false-positive results and go beyond CiD's capability (i.e., detecting compatibility issues induced by APIs' signature changes) to detect compatibility issues induced by APIs' semantic changes.

\begin{itemize}
    \item \textit{This work has led to a research paper -- \textbf{Mining Android API Usage to Generate Unit Test Cases for Pinpointing Compatibility Issues}, which has been published at the 37th IEEE/ACM International Conference on Automated Software Engineering (ASE) in 2022.}
\end{itemize}

\section{Key Thesis Contributions}

This Ph.D. thesis proposes three novel tools, i.e., {\sc SEEKER}, {\sc HiSenDroid}, and {\sc JUnitTestGen} to better detect specific security and compatibility issues in Android applications.
These three tools deal with Android app security/compatibility issues from three perspectives: (i) helping conduct sensor-focused static taint analyses directly on the Android apps' bytecode and reports not only sensor-triggered privacy leaks but also the sensor types involved in the leaks; (ii) demystifying Hidden Sensitive Operations in Android apps, helping security analysts in validating potential sensitive data operations, which would be previously unnoticed; (iii) mining Android API usage to generate unit test cases for pinpointing compatibility issues. 

Chapter3 introduces the {\sc \textbf{SEEKER}}, a tool extends the famous FlowDroid tool to detect sensor-based data leaks in Android apps, I have made the following contributions: 

\begin{enumerate}
    \item I have designed and implemented a prototype tool, SEEKER (\underline{Se}nsor l\underline{e}a\underline{k} find\underline{er}), that leverages static analysis to automatically detect privacy leaks originated from Android sensors. Although there are related works on sensor usage analysis, to the best of our knowledge, there is no other work that thoroughly analyses Android sensor leakage. Unlike previous works, our tool is the first one  to characterize all kinds of sensor leaks in Android apps.
    
    \item I applied SEEKER to analyze both malware and benign apps at a large scale. Our results show many sensor leaks that are overlooked by the state-of-the-art static analysis tool.
    
    \item I have demonstrated the effectiveness of our tool by evaluating the sensor leaks it highlights.

    \item I extended FlowDroid for supporting field sources detection (i.e., merged to FlowDroid via pull \#385 on Github\cite{FlowDroidMerge}), which I believed could be adapted to analyze other sensitive field-triggered leaks. To benefit our fellow researchers and practitioners towards achieving this, I have made our approach open source at the following Github site.
    
\end{enumerate}

Chapter4 introduces the {\sc \textbf{HiSenDroid}}, a tool for demystifying Hidden Sensitive Operations in Android apps, in this work I have made the following contributions: 

\begin{enumerate}
    \item I proposed using a static analysis approach to discover hidden sensitive operations that are not exposed to the state-of-the-art static and dynamic analysis tools in Android apps. To this end, I leverage control flow and data flow analyses to identify the unique code level characteristics of hidden sensitive operations.
    
    \item I designed and implemented a prototype tool HiSenDroid for analyzing hidden sensitive operations. I release HiSenDroid as an open source project~\cite{HiSenDroid} for supporting security analysts in their analysis needs and fostering further researches in this direction.
    
    \item I evaluated HiSenDroid on a large-scale dataset that contains 10,000 benign and 10,000 malware samples, and discovered emerging anti-analysis techniques employed by malware samples, such as fulfilling certain restrictions related to \emph{time}, \emph{location}, \emph{SMS message}, \emph{system properties}, \emph{package manager}, and other logics. 
    
    \item With the help of FlowDroid~\cite{arzt2014flowdroid}, a static taint analyzer, I further experimentally showed that HSOs have been recurrently leveraged by attackers to leak sensitive user information. 
\end{enumerate}

Chapter5 introduces the {\sc \textbf{JUnitTestGen}}, a tool for mining Android API Usage to Generate Unit Test Cases for Pinpointing Compatibility Issues, in this work I have made the following contributions: 
\begin{enumerate}
    \item I have designed and implemented a prototype tool JUnitTestGen that leverages a novel approach to automatically generate unit test cases for APIs based on their existing usages.

    \item I have set up a reusable testing framework for pinpointing API-induced compatibility issues by automatically executing a large set of unit test cases on multiple Android devices. 

    \item I have demonstrated the effectiveness of JUnitTestGen by i) generating valid test cases for Android APIs and pinpointing problematic APIs that could induce compatibility issues if accessed by Android apps, ii) outperforming state-of-the-art tools on real-world apps in detecting a wider range of compatibility issues.

\end{enumerate}

With the help of our proposed tools, (i) developers can use our tool {\sc SEEKER} to automatically detect sensor leaks; our fellow researchers and practitioners can also use our tool to analyze other sensitive field-triggered leaks; (ii) by putting the light on hidden sensitive operations, security analysts can use our tool {\sc HiSenDroid} in validating potential sensitive data operations, which would be previously unnoticed; (iii) Android app developers and google developers can use our tool {\sc JUnitTestGen} to detect zero-day compatibility issues, including not only signature-based but also semantics-based compatibility issues.

\section{Structure of the thesis}
In this thesis, Chapter 1 introduces the introduction and motivation about the thesis.
Chapter 2 surveys the related literature which includes specific security and compatibility issues in Android applications, such as sensor leaks, hidden sensitive operations, signature-based and semantics-based compatibility issues. 
In Chapter 3, I present {\sc \textbf{SEEKER}}, an approach to  automatically detect sensor leaks in Android applications.
In Chapter 4, I present  {\sc \textbf{HiSenDroid}}, an approach to demystify Hidden Sensitive Operations in Android apps.
In Chapter 5, I present {\sc \textbf{JUnitTestGen}}, a tool that mines Android API usage to generate unit test cases for pinpointing compatibility issues.
Chapter 6 gives the conclusion, some limitations and future research directions.

%% file: chapter2.tex
\chapter{Literature Review}
In this chapter, I provide the necessary background to understand the purpose, key concerns, and technical details of the three research studies I conducted in this PhD research. 
I conducted a systematic literature review to identify state-of-the-art works in detecting security and compatibility issues in Android applications. 
I categorized the relevant research into three categories: static analysis of sensor leak detection on Android apps, evasive techniques detection, and compatibility issues detection in Android apps. 
It can help the readers to get a systematic overview of the work in general and specifically to know future research directions. 

\section{Static analysis for sensor leak detection}

\textbf{Android sensor usage: } Android sensor usage has long been analyzed in software security research. Several related works \cite{zhu2013sensec, ba2020learning,xu2012taplogger,miluzzo2012tapprints,liu2015exploring,aviv2012practicality,cai2011touchlogger,owusu2012accessory,lee2015multi} have indicated that embedded sensors can be intentionally misused by malicious apps to compromise privacy. Ba et al. \cite{ba2020learning} proposed a side-channel attack that adopts accelerometer data to eavesdrop on the speaker in smartphones. Xu et al. \cite{xu2012taplogger} have shown that it is feasible to infer user's tap inputs using its integrated motion sensors. Liang Cai et al.\cite{cai2011touchlogger} revealed that confidential data could be leaked when motion sensors, such as accelerometers and gyroscopes, are used to infer keystrokes.
Also, Lin et al.\cite{lin2012new} demonstrated that the orientation sensor of the smartphone could be utilized to detect users' unique gestures used to hold and operate their smartphones.

Android Sensor misuse is one of the major causes of privacy leaks and security issues on the Android platform. Zhu et al. \cite{zhu2013sensec} collected sensor data from accelerometers, gyroscopes and magnetometers and constructs users' gesture based on these data. Their work indicates that it is feasible to get access to sensory data for personal identification and behaviour analysis. Liu et al. \cite{liu2015exploring} demonstrated the most frequently used sensors in Android devices and revealed their usage patterns through backwards tracking analysis. They further investigated sensor data propagation path for accurately characterizing the sensor usage.

\textbf{Software side-channels attacks: }
Many previous studies \cite{chang2009inferring, lester2004you, liu2009uwave, ravi2005activity, allen2006classification, schlegel2011soundcomber} explored password inference through specific sensors on smartphones. Owusu et al. \cite{owusu2012accessory} showed that accelerometer values could be used as a powerful side channel attack to figure out the password for a touchscreen keyboard. Cai et al.\cite{cai2011touchlogger} provided insights into how motion sensors, such as accelerometers and
gyroscopes, can be used to infer keystrokes. Cai et al. \cite{cai2009defending} found that mobile phone sensors are inadequately protected by the Android permission system and thus can raise serious privacy concerns. Enck et al. \cite{enck2014taintdroid} developed TaintDroid that takes sensor information (i.e., location and accelerometer) as sources to detect privacy leaks. Mehrnezhad et al. \cite{mehrnezhad2018stealing} show that orientation sensor can be stealthily listened to without requesting any permission, contributing for attackers to infer the user’s PIN. However, these works emphasize the challenges facing the detection of sensor-sniffing apps or only provided specific attacks by using sensor data. None of them can systematically characterize data leaks in all kinds of sensors.

\textbf{Static analysis of Android apps for security: }
Android users have suffered from privacy leaks \cite{li2017static, kong2018automated, samhi2021raicc, octeau2016combining}. Several solutions have been proposed for detecting such data leaks through static taint analysis~\cite{gao2020borrowing, li2015apkcombiner, yang2017characterizing}. For example, Arzt et al. \cite{arzt2014flowdroid} developed FlowDroid, a context, flow, field and object-sensitive static analysis tool for detecting potential data leaks in Android Apps. Based on Soot \cite{vallee2010soot}, FlowDroid relies on pre-defined knowledge to pinpoint taint flows between source and sink APIs. Zhang et al. \cite{zhangcondysta} developed ConDySTA, a dynamic taint analysis approach, as a supplement to static taint analysis by introducing inaccessible code and sources that help reduce false negatives. Further, Li et al. \cite{li2015iccta} presented IccTA, which can precisely perform data-flow analysis across multiple components in Android apps. Klieber et al. \cite{klieber2014android} augment the FlowDroid and Epicc\cite{octeau2013effective} analyses by tracking both inter-component and intra-component data flow in Android apps.
However, none of these tools focuses on the leaks that originate from sensors. 

The most similar work to our proposed SEEKER tool is SDFDroid\cite{liu2018discovering}, which provides sensor usage patterns through data flow analysis. As a static analysis approach, however, it focuses on different research objectives compared to SEEKER. For example, SDFDroid reveals sensor usage patterns while our work explores how and where the sensor data are leaked. On the other hand, SDFDroid applies a static approach to extract sensor data propagation path to construct sensor usage patterns through clustering analysis. In contrast to SDFDroid, SEEKER provides detailed privacy leaks caused by misuse of sensor data, which haven't been found by SDFDroid. 

\section{Evasive techniques detection}

Hidden sensitive operations have long existed in Android malware as evasive technologies have widely been used by attackers to hide malicious behaviors.
The research community has hence proposed various approaches to tackle these issues. I now discuss some of the representative works from two angles: the evasive techniques that have been proposed to hide malicious code from being identified, and the detection methods proposed to pinpoint such evasive techniques.

\textbf{Evasive Techniques: } There have been a number of research works conducted on hiding malicious behavior from detection, most of which focus on evading the dynamic test platforms such as virtual machines and emulators. Early works target the Windows platform \cite{chen2008towards}, while recently the trend has been moved to Android \cite{petsas2014rage,vidas2014evading,jing2014morpheus,diao2016evading,costamagna2018identifying,liu2021deep,liu2022explainable}. These evasive techniques detect the presence of a simulated environment by either looking into the system properties of the testing platform (e.g., system fingerprints, hardware capabilities, etc.) \cite{petsas2014rage,vidas2014evading,costamagna2018identifying}, or leveraging a reverse Turing test that examines if the app interacts with a human user \cite{diao2016evading}. For instance, Diao et al. \cite{diao2016evading} observed that programmed interaction has specific patterns of input and interaction frequency, which is different from real users. Overall, the evasive techniques usually hide malicious activities in an \emph{if-then-else statement}. The hidden malicious behavior will only be set off when certain conditions are fulfilled (e.g., not in an emulator); otherwise dummy benign operations are triggered. The prevalence of such evasive techniques motivated us to investigate the HSOs in Android apps and propose HiSenDroid to detect them.

\textbf{Detection of Evasive Techniques: } The pervasive evasive techniques (e.g., anti-emulator techniques) have motivated the research community to take countermeasures. Great effort has been spent on detecting known types of hidden behaviors that hampers the dynamic analysis process. These works include detecting anti-emulator techniques \cite{balzarotti2010efficient, lindorfer2011detecting, kirat2014barecloud,kirat2015malgene} and generic logic-bombs \cite{crandall2006temporal,brumley2008automatically,zheng2012smartdroid,fratantonio2016triggerscope,papp2019towards}. The approaches used for detecting anti-emulator techniques compare the behavioral deviation of the tested apps on the various environments when feeding them the same input. The fundamental idea is that if the app behaves differently in different environments, it is likely trying to evade one or more analysis platforms (usually referred to as bare-metal analysis in the literature) \cite{balzarotti2010efficient, lindorfer2011detecting, kirat2014barecloud,kirat2015malgene}. While these early works investigate a critical category of hidden operations (i.e., anti-emulator), their proposed methods lack generalization that cannot be applied to detect other types of hidden operations emerging recently.

Besides the detection of anti-emulator techniques, several works focused on uncovering other trigger-based behaviors. These approaches leverage symbolic execution or static code analysis and instrumentation to expose the hidden branches in an \emph{if-then-else statement} \cite{crandall2006temporal,brumley2008automatically,zheng2012smartdroid,fratantonio2016triggerscope,papp2019towards}. As examples, Zheng et al. \cite{zheng2012smartdroid} proposed to leverage a static analysis approach to retrieve all UI related events, and use dynamic testing to trigger them and log the invocation of sensitive APIs. Unlike HiSenDroid that leverages static analysis, the dynamic analysis based approach introduces significant system- and time-overhead. The coverage of the dynamic analysis is also in question.
Fratantonio et al. \cite{fratantonio2016triggerscope} proposed TriggerScope to detect hidden triggered behaviors based on the observation that certain triggers (i.e., time, location, and SMS related triggers) always involve the comparison of specific types of input (i.e., system time, system location, and received SMS). Symbolic execution is then leveraged to detect such narrow conditions. While TriggerScope is effective in detecting the above-mentioned three types of logic bombs, it cannot be generalized to detect hidden operations triggered by other types of conditions, such as system property, which has been found pervasive in Android apps.

Similar to our proposed HiSenDroid, another line of work attempts to detect unknown types of trigger-based behaviors~\cite{pan2017dark},~\cite{wang2017droid}.
A prominent example is HSOMiner~\cite{pan2017dark}, which extracts static characteristics of hidden behaviors as features and trains a machine learning model to identify the code blocks that observe similar patterns. The major differences between our work and HSOMiner are twofold. First, HSOMiner requires a large number of manually labelled training samples, which involves extensive human experts' effort. Its performance also heavily relies on the manually labelled training data, which is prone to errors. Our method, on the other hand, is an automatic process without human intervention. Second, HSOMiner, as a machine learning based approach, lacks explanations of the decisions. In contrast, our static code analysis based approach outputs the full call traces of detected HSOs, and provides more detailed information for further analysis.

\section{Compatibility issues detection}

Android API evolution is a critical issue in software maintenance ~\cite{nagappan2016future,li2018moonlightbox,martin2016survey,li2016accessing,oliveira2018android,lamothe2020a3,li2018characterising,zhou2016api,dig2005role,kapur2010refactoring,liu2022deep,sun2021characterizing}. McDonnell et al. ~\cite{mcdonnell2013empirical} have shown that the Android
system updates 115 APIs per month on average, while app developers usually adopt the new APIs at a much more slower rate. The slow adoption of API updates may raise various issues, such as security and compatibility. An empirical study on StackOverflow conducted by Linares-Vasquez et al. ~\cite{linares2014api} suggests that API updates would trigger more discussions, especially if APIs are removed from the Android system. They also revealed that users are in more favour of apps that use less fault and change-prone APIs \cite{linares2013api, bavota2014impact}, as these apps would likely introduce fewer failures, crashes and other bugs.

Android Fragmentation is one of the major causes of compatibility and security issues on the Android platform. Han et al. \cite{han2012understanding} analyzed the bug reports on two Android smartphone vendors, HTC and Motorola. Wu et al. ~\cite{wu2013impact} studied ten customized Android images across five vendors. Their findings suggest that system customization brings serious security risks through over-privileged system apps and introduced vulnerabilities. Zhou et al. \cite{zhou2014peril} further investigated the security issues in customized Android systems, and proposed a tool to detect such issues. Liu et al. \cite{liu2014characterizing} studied the characteristics of bugs in Android apps and observed that certain Android performance bugs are tied to specific devices.

Android developers have long been suffered from compatibility issues due to the fast-evolving and fragmented nature of the Android ecosystem \cite{xia2020android,kamran2016android,li2017static,nayebi2012state,zhao2022towards}. 
Researchers have proposed several solutions for detecting compatibility issues of Android APIs. Wei et al. \cite{wei2016taming, wei2018understanding} conducted an empirical study to investigate fragmentation-induced API compatibility issues and proposed a tool named FicFinder to detect such APIs. FicFinder identifies APIs with compatibility issues based on heuristic rules manually derived from a limited number of Android apps, which is expected to introduce high false negatives (i.e., missing undiscovered compatibility issues). Subsequently, several new works have been proposed to leverage data-driven techniques that automatically mine compatibility issues from various sources such as Android code base and real-world apps. Li et al. proposed a tool named CiD to detect potential compatibility issues by mining the history of the Android framework source code. CiD identifies Android APIs' lifetimes and finds if an app's declared supported versions conflict with its used APIs. He et al. \cite{he2018understanding} proposed IctApiFinder that utilizes a similar strategy to extract an API's lifetime usage and introduce inter-procedural data flow analysis to locate incompatible APIs, which reduces the false-positive results produced by CiD.

Comparable to our method, several works have been proposed to mine API usage from real-world apps. Scalabrino et al. \cite{scalabrino2019data, scalabrino2020api} considered the APIs wrapped in a version check condition (e.g., \emph{if (Build.VERSION.SDK\_INT >= 21)}) to potentially have compatibility issues and developed a tool named ACRYL to extract such APIs from real-world apps. However, ACRYL can only detect APIs whose compatibility issue is already known by the developers (i.e., they are enclosed in the version check conditions by the developers), while our method is capable of detecting zero-day compatibility issues that the app developers are not yet aware of, or even Google itself. 
Other generic test generation tools, such as EvoSuite and Randoop\cite{pacheco2007randoop}, are able to generate tests for Java classes. However, these tools do not directly aim at generating tests for APIs, and they have been demonstrated as insufficient in pinpointing compatibility issues because of the lack of API usage knowledge. 

Another work proposed by Wei et al. \cite{wei2019pivot} shares a similar idea to detect compatibility issues that may arise on specific devices introduced by the customization of the Android system by OEMs. They proposed a tool named Pivot that finds APIs enclosed in device-checking statements (e.g., \emph{if (android.os.Build.Model.equals("Nexus 6P"))}) from popular Android apps, and construct API-device correlations to imply that the API may have compatibility issues on the given device. Our method, though not specifically designed for detecting device-related compatibility issues, is also able to detect such issues. Moreover, human effort is involved in validating the API-device correlation identified by Pivot, which is prone to error. Our method, on the other hand, is fully automated without human intervention while having a true-positive rate of 100\%.

%% file: chapter3.tex
\chapter{Characterizing Sensor Leaks in Android Apps}

\begin{tcolorbox}
[width=6.2in]
\small{
\textbf{Xiaoyu Sun}, Xiao Chen, Kui Liu, Sheng Wen, Li Li, and John Grundy. 2021. Characterizing Sensor Leaks in Android Apps. The IEEE 32nd International Symposium on Software Reliability Engineering (ISSRE) 2021.
\url{https://arxiv.org/pdf/2201.06235.pdf}
}
\end{tcolorbox}

\section{Introduction}
As of 1st January 2021, there are nearly three million Android apps available on the official Google Play app store. 
The majority of them (over 95\%) are made freely accessible to Android users and cover every aspect of users' daily life, such as supporting social networking, online shopping, banking, etc.
Many of these functionalities are supported by application interfaces provided by the Android framework, essentially fulfilled by a set of hardware-based sensors~\cite{developerandroid}.
For example, Android apps often leverage  accelerometer sensors to detect the orientation of a given smartphone and user movement, and the temperature sensor to detect the device's temperature.

Despite being needed to support the implementation of many diverse Android apps, mobile phone sensors can also be abused to achieve malicious behaviors.
There have been many reports of  apps that exploit sensors in Android devices to conduct malicious activities.
For example, Adam et al.~\cite{aviv2012practicality} have experimentally shown that the accelerometer sensor could be leveraged as a side-channel to infer mobile users' tap and gesture-based input.
Xu et al.~\cite{xu2012taplogger} have also demonstrated the possibility of this attack by presenting to the community a Trojan application named \emph{TapLogger} to silently infer user's tap inputs based on the device's embedded motion sensors.
Similarly, Schlegel et al.~\cite{schlegel2011soundcomber} have provided another Trojan application called \emph{Soundcomber} that leverages the smartphone's audio sensor to steal users' private information.

These studies have experimentally shown that the leaks of Android sensor data can  cause severe app security issues.
We argue that there is thus a strong need to invent automated approaches to detect such sensor leaks in Android apps before publishing them onto app markets.
To the best of our knowledge, existing works  focus on detecting certain types of sensor usage and its corresponding suspicious behaviors. None of them are designed as a generic approach for systematically revealing data leaks in all types of Android sensors. Also, these works mainly concentrate on discovering and understanding the usage patterns of Android embedded sensors, which do not involve completed data flow analysis to pinpoint sensitive data leaks caused by sensors.

Although many generic approaches to detect privacy leaks in Android apps have been proposed, none can be directly applied to achieve our purpose, i.e., detecting generic sensor leaks in Android apps.
Indeed, the famous FlowDroid tool has been demonstrated to be effective in detecting method-based privacy leaks in Android apps.
It performs static taint analysis on Android apps' bytecode and attempts to locate data-flow paths connecting two methods, i.e., from a \emph{source} to a \emph{sink} method.
Here, \emph{source} refers to such methods that obtain and return sensitive information from the Android framework (e.g., get device id), while \emph{sink} refers to such methods that perform dangerous operations such as sending data to remote servers.
FlowDroid has been designed as a generic approach.
It has provided a means for users to pre-define the needed \emph{source} and \emph{sink} methods.
Unfortunately,FlowDroid does not allow users to configure fields as \emph{sources} so as to support the detection of privacy leaks flowing from \emph{fields} to sensitive operations (i.e., \emph{sink}).
Since sensor data in Android is mostly provided via fields, FlowDroid cannot be directly applied to detect sensor leaks in Android apps.

To address this research gap, we designed and implemented a prototype tool, SEEKER, to automatically detect sensor data leaks in Android apps.
We  extend the open-source tool FlowDroid to support field-triggered sensitive data-flow analyses. Our new
SEEKER further performs a detailed static code analysis to infer the sensor types involved in the sensitive data-flows as the leaked sensor data is not directly associated with the sensor type. (we detail this challenge in Section~\ref{subsec:type}).
We then apply SEEKER to detect and characterize sensor leaks in real-world Android apps.
Based on 40,000 randomly selected Android apps, including 20,000 benign apps and 20,000 malicious apps, our experimental results show that SEEKER is effective in detecting sensor leaks in Android apps.
We also find that malware is more interested in obtaining and leaking sensor data than benign apps, and Accelerometer and Magnetic are among the most targeted sensors by those malicious apps.

In summary, this work makes the following three main contributions:
\begin{itemize}[leftmargin=*]
\item We have designed and implemented a prototype tool, SEEKER (\underline{Se}nsor l\underline{e}a\underline{k} find\underline{er}), that leverages static analysis to automatically detect privacy leaks originated from Android sensors.

\item We apply SEEKER to analyze both malware and benign apps at a large scale. Our results show many sensor leaks that are overlooked by the state-of-the-art static analysis tool.

\item We have demonstrated the effectiveness of our tool by evaluating the sensor leaks it highlights.

\end{itemize}

The rest of this chapter is organized as follows.

Section~\ref{SEEKER:related} presents key related work on question generation and relevant techniques.
Section~\ref{SEEKER:pre} presents the motivation of this study.
Section~\ref{SEEKER:approach} presents the details of our approach for the question generation task in Stack Overflow.

Section~\ref{SEEKER:eval} presents the experimental setup, the baseline methods and the evaluation metrics used in our study.
Section~\ref{SEEKER:results} presents the detailed research questions and the evaluation results under each research question.
Section~\ref{SEEKER:discussion} presents the contribution of the paper and discusses the strength and weakness of this study.
Section~\ref{SEEKER:threats} presents threats to validity of our approach.
Section~\ref{SEEKER:con} concludes the paper with possible future work.

\section{Related Work}
\label{SEEKER:related}
\subsection{Android sensor usage}
Android sensor usage has long been analyzed in software security mechanisms. Related works \cite{zhu2013sensec, ba2020learning,xu2012taplogger,miluzzo2012tapprints,liu2015exploring,aviv2012practicality,cai2011touchlogger,owusu2012accessory,lee2015multi} have indicated that embedded sensors can be intentionally misused by malicious apps for privacy compromise. Ba et al. \cite{ba2020learning} proposed a side-channel attack that adopts accelerometer data to eavesdrop on the speaker in smartphones. Xu et al. \cite{xu2012taplogger} have shown that it is feasible to infer user's tap inputs using its integrated motion sensors. Liang Cai et al.\cite{cai2011touchlogger} revealed that confidential data could be leaked when motion sensors, such as accelerometers and gyroscopes, are used to infer keystrokes.
Also, Lin et al.\cite{lin2012new} demonstrated that the orientation sensor of the smartphone could be utilized to detect users' unique gesture to hold and operate their smartphones.

Android Sensor misuse is one of the major causes of privacy leaks and security issues on the Android platform. Zhu et al. \cite{zhu2013sensec} collected sensor data from accelerometers, gyroscopes and magnetometers and constructs users' gesture based on these data. Their work indicates that it is feasible to get access to sensory data for personalized usage. Liu et al. \cite{liu2015exploring} demonstrated the most frequently used sensors in Android devices and revealed their usage patterns through backward tracking analysis. They further investigate sensor data propagation path for accurately characterizing the sensor usage \cite{liu2018discovering}. Their findings suggest that the accelerometer is the most frequently used sensor  and the sensor data are always used in local codes.

\subsection{Software side-channels attacks.}
Many previous studies \cite{chang2009inferring, lester2004you, liu2009uwave, ravi2005activity, allen2006classification, schlegel2011soundcomber} explored password inference through specific sensors on smartphones. Owusu et al. \cite{owusu2012accessory} showed that accelerometer values could be used as a powerful side channel to figure out the password on a touchscreen keyboard. Cai et al.\cite{cai2011touchlogger} provided insights of how motion sensors, such as accelerometers and
gyroscopes, can be used to infer keystrokes. Cai et al. \cite{cai2009defending} found that mobile phone sensors are inadequately protected by permission system so that it can raise serious privacy concerns. Enck et al. \cite{enck2014taintdroid} developed TaintDroid that takes sensor information (i.e., location and accelerometer) as sources to detect privacy leaks. Mehrnezhad et al. \cite{mehrnezhad2018stealing} show that orientation sensor can be stealthily listened to without requesting any permission, contributing for attackers to infer the user’s PIN. However, these works emphasize the challenges facing the detection of sensor-sniffing apps or only provided specific attacks by using sensor data. None of them can systematically characterize data leaks in all kinds of sensors.

\subsection{Static analysis on Android apps.}
Android users have long been suffered from privacy leaks \cite{li2017static, kong2018automated, samhi2021raicc, octeau2016combining}. Several solutions have been proposed for detecting such data leaks through static taint analysis~\cite{gao2020borrowing, li2015apkcombiner, yang2017characterizing}. For example, Arzt et al. \cite{arzt2014flowdroid} developed FlowDroid, a context, flow, field and object-sensitive static analysis tool for detecting potential data leaks in Android Apps. Based on Soot \cite{vallee2010soot}, FlowDroid relies on pre-defined knowledge to pinpoint taint flows between source and sink APIs. Zhang et al. \cite{zhangcondysta} developed ConDySTA, a dynamic taint analysis approach, as a supplement to static taint analysis by introducing inaccessible code and sources that help reduce false negatives. Further, Li et al. \cite{li2015iccta} presented IccTA, which can precisely perform data-flow analysis across multiple components for Android apps. Klieber et al. \cite{klieber2014android} augment the FlowDroid and Epicc\cite{octeau2013effective} analyses by tracking both inter-component and intra-component data flow in Android apps.
However, none of these tools concerns the leaks that originated from sensors. Apart from that, our tool only takes the sensor-related code into account, which cost less time by pruning the control flow graph.

The most similar work to ours is SDFDroid\cite{liu2018discovering}, which provides the sensor usage patterns through data flow analysis. As a static approach, however, it focuses on different research object compared to SEEKER. For example, SDFDroid reveals sensor usage patterns while our work explores how and where the sensor data are leaked. On the other hand, SDFDroid applies a static approach to extract sensor data propagation path to construct sensor usage patterns through clustering analysis. In contrast to SDFDroid, SEEKER provide detailed privacy leaks caused by misuse of sensor data, which haven't been found by SDFDroid.

\section{Motivation}
\label{SEEKER:pre}
Sensors have been widely adopted for launching side-channel attacks against smart devices \cite{sikder2021survey}.   
Table ~\ref{tab:sensor_attacks} summarizes a diverse set of sensor-based attacks targeting smartphones and smartwatches. Since accessing sensitive sensor data does not require any security checks (e.g., permission check), attackers can easily trigger malicious behaviors by making use of such data. As revealed in the table, generally, sensor leakage are performed with the aim of (1) keystroke inference, (2) task inference (refers to a type of attack which reveals the information of an on-going task or an application in a smart device), (3) location inference, and (4) eavesdropping. For example, motion and position sensors can be exploited for keystroke inference, leading to severe privacy leaks such as passwords, credit card information, etc. 
Light sensor is found to eavesdrop acoustic signals in the vicinity of the device, causing private information leak. Magnetic sensors can be exploited to compromise electromagnetic emanations, which would affect the confidentiality of the devices.

\begin{table}[!h]
\small
\centering
\caption{Examples of Sensor-based Cybersecurity attacks.} 
\vspace{2mm}
\label{tab:sensor_attacks}
\resizebox{\linewidth}{!}{
\begin{tabular}{l l l } 
\hline
 Sensor Category &  Sensor Type & Attack Description \\
 \hline
 \multirow{23}{*}{Motion sensor} &
 Accelerometer & sniffing smartwatch passwords \cite{lu2018snoopy} \\
 & Accelerometer, Gyroscope & Text Inference \cite{hodges2018reconstructing}\\
 & Accelerometer, Gyroscope & Motion-based keystroke inference \cite{cai2012practicality} \\
 & Accelerometer, Gyroscope & Keystroke inference on Android \cite{al2013keystrokes} \\
 & Accelerometer & Accelerometer side channel attack \cite{aviv2012practicality} \\
 & Accelerometer & Touchscreen area identification \cite{owusu2012accessory} \\
 & Accelerometer & Decoding vibrations from nearby keyboards  \cite{marquardt2011sp}  \\
 & Gyroscope & Single-stroke language-agnostic keylogging \cite{narain2014single} \\
 & Accelerometer, Gyroscope & Inferring Keystrokes on Touch Screen \cite{cai2011touchlogger} \\
 & Accelerometer, Gyroscope & Inferring user inputs on smartphone touchscreens  \cite{xu2012taplogger}\\
 & Accelerometer, Gyroscope & Keystroke Inference \cite{bo2019know} \\
 & Accelerometer & keystrokes Inference in a virtual environment.  \cite{ling2019know}\\
 & Accelerometer, Gyroscope & Risk Assessment of motion sensor \cite{huang2019risk}\\
 & Accelerometer, Gyroscope & Infer tapped and traced user input \cite{nguyen2015using} \\
 & Accelerometer, Gyroscope & Motion-based side-channel attack  \cite{lin2019motion} \\
 & Accelerometer & Keystroke inference with smartwatch \cite{liu2015good}\\
 & Accelerometer & Motion leaks through smartwatch sensors \cite{wang2015mole}\\
 & Accelerometer & Side-channel inference attacks \cite{maiti2018side} \cite{maiti2015smart}\\
 & Accelerometer & Smartphone PINs prediction  \cite{sarkisyan2015wristsnoop} \\
 & Gyroscope & Inferring Mechanical Lock Combinations \cite{maiti2018towards} \\
 & Accelerometer, Gyroscope & Inference of private information \cite{maiti2018towards} \\
 & Accelerometer, Gyroscope & Typing privacy leaks via side-Channel from smart watch \cite{liu2019aleak}\\
 & Accelerometer, Magnetometer & Input extraction via motion sensor \cite{shen2015input} \\
 & Gyroscope & Recognizing speech \cite{michalevsky2014gyrophone} \\
 \hline
 \multirow{2}{*}{Position sensor} 
 & Magnetic & Compromising electromagnetic emanations  \cite{vuagnoux2009compromising}\\
 & Magnetic & My Smartphone Knows What You Print \cite{song2016my} \\
 & Magnetic & Location detection  \cite{block2018my} \\
\hline
\multirow{1}{*}{Environment sensor} 
 &Light Sensor &  Optical eavesdropping on displays  \cite{chakraborty2017lightspy}\\
\hline
\end{tabular} 
}
\vspace{-4mm}
\end{table}

As a concrete example, Lu et al. \cite{lu2018snoopy} revealed that sensitive intercepting password could be accessed through motion data on the smartwatch's onboard sensors. They proposed \emph{Snoopy}, a password extraction and inference approach via sensor data for PIN attack, which could affect smartwatch users in a non-invasive way.
\emph{Snoopy} extracts the segments of motion data when users entered passwords and then applies deep learning techniques to infer the actual passwords. Figure~\ref{fig:snoopy_example} gives two examples of the differences of the motion sensor data changes when the user swipes or taps a password on a smartwatch.
\emph{Snoopy} demonstrates the feasibility of sensor data leaks by intercepting password information entered on smartwatches. 
Such real-world sensor-enabled attacks motivated us to provide automatic tools for characterizing universal sensor leaks in Android Apps that have been long overlooked.

\section{Approach}
\label{SEEKER:approach}
\begin{figure*}[!h]
    \centering
    \includegraphics[width=0.7\linewidth]{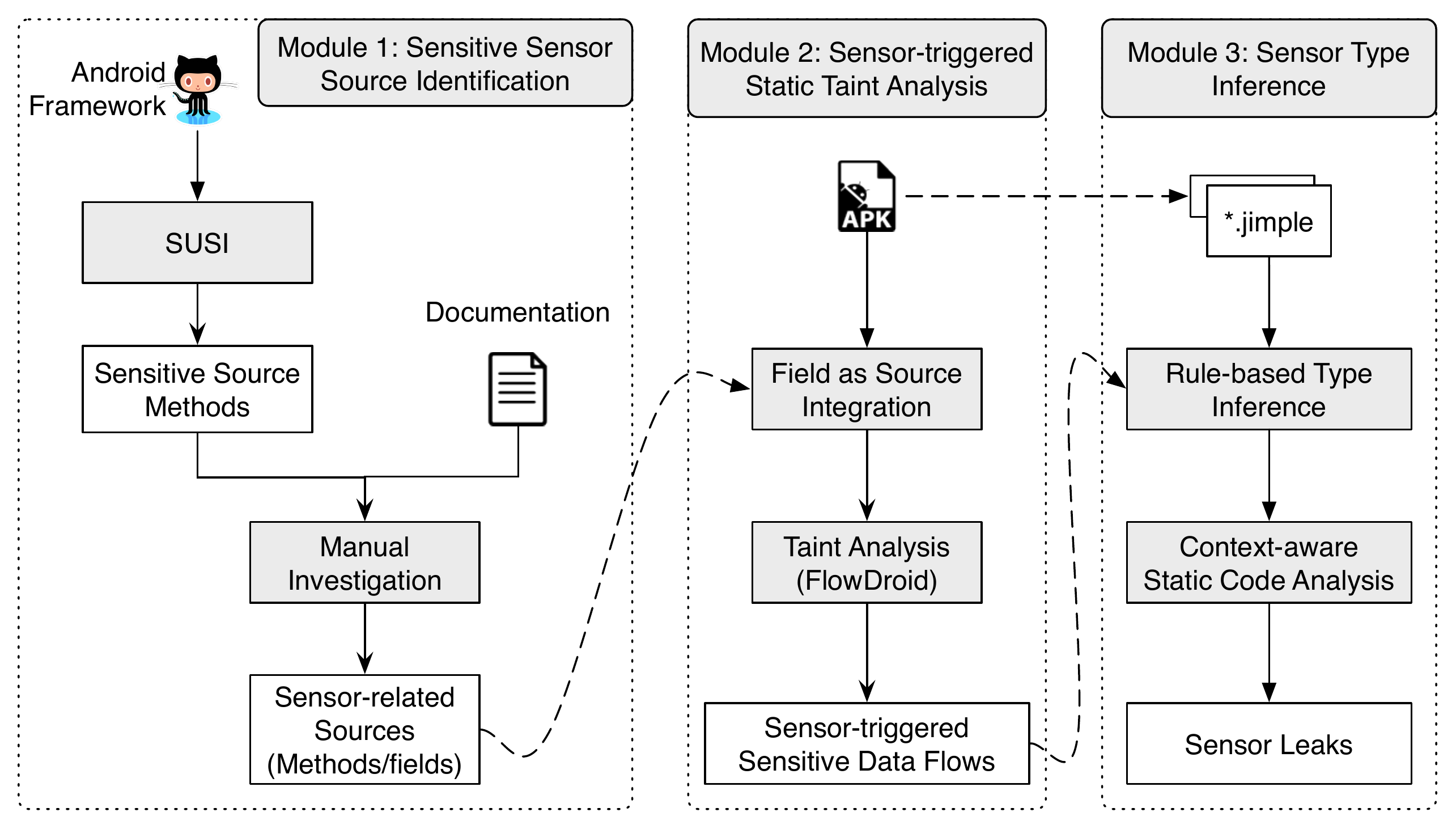}
    \vspace{-3mm}
	\caption{The working process of our approach.}
    \label{fig:methodology}
\end{figure*}

This work aims to automatically detect information leaks of onboard sensors in Android apps. To this end, we design and implement a prototype tool called SEEKER for achieving this purpose. Figure \ref{fig:methodology} describes the overall working process of SEEKER, which is mainly made up of three modules, namely Sensitive Sensor Source Identification, Sensor-triggered Static Taint Analysis and Sensor Type Inference.

\subsection{Sensitive Sensor Source Identification}
\label{subsec:source}
The first module, \emph{Sensitive Sensor Source Identification}, aims to identify sensor-related sources that access and obtain sensitive information related to the device's sensors. As reported by Liu et al.~\cite{liu2018discovering}, Android sensor data can be obtained through invoking sensor-related APIs or directly accessing local fields in which the sensor data is stored.
In this work, we take both of these types into consideration, aiming at pinpointing all the possible sensor-triggered privacy leaks.

To do this we need to identify all the sensor-related sources, including both Android methods and fields.
For Android methods, we use the well known SUSI tool~\cite{arzt2013susi} to obtain sensor-related source methods. 
SUSI is a novel machine-learning guided approach that scans Android API's source code to predict \emph{source} and \emph{sink} methods, based on a training set of hand-annotated sources and sinks.
In this work, we launch SUSI on the latest Android Open Source Project (i.e., AOSP version 11.0) and manually filter out non-sensor related source methods.

To identify sensor-related fields (as sources), there is no existing approach to achieve such a purpose. We resort to a manual process of going through the Android Developers' Documentation to identify source fields storing sensitive sensor information. The identified fields are then discussed and confirmed by the authors by measuring whether leaking such information would potentially expand the attack surface to users' privacy. Finally, we identified 79 fields and 20 methods as the sources. Table \ref{tab:sensor_sources} lists the selected sources that indeed introduce leaks in our experimental dataset. A full list of field and method sources can be found in the \emph{SourcesAndSinks.txt} file of our open-source project\footnote{https://github.com/MobileSE/SEEKER}.

\begin{table}[!h]
\small
\centering
\caption{The list of sensitive sensor sources.}\vspace{2mm}
\label{tab:sensor_sources}
{
\begin{tabular}{l c c} 
\hline
Sensor-related Source & Source Type \\
\hline
SensorEvent\#values & Field  \\
SensorEvent\#timestamp & Field \\
Sensor\#getName() & Method \\
Sensor\#getVendor() & Method  \\
Sensor\#getVersion() & Method  \\
SensorManager\#getDefaultSensor(int) & Method  \\
Sensor\#getMaximumRange() & Method  \\
SensorManager\#getSensorList(int) & Method  \\
Sensor\#getType() & Method  \\
Sensor\#getResolution() & Method  \\
Sensor\#getPower() & Method  \\

\hline

\hline
\end{tabular} 
}
\vspace{-2mm}
\end{table}

\vspace{-1mm}
\subsection{Sensor-triggered Static Taint Analysis}
The ultimate goal of SEEKER is to detect sensor-related data leaks. To this end, we implement the \emph{Sensor-triggered Static Taint Analysis} module that extends state-of-the-art tool FlowDroid \cite{arzt2014flowdroid} to facilitate sensor-related data leak detection. FlowDroid detects data leaks by computing data flows between sources and sinks. FlowDroid defined a sensitive data flow happens when a suspicious “tainted” information passes from a source API (e.g., \texttt{getDeviceId}) to a sink API (e.g., \texttt{sendTextMessage}). 

FlowDroid is a state-of-the-art tool and it provides a highly precise static taint-analysis model, especially for Android applications. However, FlowDroid only takes API statements as sources or sinks, leading to false negatives because of the lack of field-triggered sources. Thus, in this work, we extend FlowDroid by supporting field statement as sources, so as to pinpoint data leaks originated from specific field sources of interest. 

Our preliminary study discovered that certain sensor-related data leaks are sourced from data stored in class fields (e.g., android.hardware.SensorEvent\#values). We therefore implemented our own class that implements the \texttt{ISourceSinkDefinitionProvider} interface in FlowDroid for supporting the declaration of fields as sources. Also, based on the feature of class fields, we defined a new model names AndroidField extends from \texttt{SootFieldAndMethod}. After loading a specific field statement from \emph{source\&sink.txt} file, we apply a field pattern regular expression to convert it to the AndroidField model.

FlowDroid has the ability to compute data flow connections between all possible statements. In the implementation of FlowDroid, \texttt{ISourceSinkManager} interface marks all statements as possible sources and then records all taint abstractions that are passed into \texttt{getSourceInfo()}. To that end, we pass the constructed field model as a source statement to the following taint analysis process. In this way, sensitive data flow can be detected starting at given field source statements.

\subsection{Sensor Type Inference}
\label{subsec:type}
The primary goal of SEEKER is to detect data leaks from Android platform sensors.
With the help of FlowDroid's taint analysis, SEEKER's second module can detect sensor-triggered sensitive data flows.
Unfortunately for the field-triggered ones, the identified data-flows only show that there is sensor data leaked but do not tell from which sensor the data is collected.
The sensor type information is important for helping security analysts understand the sensor leaks.
Therefore, in our last module, we identify the types of sensors that are leaking information.

To identify which sensors exist on a specific Android device, we first get a reference to the sensor service by creating an instance of the \texttt{SensorManager} class via calling the \texttt{getSystemService()} method with \emph{SENSOR\_SERVICE} argument. After that, we can determine available sensors on the device by calling the \texttt{getSensorList()} method. 
The \texttt{getSensorList()} method returns a list of all available sensors on the device by specifying constant \emph{TYPE\_ALL} as the parameter. A list of all sensors from a given type can also be retrieved by replacing the parameter as the constants defined for corresponding sensor types, such as \emph{TYPE\_GYROSCOPE, TYPE\_LINEAR\_ACCELERATION}, etc. We can also determine whether a specific type of sensor exists by calling the \texttt{getDefaultSensor()} method with the target type constant (the same as the ones passed in to \texttt{getSensorList()} method). If a device has that type of sensor, it will return an object of that sensor. Otherwise, null will be returned.

We use a rule-based strategy to identify the sensor type of a leak in the case of only one sensor registered in the given app. To do this, SEEKER obtains the sensor type by looking into the type constant in the \texttt{getDefaultSensor()} statement. For instance, \texttt{getDefaultSensor(Sensor.TYPE \_ACCELEROMETER)} indicates that the Accelerometer sensor is obtained. We can then reasonably assume that all sensor-related data leaks in the class are associated with the identified sensor (because only this sensor is registered).

\begin{lstlisting}[
caption={An example of sensor type usage with switch branch.},
label=code:example_sensor_type_usage,
firstnumber=1]
public class MainActivity extends AppCompatActivity implements SensorEventListener{
@Override
public void onSensorChanged(SensorEvent sensorEvent) {
 switch(sensorEvent.sensor.getType()) {
  case Sensor.TYPE_ACCELEROMETER:
   accX = sensorEvent.values[0];
   accY = sensorEvent.values[1];
   accZ = sensorEvent.values[2];
   ...
  case Sensor.TYPE_GYROSCOPE:
   gyroX = sensorEvent.values[0] * 5;
   gyroY = sensorEvent.values[1] * 5;
   gyroZ = sensorEvent.values[2] * 5;
   ...
  case Sensor.TYPE_ROTATION_VECTOR:
   rvX = sensorEvent.values[0];
   rvY = sensorEvent.values[1];
   rvZ = sensorEvent.values[2];
   ...
}}}
\end{lstlisting}

In the case of multiple sensors registered in the given app, we further leverage context-aware static code analysis to find the connection between sensor types and the leaked field data. Firstly, we locate the invocation statement of API \texttt{android.hardware.SensorManager\\\#getDefault Sensor(int)} in the \texttt{onSensorChanged()} method. In the multiple sensors scenario, different sensor's behavior is handled in a conditional branch (e.g. if-then-else statement or switch statement). We then apply context-aware static code analysis to detect the code branch that contains the taint sensor source statement, based on which we then resolve the sensor type in the branch condition.

We further elaborate on the context-aware static code analysis with an example presented in Listing~\ref{code:example_sensor_type_usage}. The code snippet in the Listing shows an example of how multiple sensors are handled with \texttt{onSensorChanged(android .hardware.SensorEvent)} method.  Android determines the activated sensor by matching the \texttt{sensorEvent.sens or.getType()} method (line 4). For example, if \texttt{get Type()} returns \texttt{Sensor.\\TYPE\_ACCELEROMETER} (line 5), the data obtained by \texttt{sensorEvent.values} is associated with the Accelerometer sensor (lines 6-8); if \texttt{getType()} returns \texttt{Sensor.\\TYPE\_GYROSCOPE} (line 10), the data contained in \texttt{sensorEvent.values} is accordingly associated with the current activated sensor, i.e., Gyroscope (lines 11-13).

\section{Experimental Setup}
\label{SEEKER:eval}
SEEKER is designed to expose the data leak issues of sensors in Android apps.
We investigate the feasibility and effectiveness of detecting sensor leaks in Android apps with the following three research questions: 

\begin{itemize}
   \item{\bf RQ1:} {\em Can SEEKER effectively detect sensor leaks in Android apps?} This research question aims to investigate the feasibility of detecting sensor leaks in Android apps with SEEKER.

    \item {\bf RQ2:} {\em To what extent diverse sensor leaks can be identified by SEEKER?} With this research question, we explore the sensor types related to the identified sensitive data leaks, and investigate to what extent such sensor leaks are targeted by attackers. 

    \item {\bf RQ3:} {\em Is SEEKER efficient to detect the sensor leaks in Android apps?} In this study, we leverage the time costs of detecting sensor leaks to assess the efficiency of SEEKER.
\end{itemize}

To answer the aforementioned research questions, we build the experimental dataset with a \textit{malware} set and a \textit{benign} set. The \textit{malware} set contains 20,000 Android apps including malware downloaded from VirusShare repository \cite{Virusshare} that were collected between 2012 and 2020. The 20,000 Android apps in \textit{benign} set are crawled from the official Google Play store. All of the 40,000 apps are submitted to VirusTotal \cite{Virustotal}, the online scan engines aggregating over 70 anti-virus scanners (including the famous Kaspersky, McAfee, Kingsoft anti-virus engines), to check whether they contains viruses or not.
For the \textit{malware} set, we select the malware Android apps that have been labeled by at least
five anti-virus engines to ensure their maliciousness, while for the \textit{benign} set, the Android apps that are not tagged by any anti-virus engines are selected.
SEEKER is designed to detect sensor leaks, thus we filter out the Android apps without any onboard sensors by checking whether their code contains the string ``\texttt{android.hardware.sensor}''. 
The final experimental dataset used in this study consists of 6,724 malware apps and 12,939 benign apps (cf. the 3rd column of Table~\ref{tab:sensor_leak_results}).
Our experiment runs on a Linux server with Intel(R) Core(TM) i9-9920X CPU @ 3.50GHz and 128GB RAM. The timeout setting for analyzing each app with SEEKER is set 20 minutes. 

\section{Results and Analysis}
\label{SEEKER:results}
\subsection{RQ1 -- Feasibility of Detecting Sensor Data Leaks}
\begin{figure}[!t]
    \centering
    \includegraphics[width=1.0\linewidth]{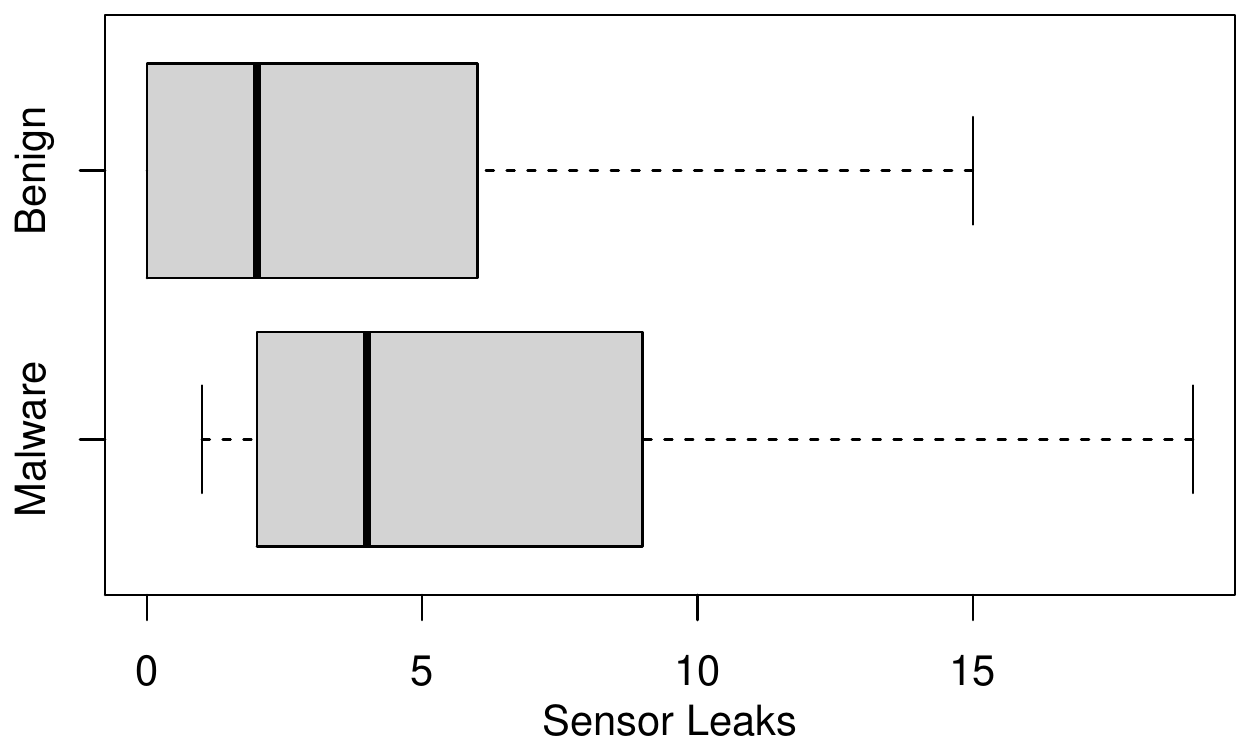}
	\caption{Distribution of sensor leaks in each app.}
    \label{fig:Android_method_field_Sensor}
\end{figure}

Our first research question evaluates the feasibility of SEEKER on detecting sensor leaks in Android apps, of which results are illustrated in Table~\ref{tab:sensor_leak_results}. 
For the quantitative aspect, 9,905 potential sensor leaks are identified by SEEKER in 1,596 apps. 
On average, one Android app could be injected with six sensor leaks.
It indicates that the sensor leaks could exist in Android apps which might have been overlooked by the security analysts of Android apps.
From the malicious aspect, 14.4\% (967 out of 6,724) malware apps are identified with sensor leaks, while 4.9\% (629 out of 12,939) benign apps are identified with such leaks.
Figure~\ref{fig:Android_method_field_Sensor} further presents the number of sensor leaks detected in each Android app, which shows that each malware app could be identified with more sensor leaks than the benign one.
It is significantly confirmed by the Mann-Whitney-Wilcoxon (MWW) test \cite{fay2010wilcoxon}, of which the resulting \emph{p-value} is less than $\alpha = 0.001$. 
All of these results imply that malware apps have a higher possibility of containing sensor leaks than benign apps.

{\bf Note that:} there is lack of the ground-truth dataset about the sensor data leaks in Android apps.
To address this limitation, we consider a sensor leak existing in an Android app if there is the data flow interaction between sensor-related sources (i.e., class fields or methods) and sinks.
With this criterion, we manually checked the 229 sensor leaks detected by SEEKER in 20 randomly selected apps (10 malware apps and 10 benign apps). 
There are only 4 false-positive identified sensor leaks among the 229 identified sensor leaks in the 20 Android apps, which are caused by inaccurate data-flow analysis results of FlowDroid (we detail this limitation cased by FlowDroid in Section \ref{SEEKER:threats}). 
Such results show that SEEKER is capable of identifying the sensor leaks in Android apps.
Simultaneously, it raises a major alarm for security analysts to pay attention to sensor leaks in Android apps that are not protected by the Android permission mechanism.

\begin{table}[!t]
\centering
\caption{Experimental results of the detected sensor leaks.}
\vspace{2mm}
\label{tab:sensor_leak_results}
{
\begin{tabular}{c c c |c c} 
\toprule
\textbf{Dataset} &
\textbf{\# apps} &
\makecell[c]{\textbf{\# selected}\\\textbf{apps}} &
\makecell[c]{\textbf{\# apps identified}\\\textbf{with sensor leaks}} &
\makecell[c]{\textbf{\# identified}\\\textbf{sensor leaks}} \\
\hline
Malware & 20,000 & 6,724 & 967 & 6,103 \\
\hline
Benign & 20,000 & 12,939 & 629 & 3,802 \\
\hline
Total & 40,000 & 19,663 & 1,596 & 9,905\\
\bottomrule
\end{tabular} }
\end{table}

\begin{tcolorbox}[title=\textbf{RQ1 \ding{43} Feasibility and Effectiveness}, left=2pt, right=2pt,top=2pt,bottom=2pt]
SEEKER is capable of automatically detecting sensor leaks in Android apps. 
Malware apps present higher possibility of committing sensor leaks than benign apps, and the sensor leaks might be ignored by security analysts.
\end{tcolorbox}

\subsection{RQ2 -- Characterization of Sensor Leaks}
\paragraph*{Sources Triggering Sensor Leaks}
The data leaks in sensor of Android apps are mainly triggered by two kinds of source: field and method (cf. Section~\ref{subsec:source}).
As presented in Table~\ref{tab:sensor_leak_results2}, $\sim$80\% (= 7941/9905) of identified sensor leaks are triggered by the method sources. For the benign Android apps, $\sim$85.8\% sensor data leaks are sourced from methods, while in the malware Android apps, $\sim$76.6\% leaks are originated from methods.
Table~\ref{tab:sensor_leak_types_method} lists the top-10 most frequent sources triggering sensor leaks identified by SEEKER, which are 8 {\tt getter} methods and 2 public fields from {\tt Sensor}-related classes. 
We observe that the most frequent leaking source is the method {\tt android.hardware.SensorManager\# getDefaultSensor(int)} that is used to get the specific sensor of a given type, which is followed by the field \texttt{values} of class \texttt{SensorEvent}.
The leaking source {\tt android.hardware.SensorManager\#getDefaultSensor(int)} occupies $\sim$89.1\% (7074 out of 7941) of the method-triggered sensor leaks,
and the field {\tt SensorEve nt\#values} occupies $\sim$95.6\% (1877 out of 1964) of the field-triggered sensor leaks.
The sensor leaks triggered by the two sources occupy $\sim$90\% of all identified sensor leaks.

\begin{table}[!ht]
\centering
\caption{Number of identified method/field-triggered sensor leaks.}
\vspace{2mm}
\label{tab:sensor_leak_results2}
{
\begin{tabular}{l c c} 
\toprule
&
\textbf{\# identified method leaks} &
\textbf{\# identified field leaks} \\
\hline
Malware & 4,677 & 1,426\\
\hline
Benign  & 3,264 & 538\\
\hline
Total &   7,941 & 1,964\\
\bottomrule
\end{tabular} }
\end{table}

\begin{table}[!ht]
\centering
\caption{Top-10 frequent leaking sources.}
\vspace{2mm}
\label{tab:sensor_leak_types_method}
\resizebox{\linewidth}{!}{
\begin{tabular}{l c c c} 
\hline
\multirow{1}{*}{\textbf{Sensor Sources}} &
\multirow{1}{*}{\textbf{Malware}} &
\multirow{1}{*}{\textbf{Benign}} &
\multirow{1}{*}{\textbf{Total}} \\
\hline
SensorManager\#getDefaultSensor(int) & 4,326 & 2,748 &  7,074\\
\hline
SensorEvent\#values & 1,358 & 519 &  1,877\\
\hline
SensorManager\#getSensorList(int) & 114 & 121  & 235 \\
\hline
Sensor\#getType() & 123 & 6 & 129 \\
\hline
Sensor\#getName() & 29 & 82 & 111 \\
\hline
Sensor\#getMaximumRange() & 19 & 73  & 92 \\
\hline
SensorEvent\#timestamp & 68 & 19 & 87 \\
\hline
Sensor\#getVendor() & 12 & 74  &  86 \\
\hline
Sensor\#getVersion() & 11 & 69  &  80 \\
\hline
Sensor\#getResolution() & 13 & 64  & 77 \\
\hline
\end{tabular} }
\end{table}

\paragraph*{Sensor Types of Field-triggered Sensor Leaks}
The sensor type is essential for deepening the understanding of sensor data leaks, i.e., knowing from which sensor the data is originally collected, as by default, this information is not given in field-triggered sensor leaks (e.g., sourced from the field variable {\tt values} in class {\tt SensorEvent}).
The last module of SEEKER is hence dedicated to infer the sensor types of such leaks. 
Overall, in the 1,964 identified field-triggered sensor leaks, SEEKER successfully infers the corresponding sensor types for 1,923 (97.9\%) of them.
After manually checking the unsuccessful cases, we find that the 41 failed cases are mainly caused by the mistaken usage of sensors which can cause the sensors unexpected functional behavior, such as lacking sensor register information.
This high success rate demonstrates the effectiveness of SEEKER in pinpointing the sensor types associated with sensor data leaks.

\begin{table}[!t]
\centering
\caption{Top-10 frequent sensor types of field-triggered sensor leaks.}
\vspace{2mm}
\label{tab:sensor_leak_types_field}
\resizebox{\linewidth}{!}{
\begin{tabular}{l c c c} 
\hline
\multirow{1}{*}{\textbf{Sensor Type}} &
\multirow{1}{*}{\textbf{Malware}} &
\multirow{1}{*}{\textbf{Goodware}} &
\multirow{1}{*}{\textbf{Total}} \\
\hline
ACCELEROMETER & 1,068 & 304 & 1,372 \\
\hline
MAGNETIC\_FIELD & 131 & 50  & 181\\
\hline
ORIENTATION & 92 & 84 &  176\\
\hline
PROXIMITY & 12 & 32 & 44\\
\hline
LINEAR\_ACCELERATION & 40 & 4  & 44 \\
\hline
STEP\_COUNTER & 14 & 9 & 23\\
\hline
TEMPERATURE & 12 & 8 & 20 \\
\hline
GYROSCOPE & 7 & 8  & 15 \\
\hline
PRESSURE  & 1 & 13 & 14 \\ 
\hline
LIGHT & 6 & 5  & 11 \\
\hline
\end{tabular} }
\end{table}

We further investigate the true-positive rate of the successfully inferred sensor types. Due to the lack of the ground-truth dataset of related sensor types for sensor leaks, we resort to a manual inspection on the source code of 20 randomly selected apps (10 malware apps and 10 benign apps), each of which is identified with at least one field-triggered leak (86 field-triggered sensor leaks in total). 
All leaks are confirmed with true-positive inferred sensor types, which implies that SEEKER is effective in inferring the sensor types of field-triggered leaks.

Table~\ref{tab:sensor_leak_types_field} presents the top 10 leaking sensor types of the identified field-triggered sensor leaks.
The type ``Accelerometer'' is the sensor type of 74.9\% and 56.5\% of identified field-triggered sensor leaks in the \textit{malware} apps and the \textit{benign} apps, respectively.
Android apps widely use the Accelerometer to monitor device motion states by measuring the acceleration applied to a device on three physical axes (i.e., x, y, and z axes). The motion data captured by the Accelerometer can be further processed or analyzed. For example, \emph{Smart-Its Friends} \cite{holmquist2001smart} pairs two devices by acquiring Accelerometer data in a shared wireless medium. Pirttikangas et al. \cite{pirttikangas2006feature} reported that the Accelerometer in smartphones can be used to track the accurate activity of users, such as brushing teeth and sitting while reading newspapers. Such information can also be utilized to steal the PIN of a device through side-channel attacks (such as \cite{lu2018snoopy} and \cite{giallanza2019keyboard}).

Apart from the Accelerometer, the other frequent sensor types of field-triggered sensor leaks include MAGNETIC\_ FIELD, ORIENTATION, PROXIMITY, LINEAR\_ACCEL- ERATION, STEP\_COUNTER, TEMPERATURE, GYROSCOPE, PRESSURE and LIGHT. These sensors are also likely to be used to harm the user's privacy. 
Biedermann et al. \cite{biedermann2015hard} stated that the magnetic field sensor can be exploited to detect what type of operating system is booting up and what application is being started. The orientation sensor can wiretap the device's orientation without requesting any permission, which can be used by attackers to infer the user's PIN. The proximity sensor data can be a trigger to automatically start a phone call recording when users hold the smartphone against their face to make a call. The individual step details can be stored by collecting data from the step counter sensor when the app runs in the background. Temperature, pressure and light sensors are also widely used in IoT devices to monitor environmental conditions, while the gyroscope sensor is utilized to verify the user's identity \cite{sikder2021survey}.

\paragraph*{Case Study}
Here we show two real-world apps that leaks the sensor data, which could be  leveraged by attacker to achieve malicious goals.

\begin{lstlisting}[style=JAVA, escapechar=\%,
caption={Example of a sensor leak in {\em com.n3vgames.android.driver}.},
label=code:Case_Study_time_stamp,
firstnumber=1]
final class a.b.b implements SensorEventListener{
 public void onSensorChanged(SensorEvent var1){
    float var5 = var1.values[0];
    float var6 = var1.values[1];
    float var7 = var1.values[2];
    Log.v("WindowOrientationListenerN3V", "Raw acceleration vector: x=" + var5 + ", y=" var6 + ", z=" + var7);
}}
\end{lstlisting}

Listing~\ref{code:Case_Study_time_stamp} showcases a typical sensor leak case in real-world apps. The code snippet is excerpted from a malicious app {\em com.n3vgames.android.driver}.
This app collects raw accelerometer data from the class field \texttt{SensorEven\#values[]} (lines 3-5), and then leak them through invoking \emph{android.util.Log} API (line 6). The app is flagged as a Trojan that downloads additional executable content from a remote server. While leaking such information may not direct link to its malicious behaviour, it expands the attack surface to the attackers. For example, the sensor information can be used to predict the device's motion state, which may lead to a stealthier attack (e.g., downloading malicious content when the device is not in use). It is worth noting that Zhang et al.~\cite{zhang2019using} have demonstrated the possibility of using the sensor information to launch stealthy attack for taking control of an Android phone via Google's voice assistant.

\begin{figure}[t!]
    \centering
    \includegraphics[width=0.85\columnwidth]{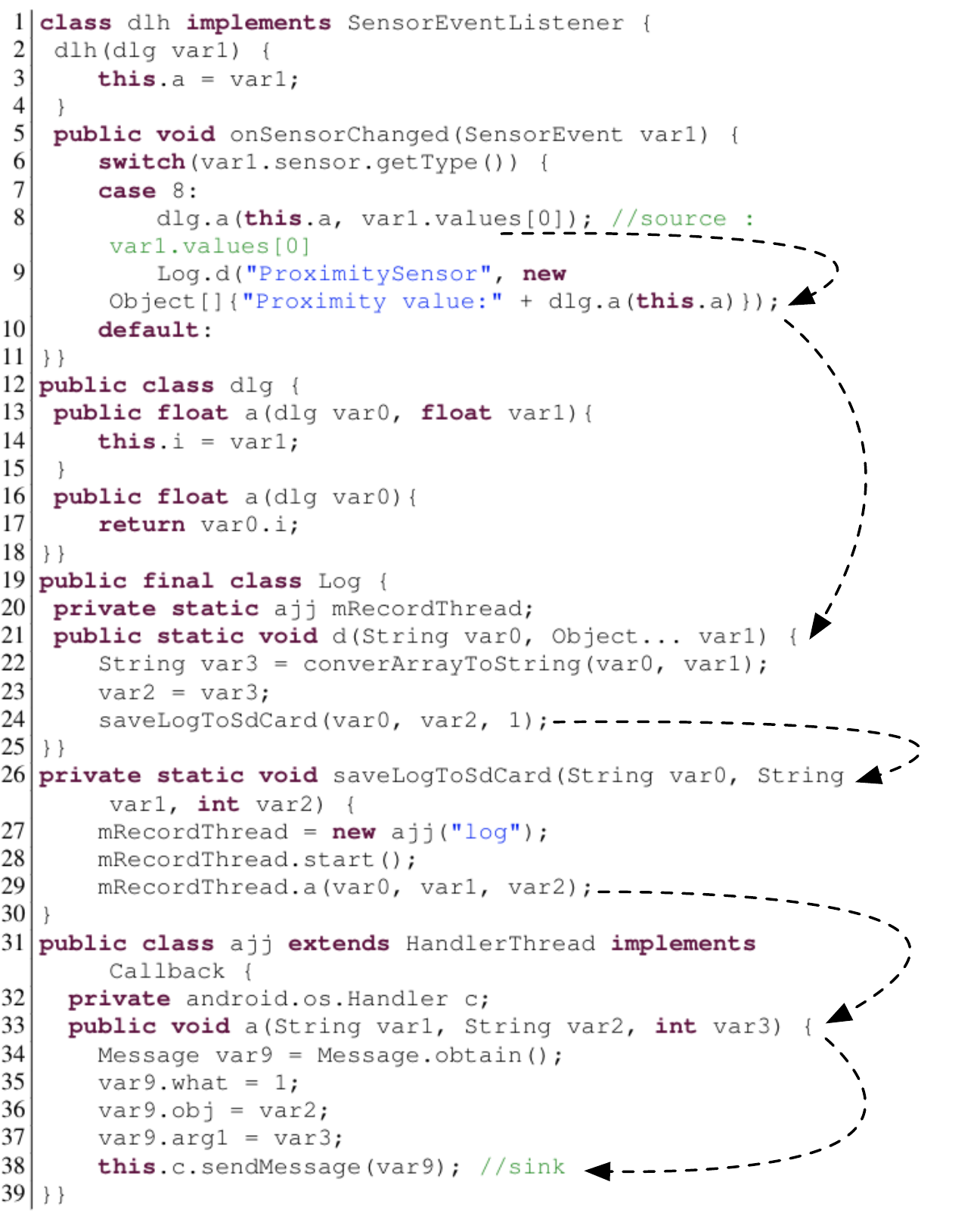}
	\caption{Code snippet of sensor value leak excerpted from \textit{com.tencent.pb}.}
    \label{fig:case_study_2}
\end{figure}

Figure \ref{fig:case_study_2} shows another example derived from a phone book app \emph{com.tencent.pb}. It collects the Proximity sensor data (line 8) and eventually send it out through {\tt sendMessage} method (line 41) in a asynchronous thread. The  Proximity sensor data was passed as a parameter of method {\tt dlg.a(dlg, float)} (line 8), then the data was stored in the class field {\tt i} of object \texttt{dlg}. The data flows through the method \emph{Log.d(String, Object...)} (line 9), which obtains the field variable \texttt{i} of object \texttt{dlg} as the second parameter via \texttt{dlg.a(dlg)} (line 23). Finally, the tainted parameter passed on to the method \texttt{Log.saveLogToSdCard(String, String, int)}, which creates a new thread (line 29-33) and send the sensor data out (line 35-42).

\begin{tcolorbox}[title=\textbf{RQ2 \ding{43} Characterizing Sensor Leaks}, left=2pt, right=2pt,top=2pt,bottom=2pt]
SEEKER is capable of inferring sensor types and pinpointing the corresponding  source sensors for data leaks. Our results show that the Accelerometer leaks the most sensor data, both in malware samples and benign apps. The most leaking sources are the method \textit{SensorManager\#getDefaultSensor(int)} and the field \textit{SensorEvent\#values}, the latter of which has been frequently leveraged in malicious behaviours such as inferring user's PIN.
\end{tcolorbox}

\begin{figure}[!h]
    \centering
    \includegraphics[width=0.85\linewidth]{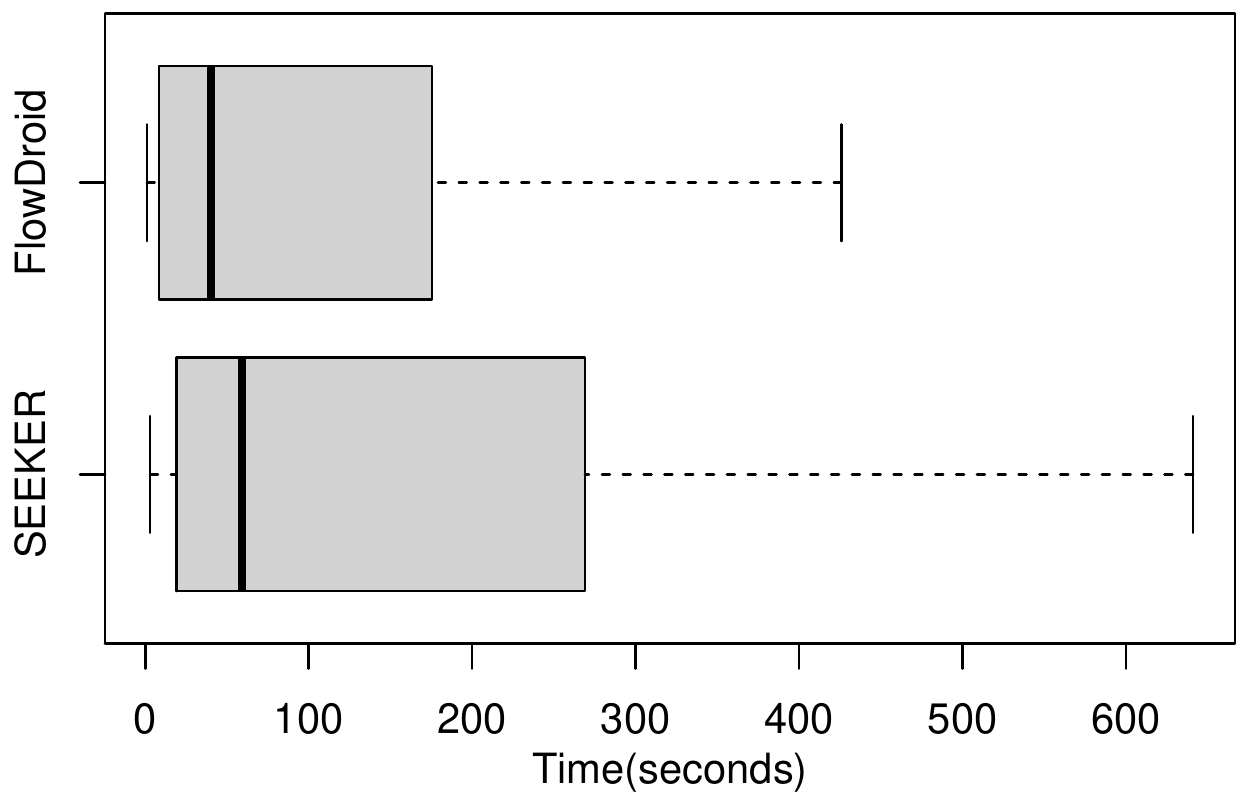}
	\caption{Distribution of time Performance spent to analyze an app by FlowDroid and SEEKER, respectively.}
    \label{fig:time_performance}
\end{figure}

\subsection{RQ3 -- Runtime Overhead}
SEEKER extends FlowDroid to detect sensor-related data leaks and for inferring the sensor types involved in the leak. We evaluate the runtime overhead of SEEKER and compare it with the original FlowDroid. Figure~ \ref{fig:time_performance} shows the time consumed by FlowDroid and SEEKER, respectively.
On average, it takes 177.09 seconds for SEEKER to process an app in our dataset, which is comparable to that of the original FlowDroid (i.e., on average, 132.74 seconds to process an app).
As experimentally demonstrated by Avdiienko et al.~\cite{avdiienko2015mining}, by increasing the capacity of the execution server, the performance of FlowDroid could be further improved.
This improvement should also be applicable to SEEKER, making it also possible to analyze real-world apps in practice.
The fact that the time difference between SEEKER and FlowDroid is relatively small suggests that it is also capable of applying SEEKER to analyze (in parallel) large-scale Android apps, as what has been experimentally demonstrated to be true for FlowDroid.

\begin{tcolorbox}[title=\textbf{RQ3 \ding{43} Efficiency}, left=2pt, right=2pt,top=2pt,bottom=2pt]
The time consumption of SEEKER is acceptable for real-time sensor leak detection, with on average 177.09 seconds for one app  without a high increase when comparing with FlowDroid, which is suitable for real-time app analysis.
\end{tcolorbox}

\section{Discussion}
\label{SEEKER:discussion}

We now discuss the potential implications and limitations of this work.

\subsection{Implications}

\textbf{Beyond smartphone apps.}
The motivating example presented in Section~\ref{SEEKER:pre} is extracted from an attack targeting smartwatches, which also supply sensors to support client apps to implement advanced features.
These sensors could be abused by smartwatch app developers, especially malicious attackers.
We argue there is also a strong need to characterize sensor leaks in smartwatch apps, not just smart phones.
Our preliminary experiment has shown that SEEKER can be directly applied to correctly pinpoint the sensor leaks in the Android-based smartwatch apps that sniff passwords~\cite{chen2021comparative}.

Android has been used on more and more devices, such as TVs, home appliances, fitness machines and cars.
The apps in these devices could also all be compromised to leak end-users' sensitive data and hence should also be carefully analyzed before releasing them to the public.
SEEKER could also be useful to characterize data leaks for such Android devices and we will examine some of these in our future work.

\textbf{Beyond sensor leaks.}
As argued by Zhang et al.~\cite{zhangcondysta}, tainted values of string type could be organized as fields in objects, which cannot be detected by state-of-the-art static taint analysis tools such as  FlowDroid. This is because FlowDroid only supports methods as sources. Thus, sensitive field sources are overlooked by FlowDroid, giving rise to many false negatives. 

Our SEEKER extends FlowDroid to mitigate this research gap by introducing field-triggered static taint analysis. 
It is worth highlighting that SEEKER is capable of not only detecting sensor leaks but also pinpointing general privacy leaks, either triggered by source methods or fields.
To help users experience this feature, we have committed a pull request to the original FlowDroid on GitHub so that users can easily access Field-triggered Static Taint Analysis by simply configuring their interested field sources in \emph{SourcesAndSinks.txt} file. 

\textbf{Automated approaches for discovering sensitive source fields.}
In this work, the sensor-related sensitive source fields are identified through manual effort. These are well known to be time-intensive and error-prone.
Hence, our current SEEKER approach is not directly applicable for detecting general field-triggered privacy leaks.
To achieve this, we need to go through all the fields defined in the Android framework to identify sensitive ones. This is non-trivial as Android is now one of the largest community software projects and contains nearly 10K classes.
There is a need to invent new automated approaches to discover sensitive source fields.
One possible solution would be to extend the machine learning approach applied in the SUSI tool to support the prediction of sensitive source fields.

\section{Threats to Validity}
\label{SEEKER:threats}

\textbf{Limitation of static analysis.}
One major limitation of our tool lies in the intrinsic vulnerability of static code analysis when encountering code obfuscation, reflection, native code, etc. These lead to the unsoundness of our approach. However, these challenges are regarded as well known and non-trivial issues to overcome in our research community.  In our future work, we want to integrate other useful tools developed by our fellow researchers to overcome these shortcomings. For example, we plan to leverage DrodRA \cite{sun2021taming, li2016droidra} to reduce the impact of reflective calls on our static analysis approach.

As explained in Section \ref{subsec:type}, our sensor type inference approach can not trace the sensor type in method-triggered leaks when multiple sensors are available on a device. This is because the actual calling object of a method can only be obtained at run-time. We plan to overcome this limitation in our future work by incorporating dynamic analysis approaches to obtain the required run-time values.

\textbf{Limitations inherited from FlowDroid.}
Since our SEEKER approach directly extends FlowDroid to detect sensor-triggered privacy leaks, it also has all of the limitations of FlowDroid.
For example,  FlowDroid may yield unsound results because it may have overlooked certain callback methods involved in the Android lifecycle or incorrectly modelled native methods accessed by the app.
FlowDroid is also oblivious to multi-threading and it assumes threads to execute in an arbitrary but sequential order, which may also lead to false results.

\textbf{Limitations inherited from SUSI.}
The sensor-related sensitive source methods are collected based on the results of the state-of-the-art tool SUSI. This is also the tool leveraged by the FlowDroid to identify source and sink methods.
However, the results of SUSI may not be  completely correct -- some of its identified sources may not be truly sensitive.
However, this threat has no impact on our approach but only on our experimental results.
This limitation could be mitigated if a better set of source and sink methods are configured.

Apart from these technical limitations, our work also involves some manual efforts. For example, the sensor-related sensitive source fields are summarized manually by reading the Android developers' documentation.
Such manual processes may also introduce errors of their own.
To mitigate this threat, the authors of this paper have cross-validated the results, and we release our tool\footnote{https://github.com/MobileSE/SEEKER} and dataset\footnote{https://zenodo.org/record/4764311\#.YJ91jJMzadZ} for public access.

\section{Summary}
\label{SEEKER:con}
We have presented a novel tool, SEEKER, for characterizing sensor leaks in Android apps. Our experimental results on a large scale of real-world Android apps indicate that our tool is effective in identifying all types of potential sensor leaks in Android apps. Our tool is not only capable of detecting sensor leaks, but also pinpointing general privacy leaks that are triggered by class fields. 
Although there are related works on sensor usage analysis, to the best of our knowledge, there is no other work that thoroughly analyses Android sensor leakage. Unlike previous works, our tool is the first one  to characterize all kinds of sensor leaks in Android apps. We extend FlowDroid for supporting field sources detection (i.e., merged to FlowDroid via pull \#385 on Github\cite{FlowDroidMerge}), which we believe could be adapted to analyze other sensitive field-triggered leaks.
To benefit our fellow researchers and practitioners towards achieving this, we have made our approach open source at the following Github site.

\begin{center}
   \url{https://github.com/MobileSE/SEEKER} 
\end{center}

%% file: chapter4.tex
\chapter{Demystifying Hidden Sensitive Operations in Android apps}

\begin{tcolorbox}
[width=6.2in]
\small{
\textbf{Xiaoyu Sun}, Xiao Chen, Li Li, Haipeng Cai, John Grundy, Jordan Samhi, Tegawend\'e F. Bissyand\'e and Jacques Klein, 2022, Demystifying Hidden Sensitive Operations in Android apps. ACM Transactions on Software Engineering and Methodology (TOSEM) 2022.
\url{https://arxiv.org/abs/2210.10997}
}
\end{tcolorbox}

\section{Introduction}
\label{HiSenDroid:intro}
Android is the most adopted mobile operating systems in terms of users, applications and developers~\cite{IDCReport}. However, its popularity means that legitimate developers must co-exist with malware writers. Reports on many different kinds of attacks are presented in the technology and lay media. For example, security researchers have reported a malicious ``clicker trojan''\footnote{Such as the \emph{Android.Click.312.origin} trojan and its modified variant {Android.Click.313.origin} trojan. This aims to generate fraudulent click-through and subscription revenues.} which has been bundled into 34 different Google Play apps that have already been installed more than 100 million times\footnote{\url{https://www.forbes.com/sites/zakdoffman/2019/08/13/android-warning-100m-users-have-installed-dangerous-new-malware-from-google-play/\#1956f51c22a9}}. On a larger scale, antivirus engines have been flagging a large number of apps as potential threats. For example, as of October 2020, the popular AndroZoo dataset~\cite{allix2016androzoo} has recorded more than 226,000
Android GooglePlay apps than have been flagged as adware/malware by at least 5 Antivirus products, and this number is still growing.
Those adware/malware often not work along but collaborate with many third parties over the internet. Some of the representative malicious behaviors include leading users to malicious websites through devious advertisements~\cite{liu2020maddroid, 9282795, dong2018frauddroid, dong2018mobile}, distributing malicious apps in the mobile network through drive-by downloads~\cite{cova2010detection}, leaking users' sensitive data to web servers through HTTP connections~\cite{sun2021characterizing,liu2021first,li2015potential,gao2020borrowing}, etc.

To protect Android users against the rapid spread of malware, the research and practice communities have implemented a variety of measures and proposed several approaches to detect malware~\cite{arp2014drebin, mariconti2016mamadroid, zhao2021impact,li2017understanding,sun2022mining,xu2022lie}. These include static code analysis-based approaches~\cite{li2017static,li2019rebooting}, dynamic testing based approaches~\cite{kong2018automated}, and learning-based approaches~\cite{liu2022deep}. 
Unfortunately the emergence of many different malware detection techniques has also stimulated malware attackers into being more innovative to increasingly better hide malicious behaviour, in order to bypass static code analysis (e.g., via obfuscation) and even dynamic detection (e.g., sensing of sandbox execution). 
In practice, sophisticated code obfuscation techniques \cite{moser2007limits} are being leveraged by attackers to hide their malicious program behavior, leading to false negatives in most static analyses thus resulting in imprecise and unsound results.
Camouflage techniques have been frequently leveraged by attackers to evade dynamic testing approaches~\cite{rasthofer2017making,egele2008survey}.
Attackers often introduce a so-called logic bomb or time bomb to set off malicious functions only after certain conditions are met.
For instance, after knowing that Google  applies a dynamic analysis tool called \emph{bouncer} to scan every app submitted to Google Play for five minutes, as revealed by Oberheide et al.~\cite{oberheide2012dissecting}, a bunch of malicious apps has been created and been demonstrated to be capable of penetrating Google's bouncer vetting system by simply waiting five minutes before triggering their malicious behavior. 

To cope with such hidden malicious behaviors, researchers have devised new detection approaches.
For example, Fratantonio et al.~\cite{fratantonio2016triggerscope} have proposed an approach called TriggerScope to detect hidden behaviors triggered by predefined circumstances such as events related to location, time, and SMS. 
However, TriggerScope is not capable of detecting such malicious activities hidden behind other trigger types, such as the existence of other services (i.e., other than location, time and SMS).
In line with this research, Pan et al.~\cite{pan2017dark} have proposed a machine learning-based approach aiming to discover unknown trigger types.
Their approach, however, needs to manually label a dataset for training, which is known to be resource-intensive and error-prone.

Static analyzers suffer less than dynamic approaches from evasion techniques such as logic bomb or time bomb. 
In particular, regarding sensitive flow detection (also called privacy leak detection), numerous static analysis tools have been proposed such as FlowDroid~\cite{arzt2014flowdroid} (and its extension \textsc{IccTA}~\cite{li2015iccta}), \textsc{Amandroid}~\cite{wei2014amandroid},  or \textsc{DroidSafe}~\cite{gordon2015information}. Although these tools are able to track sensitive flows (which are often hidden) by bringing  key new contributions to the research community, they still face some well-known limitations~\cite{AnalyzingAnalyzers-ISSTA2018}: their inherent over-approximations inevitably lead to false alarms, which, for some analyzers, occur at a high rate, making them impractical. Consequently, when building on static analysis, manual investigation is often required.
Unfortunately, such efforts cannot scale. Dynamic validation then appears as an alternative. Unfortunately, runtime execution often misses hidden sensitive flows due to the implementation of evasion techniques by attackers.
While some effort (e.g.,~\cite{fratantonio2016triggerscope,pan2017dark}) has been put to \emph{characterize} Hidden Sensitive Operations (HSOs) in Android apps, our community has not yet proposed dedicated approaches to \emph{detect and explain} such operations, allowing attackers to achieve malicious behaviors while bypassing certain security vetting mechanisms.

We fill this research gap in this work by proposing a new prototype tool, HiSenDroid, which deploys an automated static app analyzer tailored for detecting \textit{hidden} sensitive operations. 
HiSenDroid performs a sequence of static analyses, including call graph analysis, forward data-flow analysis, inter-procedural backward data-flow analysis, etc. 
For exposed HSOs, HiSenDroid further goes one step deeper to record detailed information for explaining why these HSOs should be flagged as such.

To summarize, key contributions of our work include:
\begin{itemize}
    \item
    We propose using a static analysis approach to discover hidden sensitive operations that are not exposed to the state-of-the-art static and dynamic analysis tools in Android apps. To this end, we leverage control flow and data flow analyses to identify the unique code level characteristics of hidden sensitive operations.
    \item
    We designed and implemented a prototype tool HiSenDroid for analyzing hidden sensitive operations. We release HiSenDroid as an open source project \cite{HiSenDroid} for supporting security analysts in their analysis needs and fostering further researches in this direction.
    \item We evaluated HiSenDroid on a large-scale dataset that contains 10,000 benign and 10,000 malware samples, and discovered emerging anti-analysis techniques employed by malware samples, such as fulfilling certain restrictions related to \emph{time}, \emph{location}, \emph{SMS message}, \emph{system properties}, \emph{package manager}, and other logics. 
    \item With the help of FlowDroid~\cite{arzt2014flowdroid}, a static taint analyzer, we further experimentally show that HSOs have been recurrently leveraged by attackers to leak sensitive user information. 
\end{itemize}

The rest of the paper is organized as follows: Section \ref{HiSenDroid:motivation} defines HSO and presents the motivation of our research, i.e., why there is a strong need to demystify HSO.
Section~\ref{HiSenDroid:approach} depicts the design and implementation of the proposed approach. Section~\ref{HiSenDroid:commonHSO} and Section~\ref{HiSenDroid:SuspiciousHSO} respectively describe the characteristics of common and susipious HSOs detected by our approach from a large-scale dataset.
Section~\ref{HiSenDroid:implication} presents a practical implication of our approach by characterizing sensitive data leaks triggered by HSOs.
Section \ref{HiSenDroid:limitations} discusses the limitations of the tool. 
Section \ref{HiSenDroid:related_work} reviews the related works, and finally Section \ref{HiSenDroid:summary} concludes this paper. 

\section{HSO Definition and Motivation}
\label{HiSenDroid:motivation}
We conducted an exploratory study to understand the characteristics of \emph{Hidden Sensitive Operations} (HSO) in Android apps. We first dumped operations in a set of real-world Android malware. Then, we manually examined those operations to observe the characteristics of such operations that could be considered as hidden-triggered operations.
Based on our manual summarization, we found that
(1) \emph{if statement} and the notion of \emph{branch} are key in the definition of HSO; 
(2) the \emph{if statement} contains a specific \emph{operation} that triggers the hidden sensitive flows, and this trigger condition is related to Android API.

Let $B$ denote one of the two branches of an \emph{if-then-else statement}, or the branch of an \emph{if statement} where the \emph{else} branch is considered empty.

\textbf{Definition 1 [Hidden Sensitive Branch (HSB)]:} 
$B$ is an HSB if it fulfills the following \textbf{rules}: 
\begin{enumerate}
    \item $B$ contains sensitive Android APIs, and these APIs are different from those contained in the other branch involved in the \emph{if-then-else statement}.
    The rationale behind this condition is that a hidden branch is supposed to achieve some sensitive behaviors that are different from those of the "normal" branch (i.e., non-HSB), which per se might also access sensitive APIs as part of the app's expected behaviors.
    \item $B$ does not involve any of the variables appearing in the \emph{condition expression} of the \emph{if-then-else statement}. The rationale behind this is that the branch is triggered by conditions that are also different from its (sensitive) behaviors.
\end{enumerate}

Less formally, an HSB could be defined as an "if branch" which accesses sensitive APIs, and which is fully "independent" of the \emph{if condition} and the other branch of the  \emph{if statement}. 

Let $C$ denote the $condition$ of an \emph{if statement}.

\textbf{Definition 2 [Hidden Sensitive Operation (HSO)]:} 
An HSO is an HSB that is triggered by a condition $C$ containing values obtained via (or directly impacted by) Android system APIs or system properties (i.e., attributes of system classes). This may return different values when being executed under different circumstances, so as to triggering hidden sensitive operations.

\begin{lstlisting}[
caption={An example of a real-world hidden sensitive data flow.},
label=code:emu,
float=t,
firstnumber=1]
public class MainActivity extends AppCompatActivity {
 protected void onCreate(Bundle savedInstanceState) {
  SmsManager smsManager = SmsManager.getDefault();
  ED ed = new ED(this);
  StringBuilder message = new StringBuilder();

  if(ed.checkPackageName()) {
   TelephonyManager tm = (TelephonyManager)     getSystemService(Context.TELEPHONY_SERVICE);
   String imei = tm.getDeviceId();
   String phoneNumber = tm.getLine1Number();
   String subscriberId = tm.getSubscriberId();
   message.append(imei);
   message.append(phoneNumber);
   message.append(subscriberId);
   smsManager.sendDataMessage("+115800763861", null, (short)1001, message.toString().getBytes(), null, null);
  } else {
   //benign string operations
}}

public class ED {
 public ED(Context pContext) {
  mContext = pContext;
  mListPackageName.add("com.google.android...genymotion");
  mListPackageName.add("com.bluestacks");
  mListPackageName.add("com.bignox.app");
 }
 public boolean checkPackageName() {
  if (!isCheckPackage || mListPackageName.isEmpty()) {
   return false;
  }
  final PackageManager packageManager = mContext.getPackageManager();
  for (final String pkgName : mListPackageName) {
   final Intent tryIntent = packageManager.getLaunchIntentForPackage(pkgName);
   if (tryIntent != null) {
    final List<ResolveInfo> resolveInfos = packageManager.queryIntentActivities(tryIntent, PackageManager.MATCH_DEFAULT_ONLY);
    if (!resolveInfos.isEmpty()) {
     return true;
    }
   }
  }
  return false;
}
\end{lstlisting}

Listing 1 exemplifies a simplified code snippet illustrating these definitions in practice. Note that Listing 1 presents the typical characteristics of an HSO in many real-world apps that we have manually analyzed. 
At line 7, the app firstly checks if it is running on one of the popular Android emulators (i.e., \emph{genymotion, bluestacks, and bignox}). If not, the app reads the device information and sends it to a hard-coded phone number through an SMS. Otherwise, if an emulator environment is detected, it will only perform some unharmful string operations (ignored).
In this example, three private data -- namely the device's IMEI, IMSI, and phone number -- are retrieved in lines 9-11 and sent to a hard-coded phone number via SMS (line 15).
All of these three leaks are hidden behind the trigger condition \emph{ed.checkPackageName()} (line 7). The trigger condition checks the return value of a self-defined method \emph{checkPackageName()} (line 30), which is determined by several other \emph{if-conditions} defined in the invoking method (lines 31,37,39). Finally, the trigger condition in the HSO is traced back to a system API \emph{PackageManager.queryIntentActivities()} (line 38) (\textbf{cf. Definition 2}).
This trigger condition examines whether popular Android emulator packages (lines 26-28) are available in the device, i.e., checking if the app is running on these emulators. If the running environment is not one of the hard-coded emulators, the HSO will be performed.
Otherwise, benign string operations are executed (lines 17-19) (\textbf{cf. Definition 1}).

\section{Approach}
\label{HiSenDroid:approach}
\begin{figure}[!h]
    \centering
    \includegraphics[width=1.0\linewidth]{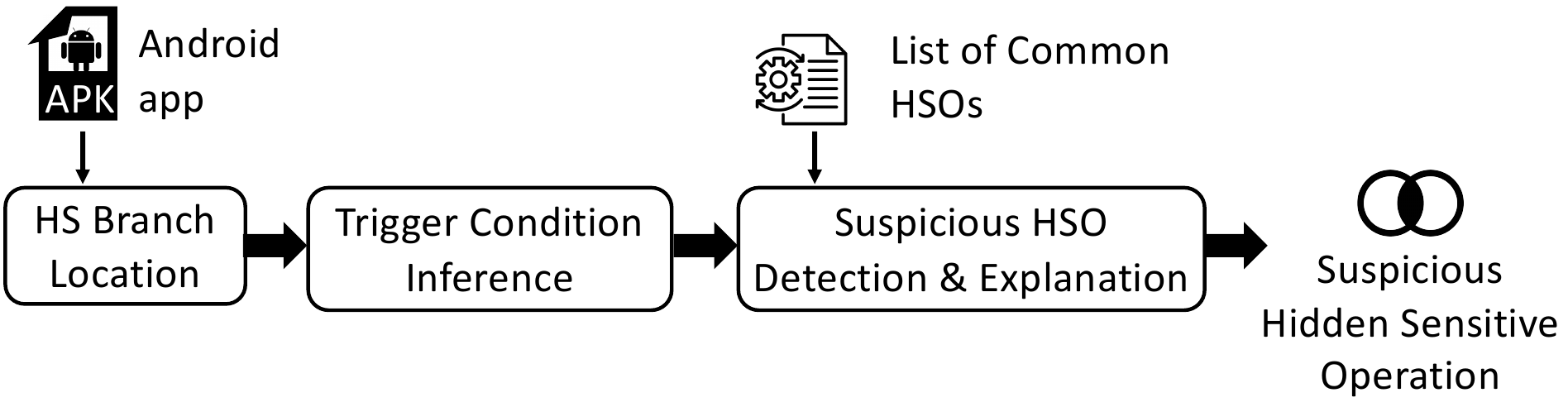}
	\caption{The working process of HiSenDroid.}
    \label{fig:overview}
\end{figure}

To better help security analysts understand Hidden Sensitive Operations (HSO) placed in Android apps, we designed and implemented a prototype tool, named HiSenDroid, to automatically locate such operations in Android apps.
HiSenDroid takes as input an Android app and outputs a set of hidden sensitive operations.
Fig.~\ref{fig:overview} illustrates the working process of HiSenDroid. It achieves the aforementioned goal through three main modules, namely: (1) Hidden Sensitive Branch Location; (2) Trigger Condition Inference; (3) Suspicious HSO Detection and Explanation. 
We now respectively detail these three modules.

\subsection{Hidden Sensitive Branch Location}
\label{subsec:hbl}
The first module of HiSenDroid is responsible for locating hidden sensitive branches (HSBs) in Android apps (i.e., fulfilling the rules in Definition 1).
Towards locating HSBs, this module first statically goes through all the methods that appeared in the DEX file of the input APK.
For each method, this module then constructs an intra-procedural control-flow graph (CFG) and traverses the graph to locate \emph{if-then-else statements}.
Once an \emph{if-then-else statement} is located, it further extracts the sensitive APIs accessed by the two branches 
(hereinafter referred to as \emph{if-branch} and \emph{else-branch}).
Sensitive APIs are such methods that are protected by Android permissions, which are classified following the latest Android API-permission mappings PSCout~\cite{au2012pscout}, Axplorer~\cite{backes2016demystifying}, Arcade~\cite{aafer2018precise}, and NatiDroid~\cite{chaoran2022cross}.
Any of the two branches will be considered a 
potential HSO if it has indeed accessed sensitive APIs that
are different from the APIs accessed by the other branch. 

When extracting sensitive APIs, in order to obtain a \emph{soundy} result~\cite{livshits2015defense} (e.g., including all the sensitive APIs accessed by a potential HSB), this module traverses not only the methods directly presented in the HSB but also all the methods that could be reached from the branch.
This process is made possible by first constructing a call graph (CG) for the input APK.
Unfortunately, as discussed by many existing works, Android apps do not have a single entry point (e.g., \emph{main()}) that connects other parts of the application code, making static analyses challenging to cover all the app code. Fortunately, this challenge has been well addressed by the state-of-the-art by artificially creating a so-called dummy main method, connecting together all the separated code parts, including system-driven lifecycle methods and event-driven callback methods ~\cite{arzt2014flowdroid}. 
Based on our observation and the findings of previous work~\cite{pan2017dark}, the connection between trigger conditions and the operations along its paths is often weak. Indeed, the variables appearing in triggers typically do not propagate data flow to its following paths. Take Listing 1 as an example, the app checks if it is running inside Android emulators at line 7, where the trigger condition code itself is not supposed to steal private data and is only meant to determine the right situation for running hidden code. To leverage this property, we attempt to check whether variables appearing in the HSB have data dependency with any variable within the condition expression.
Thus, for a given potential HSB, this module goes one step further to check if any of the variables appeared in the HSB's \emph{condition expression} has been leveraged by the HSB code.
If so, this HSB will not be considered as a true HSB and thereby will be excluded from further analyses.
This module achieves this by conducting a simple intra-procedural control-flow analysis.
In a case of true HSB, there should not be intersections between the set of variables that appeared in its \emph{condition expression} and those within the branch.

\subsection{Trigger Condition Inference}
After locating HSBs, the second module goes one step deeper to infer hidden sensitive operations (HSOs) so as to fulfill Definition 2.
Given a true HSB, the idea of detecting HSOs is to infer the detailed trigger conditions that lead to the execution of the HSB.

We began with a preliminary study to understand what kinds of trigger conditions have been used to hide suspicious APIs, as identified from the literature~\cite{pan2017dark, petsas2014rage, chen2008towards, vidas2014evading, jing2014morpheus, diao2016evading, costamagna2018identifying, Android.hehe, Hackingteam, tamperingdetection, norboev2017robustness} on trigger conditions. 
For example, Petsas et al.~{\cite{petsas2014rage}} investigated anti-analysis techniques that can be employed by Android apps to evade detection, including pre-initialized static information(e.g., IMEI value), dynamic information that does not change (e.g., Sensors data) and VM instruction emulation (e.g., hardware variable). 
In their paper, they demonstrated how dynamic analysis could be evaded by the aforementioned trigger conditions in an emulated environment. 
Pan et al.~{\cite{pan2017dark}} further summarize that almost all the trigger conditions of HSOs can be characterized by \textbf{System Properties} (e.g., OS or hardware traces of a mobile device) or \textbf{Environment Parameters} (time, locations, SMS, etc.).
To the best of our knowledge, the values in both types can be obtained through Android system APIs.
In other words, an HSO trigger condition is expected to involve, directly or indirectly,
one or more system API calls for interacting with the Android operation system.
Therefore, since it is very important to identify all possible trigger conditions, we propose considering all the condition checks to infer HSO's trigger conditions\footnote{We remind the readers that state-of-the-art studies (e.g., by Moser et al.~\cite{moser2007limits} and Zeng et al.~\cite{zeng2018resilient}) have further revealed that obfuscation (via reflective calls or opaque predicates) could be leveraged to complicate the inference of trigger conditions (e.g., changing the way how a system property is obtained from the system). We do not take obfuscation as a type of trigger condition but will only consider it as a technique that complicates the process of identifying trigger conditions. We will discuss the impact of obfuscation on our approach at the end of Section~\ref{HiSenDroid:SuspiciousHSO}.} as long as they involve system properties, environment parameters, and any other values yielded by system APIs.

In this work, we follow the same criteria to infer HSOs (i.e., the trigger conditions involves values obtained through Android system APIs).
Specifically, to infer the trigger conditions, for each of the variables that appeared in the HSB's \emph{condition expression}, there is a need to conduct backward data-flow analyses to locate its definition statement.
The code block between the definition statement and the \emph{if-then-else statement} is then referred to as a Condition Triggering Block (CTB).
Then, given a potential HSO, we check whether a system API is involved in the \emph{definition statement} of the HSO's CTB.
If not, we will regard this HSO as a false result and consequently will not consider it for further analyses.

\begin{figure}[!h]
    \centering
    \includegraphics[width=0.6\textwidth]{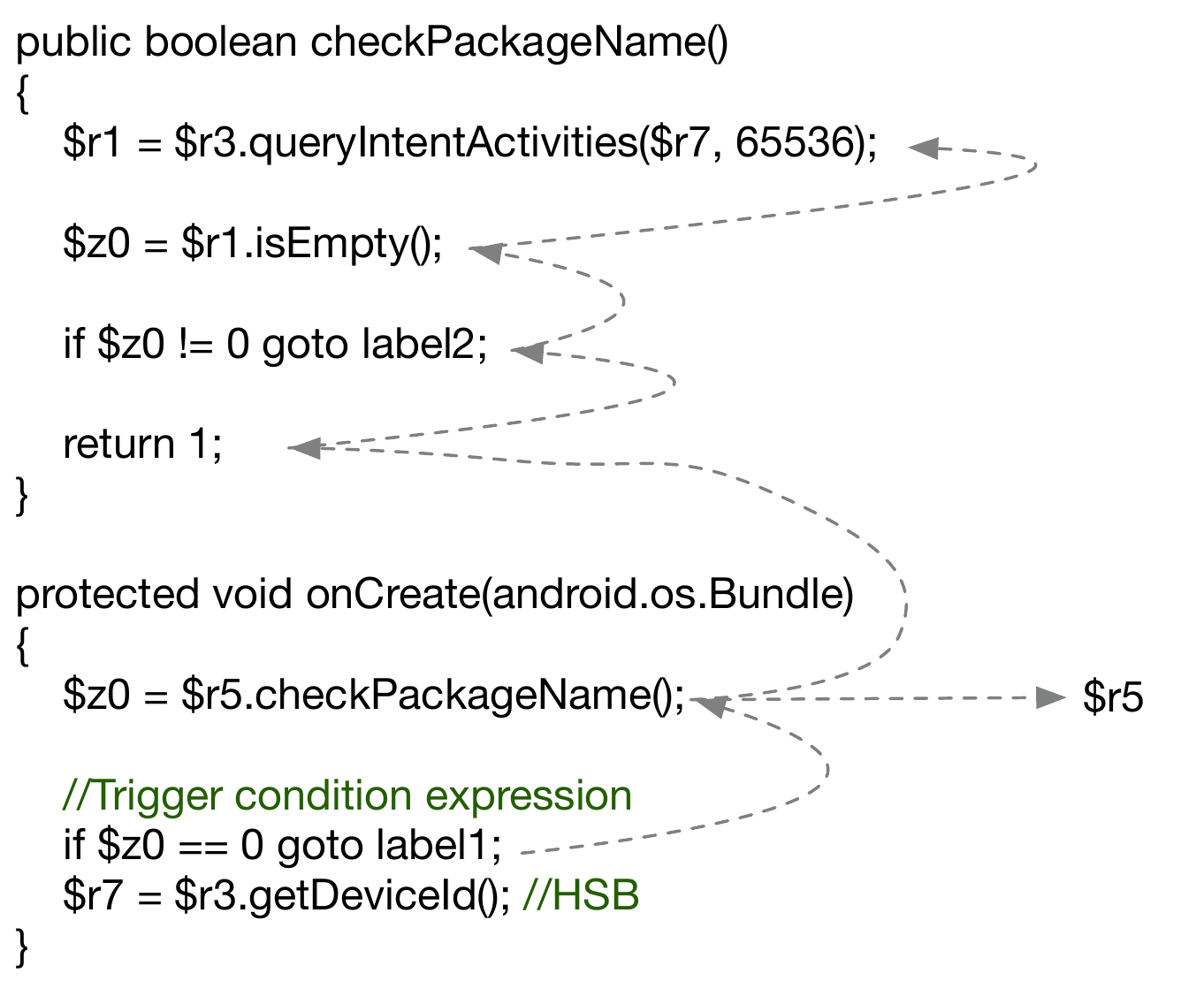}
	\caption{The simplified working process of the trigger condition inference module. The code is presented in simplified Jimple, which is an intermediate representation of Soot~\cite{lam2011soot}. Soot is the underline static analysis framework leveraged by HiSenDroid to achieve the backward data-flow analysis.}
    \label{fig:triggercondition}
\end{figure}

When inferring the \emph{definition statement}, inter-procedural analysis needs to be taken into account because the trigger conditions can be defined in other methods and transferred to the HSB via callee's returned values or caller's parameter values.
Indeed, take the code snippet shown in Listing~\ref{code:emu} as an example,  the trigger condition is actually defined in method \emph{checkPackageName()} despite the HSB is seated in the \emph{onCreate()} method.
Fig.~\ref{fig:triggercondition} illustrates the backward tracking flow showing how our approach identifies the trigger condition.
When there is a method involved in the backward tracking flow, our data-flow analysis will keep tracking the method's caller object as it may be relevant to the definition of the trigger condition.
For example, our analysis will keep tracking \$r1 when statement \emph{\$r1.isEmpty()} is reached.
If the method is a user-defined function, our data-flow analysis will further jump into the method and keep tracking its returned variables (all variables will be tracked if there are several return statements).
The backward data-flow analysis will terminate if System APIs are identified, or Android's entry-point methods (such as UI callback methods or components' lifecycle methods) are reached.
The analysis will also stop if the condition is linked to a constant value that is further not originated by \emph{if-statements}.

\subsection{Suspicious HSO Detection}
\label{sec:common}
The last module takes the outputs of the previous module to detect hidden sensitive operations, following the rules presented in Definition 1 and Definition 2 (cf. Section~\ref{HiSenDroid:motivation}).
Unfortunately, these rules are not perfect and 
may introduce false-positive results that have similar characteristics of HSOs but are actually user intended behaviors. Indeed, for the same operations, under different circumstances, they could be flagged as conventional usages or suspicious operations and could lead to benign or user intended malicious behaviors.
These false results include common programming patterns used in legitimate \emph{if-then-else statements} (hereinafter referred to as \emph{conventional usages}), which should be excluded by HiSenDroid . Therefore, we resort to building a list of conventional usages (or whitelist) and based on it, in the last module of HiSenDroid, we filter out non-malicious HSOs and only keep suspicious HSOs. 

Nevertheless, we argue that it is non-trivial to understand the developer's intention behind the operations.
Therefore, in this last module, in addition to automatically detect suspicious hidden sensitive operations, HiSenDroid goes one step deeper to also provide adequate details to explain why an suspicious HSO is flagged as such, i.e., what is the trigger condition, what is the logic of the \emph{if condition}, and what are the sensitive behaviors triggered if the logic is fulfilled.
This function is provided for helping security analysts understand whether the flagged HSOs should be regarded as malicious or not.

By leveraging HiSenDroid, in Section~\ref{HiSenDroid:commonHSO}, we study and collect \emph{conventional usages} in large sets of Android apps, whereas in Section~\ref{HiSenDroid:SuspiciousHSO}, we put the emphasize on \emph{suspicious} HSOs.

\section{Conventional Usage}
\label{HiSenDroid:commonHSO}

The overall goal of this work is to detect hidden sensitive operations so as to unveil the evasive technologies that are frequently leveraged to hide malicious behaviors. 
In this section, we evaluate our approach based on a large set of Android apps towards checking if our approach HiSenDroid is capable of fulfilling this goal.
Specifically, in this section, we conduct an exploratory study of recent hidden sensitive operations aiming to understand the current status quo of conventional usages and build a comprehensive list of conventional usages (to be used by HiSenDroid to discriminate suspicious HSOs from conventional usages).

Recall that our approach, in its last working step, takes as input a customizable list of conventional usages to filter out non-suspicious HSOs, which subsequently helps in saving significant security analysts' efforts as they now only need to scrutinize the retained small number of likely suspicious HSOs.
Towards identifying such conventional usages, we apply a semi-automatic process to summarize based on their frequency of occurrence. The conventional usage whitelist is built based on reasonable assumptions that legitimate HSOs frequently appear in Android apps, including both malware\footnote{Malware is included because often not all of its code is malicious. It might contain a malicious payload but the other code could still remain benign.} and goodware.
We manually inspected the trigger APIs that have appeared more than 50 times in our dataset and determined if they should be categorized as a conventional usage. By doing so, we defined seven major categories of conventional usages. Also, the results of our manual analysis are cross-validated by two authors. The two authors first independently conduct the manual analysis (to discover knowledge with support evidence from various software artifacts). They then had physical meetings to discuss, merge, and finalize the results.

\textbf{\emph{Experimental\ Setup}}.
We applied HiSenDroid (with the list of conventional usages set to be empty\footnote{The experimental results should contain both conventional usages and suspicious HSOs.}) on a dataset that contains 10,000 malware samples (referred to as \textit{malware set}) and 10,000 benign apps (referred to as \textit{benign set}).
The \textit{malware set} was collected from VirusShare \cite{Virusshare} from 2012 to 2020. To better reflecting recent trends on the deceptive techniques used in malware samples, we only include the samples whose first seen date was on or after 2016. The malware samples were submitted to VirusTotal\footnote{https://www.virustotal.com} for screening, and only the ones that have been labelled by more than five anti-virus engines (VirusTotal has hosted over 70 anti-virus scanners) were selected.

The \textit{benign set} was randomly selected from a pool of more than 100,000 apps crawled from Google Play in 2019, which are further scanned to ensure non of them are tagged by VirusTotal.

Our tool has identified 45,342 HSOs (35,974 in the \textit{malware set}, and 9,368 in the \textit{benign set}) triggered by 54,152 conditions. Note that some HSOs may be triggered by more than one condition (e.g., multiple conditions in a CTB that are connected by \emph{AND} or \emph{OR} operators). 
Towards evaluating the precision of HiSenDroid, i.e., the identified HSOs meet our previous rule definitions, we manually examine 20 randomly selected APKs from the total 8,107 apps that have been identified to contain at least one HSO. From these apps, our approach identified 157(with a confidence level of 95\% and a confidence interval of 7.81\%) HSOs in total, among which 155 of them are eventually confirmed to be true HSOs, giving an precision of 98.7\%. This result suggests that HiSenDroid is capable of identifying HSOs in Android APIs.

\begin{figure}[!t]
    \centering 
    \includegraphics[width=0.65\textwidth]{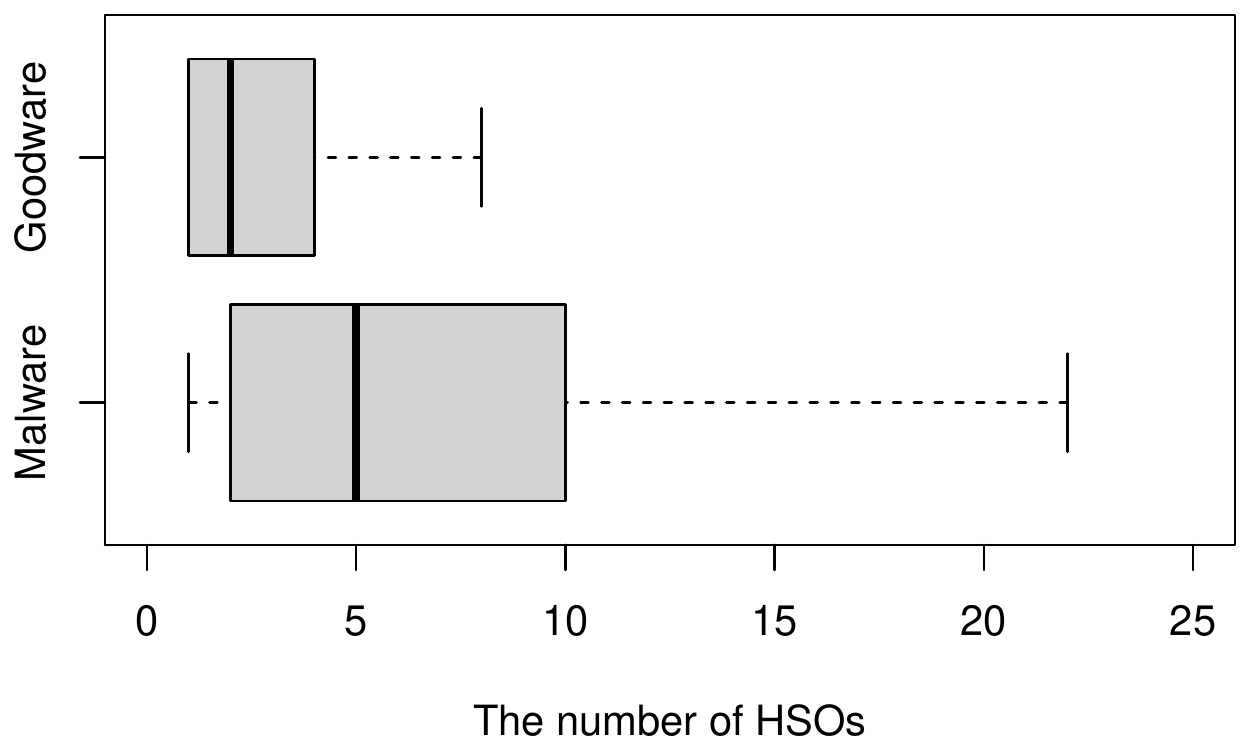}
    \caption{Distribution of the number of HSOs in \textit{benign set} and \textit{malware set}.}
    \label{fig:HSO_common_cases}
\end{figure}

Figure~\ref{fig:HSO_common_cases} further presents the distribution of the number of HSOs detected in the apps from \textit{benign set} and \textit{malware set}.
Expectedly, malware samples involve significantly more HSOs than that of benign apps, as confirmed by the \emph{p-value} a Mann-Whitney-Wilcoxon (MWW) test at a significance level\footnote{Given a significance level $\alpha = 0.001$, if p-value $< \alpha$, there is one chance in a thousand that the difference between the two datasets is sue to a coincidence.} at 0.001~\cite{fay2010wilcoxon}.
This result suggests that HSOs are more favored by malware than benign apps.
Hence, our community should pay more attention to the appearance of HSOs to help security analysts better dissect malicious apps.

Based on the previous experimental results, we manually analyzed the trigger conditions and the corresponding hidden operations to identify \emph{conventional usages}, i.e., HSOs (at least based on our definition) that are actually legitimate and occur relatively often in Android apps. We first inspected the trigger APIs that have appeared more than 50 times in our dataset and determined if it is a \emph{conventional usage}. By doing so, we identified seven major categories of \emph{conventional usages}. Then we reviewed each of the rest of the cases to further filter out the other \emph{conventional usages}.
Finally, 43,141 \emph{conventional usages} have been identified, out of which 40,412 cases belong to the seven major categories.
As the whitelist is generated by manual analysis, it cannot cover all possible \emph{conventional usages}. However, we believe that the majority of the \emph{conventional usages} have been identified (i.e., from the seven categories), new special cases can always be added to the list and incorporated into HiSenDroid in the future.  
We now elaborate on the seven major categories, each with an example code snippet presented in Listing \ref{code:commoncases}.

\begin{lstlisting}[
caption={Examples of conventional usages.},
label=code:commoncases,
firstnumber=1,float=t]
// conventional usages - SDK version check
public void addJavascriptInterface(Object var1, String var2) {
 if (VERSION.SDK_INT < 17) {
  TaoLog.e("HybridWebView", "addJavascriptInterface is disabled before API level 17 for security.");
 } else {
  super.addJavascriptInterface(var1, var2);
}}

// conventional usages - User Interface
class ClickEvent implements View.OnClickListener {
    public void onClick(View view) {
        if(view.getId() == backButton.getId()){
            webView.goBack()
        }
        else if (view.getId() == reloadButton.getId()){
            webView.reload();
}}}

// conventional usages - File Handling
public static File inputstreamtofile(InputStream ins) {
 File SDFile = Environment.getExternalStorageDirectory();
 File desDir=new File(SDFile.getAbsolutePath());
 File newFile=new File(desDir.getAbsolutePath() + File.separatorChar+"myPaint.png") ;
 if(desDir.exists()){
  OutputStream os = new FileOutputStream(newFile);
  while ((bytesRead = ins.read(buffer, 0, 8192)) != -1) {
   os.write(buffer, 0, bytesRead);
}}}
// conventional usages - Permission Check
public static String getDeviceInfo(Context context) {
 if (checkPermission(context, Manifest.permission.READ_PHONE_STATE)) {
  String device_id = tm.getDeviceId();
 } else {
  requestPermissions(context, new String[] {Manifest.permission.READ_PHONE_STATE}, REQUEST_CODE)
}}

// conventional usages - Network
public String g() {
var1 = ((ConnectivityManager)a.getSystemService("connectivity")) .getActiveNetworkInfo();
var9 = var1.getType();
if(var9 == 1){
 var10 = ((WifiManager)a.getSystemService("wifi")).getDhcpInfo();
}}

// conventional usages - Intent Management
public void onReceive(final Context context, Intent intent) {
 String action = intent.getAction();
 if (action.equalsIgnoreCase ("android.net.conn.CONNECTIVITY_CHANGE") {
  connectivityManager.getActiveNetworkInfo();
}}

// conventional usages - SharedPreferences
public class VpnAddressIp{
 public SharedPreferences sp;
 public String VPNAddress() {
  sp = context.getSharedPreferences("SP", Context.MODE_PRIVATE);
  VPNflag = sp.getInt("VPNFlag", 1);
  VPNAddress vpnaddress = new VPNAddress(context);
  if (VPNflag == 1) {
   VPNAddressBean bean = vpnaddress.queryVPN(1);
   networkaddress = bean.getNetwork();
  } 
  return networkaddress;
}}
\end{lstlisting}

\textbf{\emph{SDK Version.}}
With Android system update, new APIs are defined to replace old ones. To maintain the compatibility of apps across different Android versions, it is a common practice to check the SDK version before deciding the right API to use. 
Lines 2\textasciitilde7 show a simplified code snippet of a legitimate conventional usage that fulfills all the rules we defined for an HSO. The code checks whether the Android version is newer than \emph{Android level 17} (line 3), if so, the app leverages the \emph{addJavascriptInterface()} API to inject Javascript into the \emph{WebView} (line 6), otherwise, it logs an error message (line 4) as the API is not available in the Android version lower than 17. While SDK version check commonly exists in both malware and benign apps (with 10,303 and 3,223 cases, respectively), this check does not intend to hide the behaviors within the \emph{if-then-else} statement and hence should be excluded from the HSO results.

\textbf{\emph{User Interface.}}
When the user interacts with UI widgets (e.g., press a button), it retrieves and compares the UI widget's \emph{id} (i.e., a system API) to determine which widget has been fired. If there happened to be a sensitive API invoked in one of the branches' statements, this code block will be misidentified as HSO. Lines 10\textasciitilde17 show an example of a button's callback method, which checks the ID of the buttons (lines 12,15) and either go back to the previous webpage (line 13) or reload the current page (line 16). User Interface has 8,052 instances in the \textit{malware set} and 2,426 instances in the \textit{benign set}.

\textbf{\emph{File.}}
The existence of a file or a directory is usually checked before file operations, such as reading and writing files. The code for checking file existence typically put the subsequent actions in one branch (where the file does exist), and show an error message in the other branch (where the file does not exist). In some cases, it even has only the \emph{if-branch}. Therefore, it satisfies the rules mentioned above and will be mistakenly identified as an HSO. Our results have observed 7,217 and 1,701 cases in our \textit{malware set} and \textit{benign set}, respectively.
A file checking example can be found in lines 20\textasciitilde28, where the code checks the existence of an external storage (line 24), and copy an image there (lines 26,27). 

\textbf{\emph{Permission.}}
Since Android 6.0, the dangerous-level permissions need to be explicitly checked and requested before accessing the APIs protected by these permissions. The example code for checking permission can be found in lines 31\textasciitilde36. It first checks whether the app has been granted \emph{READ\_PHONE\_STATE} permission (line 32). Then, the app either invoke the permission protected API (line 33) or request the missing permission (line 35) based on the check result. Even though a sensitive API \emph{getDeviceId()} is called in one branch, which behaves quite differently than the other branch, it does not mean to hide this behavior. Therefore, it is regarded as a \emph{conventional usage}. Permission check has appeared 6,727 and 936 times in the HSOs identified in the \textit{malware set} and \textit{benign set}, respectively.

\textbf{\emph{Network.}} Network information (e.g., network type, connection status, etc.) is always checked before performing network-related behaviors, ensuring that the network status is suitable for accomplishing the subsequent tasks. For example, the network type (e.g., WiFi, cellular, etc.) is checked before downloading large files, and if it is on the cellular network, the download will be suspended. Another example demonstrated in lines 39\textasciitilde44 examines the type of connected network (line 41) and get its DHCP information if the phone is connected to WiFi (1 is the value of ConnectivityManager\#TYPE\_WIFI). There are 5,224 an 744 identified HSO cases that are related to the \emph{Network} in the \textit{malware set} and the \textit{benign set}, respectively.

\textbf{\emph{Intent.}}
\emph{Intent} is a crucial mechanism to assist the communication between different components in the Android system. \emph{Intent} has various legitimate usages, including starting activities and services, passing data and properties, etc. Lines 47\textasciitilde51 demonstrate a legitimate example of handling the callback method of receiving an \emph{Intent}. In this example, it checks the \emph{action} defined in the received \emph{Intent}, and calls \emph{getActiveNetworkInfo()} method (i.e., a sensitive API) if the \emph{action} is \emph{CONNECTIVITY\_CHANGE}. There are 1,911 an 1,011 identified HSO cases that are related to the \emph{Intent} in the \textit{malware set} and the \textit{benign set}, respectively.

\textbf{\emph{SharedPreferences.}}
In Android system, data can be saved as <key, value> pairs and stores as a \emph{SharedPreferences} object in a file that can be accessed by \emph{getSharedPreferences()} interface. It provides a lightweight and easy-access data store mechanism, which is widely used in storing small collection of data, such as configurations of the app. Reading the values from the \emph{SharedPreferences} and action accordingly is considered a legitimate behavior.  
Lines 54\textasciitilde65 illustrate an example of using \emph{SharedPreferences}, where the code retrieves the value of a configuration item ``VPNFlag'' (line 58) from a \emph{SharedPreferences} object named ``SP'' (line 57), and query the corresponding VPN services accordingly (lines 60\textasciitilde62). There are 878 and 358 cases involving the usage of \emph{SharedPreferences} in our \textit{malware set} and \textit{benign set}, respectively. 

\textbf{Completeness of conventional usages.}
\label{subsec:Completeness_conventional_usages}
Since the conventional usage categories are summarized with manual efforts on a given set of apps, they may not be representative and thereby may not cover all possible cases.
Therefore, in this work, we go one step deeper to further investigate the completeness of all the seven categories of conventional usages by applying our approach to another set of randomly selected 10,000 malware and 10,000 benign apps from AndroZoo~\cite{liu2020androzooopen}. 
We remind the readers that AndroZoo includes over 10 million Android apps that were collected from both the official Google Play store and several third-party app markets. To avoid potential biases in our results, we made additional efforts to remove potentially duplicated apps (i.e., different versions of the same app), and only the latest version is retained. For the 20,000 apps, we apply HiSenDroid to analyze these apps and inspect the trigger conditions that have appeared more than 50 times. We then manually determine if they are conventional usage. To do this, two of the authors spent ten person-days manually summarizing conventional usages (e.g., API-API or Key-API pairs). After manually checking the experimental results and the bytecode of apps, we have totally picked up 41,035 conventional usages, among which 38,920 cases fall into the predefined whitelist (with a success rate of 94.8\%). This result shows that, despite testing on different apps, our whitelist is still quite stable and effective in eliminating conventional usages.

Apart from the aforementioned commonly appeared conventional usages, we further look into some of the uncommon conventional usages. Our manual observation confirms that those uncommon conventional usages are indeed legitimate HSOs that do not appear frequently in Android apps. We present two concrete examples of uncommon conventional usages to illustrate this concept. One example is that an app first checks if the directory of downloads exists (i.e., a standard directory to place files that have been downloaded by the user), and then automatically starts the download using the \emph{android.app.DownloadManager\#enqueue} API once the download manager is ready and connectivity is available. We consider it a conventional usage because it is against the second principle of suspicious HSO's definition: the user does not intend to hide such behavior. In addition, given that there exist several substitute ways of downloading files (e.g., Http request, URLConnection, BufferedInputStream, FileOutputStream, etc.), the native APIs lie in \emph{android.app.DownloadManager} are not that commonly used by app developers. Thus, we regarded it as an uncommon conventional usage. As another example, the sensitive behavior of vibration could be triggered only when a user clicks a certain button. We consider it also a conventional usage because it involves non-hidden behaviors. In fact, if app developers intend to hide sensitive behaviors, it would be obvious that they won't use vibration functionality to notify users. Moreover, the usage of vibration is less common because it would annoy Android users, leading to a poor user experience. Such cases appear less than 50 times in our dataset and thus we regard them as uncommon conventional usage as well.

\section{Suspicious HSO Analysis}
\label{HiSenDroid:SuspiciousHSO}

\begin{figure}[h!]
    \centering 
    \includegraphics[width=0.85\textwidth]{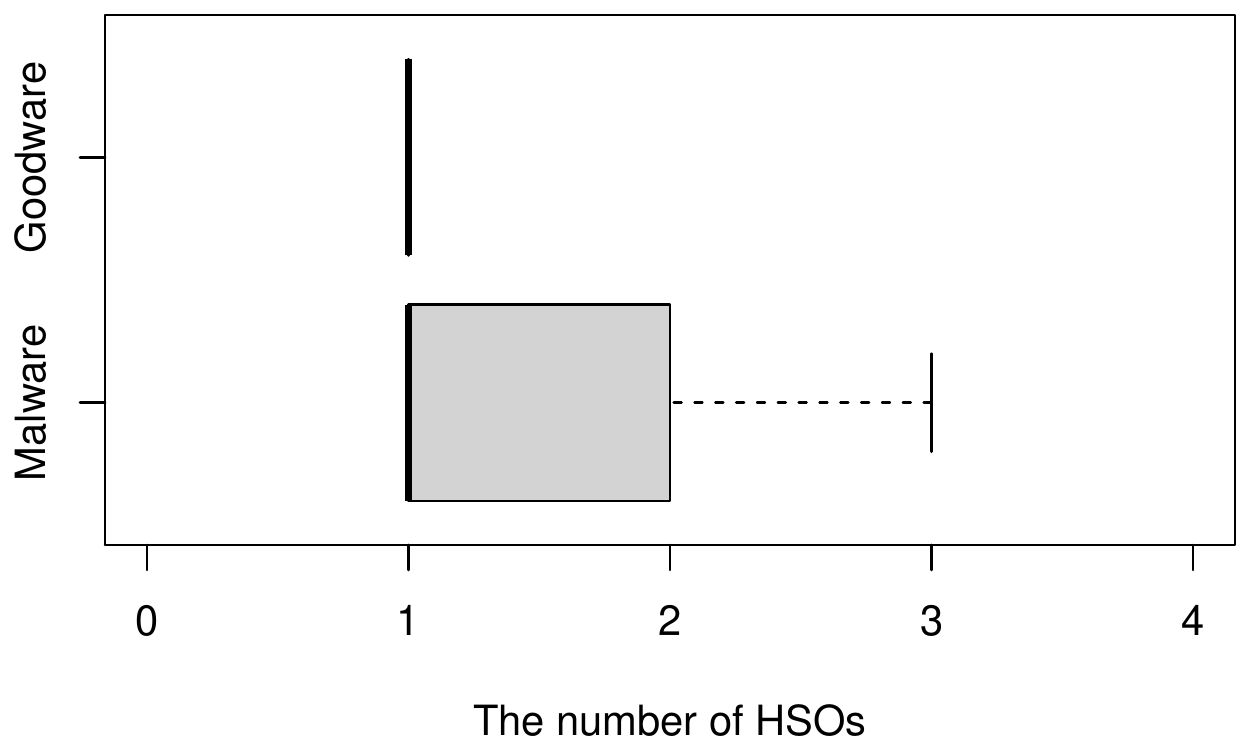}
    \caption{Distribution of the number of \textit{suspicious} HSOs in \textit{benign set} and \textit{malware set}.}
    \label{fig:HSO_no_common_cases}
\end{figure}

After eliminating conventional usages, all the remaining ones will be reported as suspicious HSOs.
Among the 20,000 apps considered in this work, 1,304 of them, including 982 malware and 322 benign samples, were retained.
These apps have been reported to contain in total 2,201 suspicious HSOs, with 1,790 and 441 from malware and benign apps, respectively.
These numbers are recapped in Table~{\ref{tab:numberOfSuspiciousHSOs}}. 
This experimental result shows that suspicious HSOs are widely present in real-world Android apps. 
Figure {\ref{fig:HSO_no_common_cases}} further illustrates the distribution of suspicious HSOs in our dataset. On average, there are 2.0 and 1.4 HSOs in each malware sample and benign app, respectively.

\begin{table}[!h]
\setlength{\belowcaptionskip}{8pt}
\centering
\caption{Number of suspicious HSOs.} 
\label{tab:numberOfSuspiciousHSOs}
\resizebox{0.6\linewidth}{!}{
\begin{tabular}{r| l l } 
Initial Dataset   & \# HSOs & \# Suspicious HSOs \\

\hline
\hline
10,000 benign apps  & 9,368 (in 3,071 apps)  & 441 (in 322 apps)\\

\hline
10,000 malicious apps & 35,974 (in 5,036 apps) &  1,790 (in 982 apps)\\ 

\hline
Total & 45,342 (in 8,107 apps)  &  2,201 (in 1,304 apps)\\ 
\end{tabular} 
}
\end{table}

In this work, the elimination of conventional usages is based on a pre-defined whitelist, which only includes recurrently presented HSOs in benign apps. 
Some less frequent yet still legitimated HSOs could have been overlooked and hence result in suspicious ones.
Indeed, the remaining suspicious HSOs may not always be true positives (i.e., may contain a small number of false positives).
To this end, we go one step further to calculate the precision of our approach in pinpointing suspicious HSOs in Android apps.
Unfortunately, there is no known ground truth available for evaluating HSO usage in Android apps.
Thus, we resort to a manual process to calculate the precision. In this work, we manually inspected the bytecode of apps to see if HiSenDroid correctly and precisely identified the suspicious trigger rather than those commonly appeared code blocks for normal usage. Here, we identify truly suspicious behavior (i.e., confirmed to be true positive) only when the HSO is security-relevant and potentially brings harm to Android users.
Specifically, we rely on two principles to identify truly suspicious HSOs: (1) the hidden behavior involves security-relevant APIs that are protected by Android permissions, classified following the latest Android API-permission mappings (cf. Section 3.1), and (2) the sensitive APIs are intentionally hidden under dedicated trigger conditions. As a result, we count those who meet the two aforementioned principles as true positives. For example, if an app first intends to retrieve Device ID, and when unsuccessful, tries to read the MAC address, we will consider it as a false positive because it is against the second principle: does not intend to hide such behavior.  As another example, an app checks the build's fingerprint to see if it is running on popular emulators, and the sensitive behavior of retrieving subscriberId would be triggered only when it is not running in an emulator. We consider it as a true positive because it involves security-relevant APIs and there is sensitive behavior that is clearly hidden under trigger conditions.
In our dataset, 1,304 apps have been reported to contain at least one HSO. Among the 1,304 apps, HiSenDroid has identified 14,394 HSOs, for which 2,231 of them are regarded as suspicious HSOs. By manually looking at each of those reported suspicious HSOs, we are able to confirm that 1,938 out of 2,231 of them are true positive results (or 293 of them cannot be confirmed without deeply examining the code), giving a precision of 86.8\%.

Recall that the conventional usages are excluded in this work through a whitelist built through empirical evidence, and the whitelist is only considered as a configuration option to our approach.
We believe that the performance of detecting suspicious HSOs could be further improved if we are able to construct a better whitelist of legitimate HSOs. This is nevertheless outside the scope of this work. We hence consider it as our future work.

\subsection{Trigger Conditions}
Known trigger types such as time-bomb and anti-emulator techniques have been broadly studied, specific algorithms for detecting such known trigger types have been developed~\cite{balzarotti2010efficient, kirat2015malgene, fratantonio2016triggerscope}. Nevertheless, the community still lacks the understanding of unknown trigger types. We therefore investigate the most frequent triggering conditions in the suspicious HSOs detected by HiSenDroid. 

HiSenDroid has identified 168 unique APIs that have been traced as the source of the triggers in detected suspicious HSOs. To make it much clearer, we present all these system properties trigger conditions and environment parameters trigger conditions in the artefact package ~\footnote{https://bitbucket.org/se\_anonymous/hisendroid/src/master/experiments\_results/}. We then manually categorize them according to the types of objects they accessed. Table~\ref{tab:top10trigger} illustrates the trigger condition categories and examples of system APIs that are frequently leveraged to fulfill the trigger conditions discovered in HSOs.
The top trigger condition categories include time (e.g., at a certain time of a day), SMS (e.g., when receives SMS of certain formats), location (e.g., if the device is in certain countries), 
system property (e.g., checks the device's manufacturer), and package manager (e.g., if specific apps are installed). 

\begin{table}[!h]
\setlength{\belowcaptionskip}{5pt}
\centering
\caption{Categories of Trigger Conditions in HSO} 
\label{tab:top10trigger}
\resizebox{0.65\linewidth}{!}{
\begin{tabular}{| r | l |} 
\hline
Category &  Most Frequent Trigger API Examples \\
 \hline
\multirow{3}{*}{Time}       & 
 util.Calendar\#get    \\
 &  util.Date\#getTime  \\
 &  util.Calendar\#getTimeInMillis\\
\hline
\multirow{3}{*}{System Properties}     & 
 os.Build\#MODEL \\
&  telephony.TelephonyManager\#getSubscriberId   \\
&  telephony.TelephonyManager\#getDeviceId  \\

\hline
\multirow{3}{*}{Location}       & 
 telephony.TelephonyManager\#getSimCountryIso \\
 &  telephony.TelephonyManager\#getCellLocation \\
 &  location.LocationManager\#getLastKnownLocation \\
\hline
\multirow{3}{*}{SMS Message}     & 
telephony.SmsManager\#divideMessage \\
 &  telephony.SmsManager\#getDefault \\
 &  telephony.SmsManager\#getData \\  
\hline
\multirow{3}{*}{Package Manager}     &  
 content.Context\#getPackageManager \\
 &  content.pm.PackageManager\#getPackageInfo  \\
 &  content.pm.PackageManager\#getApplicationInfo \\
\hline
\multirow{3}{*}{Miscellaneous}     & 
 android.widget.CheckBox\#isChecked \\
 & android.app.KeyguardManager\#isKeyguardLocked  \\
 &  java.net.NetworkInterface\#getHardwareAddress \\
\hline

\end{tabular} 
}
\end{table}

Here we elaborate on each trigger condition category with real-world suspicious HSO cases identified in our dataset. 

\textbf{\emph{Time Triggers}} compare time-related properties (such as current system time, time zone, etc.) with hard-coded values to determine whether or not to execute the hidden sensitive behaviors. Listing \ref{code:Time} demonstrates a code snippet from app \emph{com.wukongtv.wukongtv}\footnote{SHA-256:3397079daa388bdbcdcc42b6834d3c792bf5c80ad24491e3893de7cfc2b11db7}, which leverages time-related triggers to hide suspicious behaviors. When the first time the app launches, it writes the timestamp into the SharedPreferences (i.e., \textit{var0}). It then compares the current system time with the first launch time (line 6); if the time interval is greater than one day, it triggers the sensitive method \textit{bq.f()} (line 7) that retrieves the information (e.g., package name and process name) of running tasks (lines 10\textasciitilde15). Doing so conceals the suspicious behaviors from automatic dynamic detection, which usually starts testing immediately after the app is installed.

\begin{lstlisting}[style=JAVA, escapechar=\%,
caption={Code Example of Time Trigger.},
label=code:Time,
firstnumber=1]
static void c(Context var0) {
 Calendar var10000 = Calendar.getInstance();
 int var2 = var10000.get(6) * 100;  //day_of_year
 var2 += var10000.get(11);          //hours_of_day
 // var0 is retrieved from SharedPreferences
 if %\underline{(Math.abs(var0/100L - (long) (var2/100)) >= 1L)}% {
  ab.h = bq.f(var0);
}}}

public static Long[][] f(Context var0) {
 var31 = var3.getRecentTasks(10, 1);
 while(var31.iterator().hasNext()){
  // get package name and process name of recent tasks
  ...
}}
\end{lstlisting}

\textbf{\emph{System Property Triggers}} leverage system properties, such as the phone model, the phone number, and hardware information, to limit the sensitive behaviors within specific device brands (e.g., Samsung) or types (e.g., real device). These triggers are also commonly adopted by anti-emulator techniques to detect the presence of emulators. Listing \ref{code:System_Properties} demonstrates an anti-emulator example extracted from app \emph{com.gwsoft.imusic.controller} \footnote{SHA-256:8c679a7c57a7fbb355fb363d3784cc8380655701d482837869edd95f3a3ea470}, which checks if the build's fingerprint contains specific strings that indicate popular emulators (line 2). The sensitive behavior of retrieving \textit{subscriberId} (line 5) is only executed if it does not run in an emulator. 

\begin{lstlisting}[style=JAVA, escapechar=\%,
caption={Code Example of System Property Trigger.},
label=code:System_Properties,
firstnumber=1]
private static boolean a(Context var0) {
 if %\underline{(Build.FINGERPRINT.contains("vbox86p/vbox86p")\&\&} \underline{!Build.FINGERPRINT.contains("ttVM\_Hdragon/ttVM\_Hdragon")\&\&} \underline{!Build.FINGERPRINT.contains("generic/sdk/generic")\&\&} \underline{!Build.FINGERPRINT.contains("generic\_x86/sdk\_x86/generic\_x86")}%
 // omit other strings that fingerprints popular Android emulators
 ){
  var2 = ((TelephonyManager)var0.getSystemService("phone")) .getSubscriberId();
  var11.put("imsi", var2);}}
\end{lstlisting}

\textbf{\emph{SMS Triggers}} Utilize the content, type, and phone number of received SMS messages to hide sensitive behaviors. An example derived from app \emph{com.fingersoft.hillmotor}\footnote{SHA-256:95e1cf498dec79351a9d104f5e9fb0110c267e9eff0099ada475d8832a2afb7302521} is shown in Listing \ref{code:SMS_Message}. When an SMS message is received, it checks the originating address of the message (line 4). If it matches a pre-defined value (e.g., 10 or 11 etc in this example), the behavior that repeatedly sends a message (line 6) to the same number via a text message service.

\begin{lstlisting}[style=JAVA, escapechar=\%,
caption={Code Example of SMS Trigger.},
label=code:SMS_Message,
firstnumber=1]
//var2 is the originating address retrieved from SMS
//var3 is the message body
public boolean repeat(Context var1, String var2, String var3) {
if ((var2.startsWith("10") || var2.startsWith("11") || var2.startsWith("12")) && !var2.equals("114") && !var2.equals("12306") && !var2.equals("116114") && !var2.equals("12580")) {
SmsManager var13 = SmsManager.getDefault();
var25.sendTextMessage(var2, (String)null, "Y", var16, var12);
}}
\end{lstlisting}

\textbf{\emph{Location Triggers}} obscure sensitive behaviors with fine grained (e.g., latitude and longitude) and coarse grained (e.g. country) location information. Listing \ref{code:Location} shows an example derived from \textit{com.inter.\\apps.patqut.apk} \footnote{SHA-256:22c9d7738073a7ac8f9b58029057c2741e89faac76b623837db2f3a8bb2d93c5} which queries the country code of the device (saved as \textit{var1}), and checks if it is in Malaysia (line 4). If so, the app then triggers the \textit{postLoginData2()} method (line 5), which retrieves the device's id (line 9) and hands it over to another activity for further malicious behaviors. 

\begin{lstlisting}[style=JAVA, escapechar=\%,
caption={Code Example of Location Trigger.},
label=code:Location,
firstnumber=1]
TelephonyManager var3 = (TelephonyManager)this.getSystemService("phone");
String var1 = var3.getSimCountryIso().toUpperCase();
public void getin(String var1) {
 if %\underline{(var1.equals("MY"))}% {
  this.postLoginData2();
}}

public void postLoginData2() {
 String var2 = ((TelephonyManager)this.getSystemService("phone")) .getDeviceId();
 // hand over the obtained DeviceId to a new activity
 ...
}
\end{lstlisting}

\textbf{\emph{Package Manager Triggers}} scan the list of installed apps and inspect if specific apps (usually anti-virus tools) are installed before conducting any sensitive behaviors.
Listing~\ref{code:Check_PackageName} shows a code snippet taken from \emph{flash15.1.apk} \footnote{SHA-256:fdaba7f032ee7ff9adf799713b25d4c2fef86ddbbe8709bf6ec021505b8f1d0d} which searches for \emph{AhnLab V3 Mobile Plus 2.0} (i.e., an anti-virus tool) in the list of installed apps (line 1\textasciitilde7). If the anti-virus tool is not installed (line 10), it then starts its malicious behaviors. Specifically, it gets the package name of the current active activity (lines 11, 12), and puts it into sleep if it is a bank app. After that, it launches a new activity that contains a phishing web page to steal user's bank credentials.

\begin{lstlisting}[style=JAVA, escapechar=\%,
caption={Code Example of Package Manager Trigger.},
label=code:Check_PackageName,
firstnumber=1]
private boolean judgeAV() {
 this.pm = this.getPackageManager();
 this.listAppcations = this.pm.getInstalledApplications(8192);
 for(int v = 0; v < listAppcations.size(); ++v) {
  if(listAppcations(v).name.equalsIgnoreCase("AhnLab V3 Mobile Plus 2.0")){
   return true;}
  return false;}
    
public void run() {
if %\underline{(!AutBan.this.judgeAV())}% {
 List var2 = ((ActivityManager)AutBan.this.getSystemService ("activity")).getRunningTasks(1);
 String var1 = ((RunningTaskInfo)var2.get(0)).topActivity .getPackageName();
 // if the top activity is a bank app, it puts the activity into sleep and start a phising page
 ...
}};
\end{lstlisting}

\textbf{\emph{Other Triggers.}} Besides the most frequent trigger categories, we also observed some sophisticated triggers specifically designed to counter automated dynamic testing approaches. 
Listing~\ref{code:Item_Click} shows an example taken from a music player app \emph{com.gwsoft.imusic.controller}\footnote{SHA-256:8c679a7c57a7fbb355fb363d3784cc8380655701d482837869edd95f3a3ea470}. The app hides sensitive behaviors that retrieve the device's information (lines 8\textasciitilde12) behind a trigger that will only be fired when an item on the song list (i.e., \textit{mCatalogSongsList}) is clicked (line 2). The trick here is that automated dynamic testing tools running on an emulator are likely not to have any music files and, therefore, will have no items on the list to click. Hence, only legitimate users who intend to use it to play music will have the chance to trigger the sensitive behavior. 

\begin{lstlisting}[style=JAVA, escapechar=\%,
caption={Code Example of Other Trigger.},
label=code:Item_Click,
firstnumber=1]
//contains at least one song in the list
public void onItemClick(AdapterView<?> var1, View var2, int var3, long var4) {
 if %\underline{(mCatalogSongsList != null \&\& var3 + -1 >= 0 \&\& var3 + -1} \underline{< mCatalogSongsList.size())}%{
  CountlyAgent.onEvent(CuttingActivity.this,  "activity_diy_do_re", String.valueOf(var3));
}}

public static void onEvent(Context var0, String var1, String var2) {
 HashMap var3 = new HashMap;
 var3.put("phone", getIMSI());
 var3.put("ip", getLocalIpAddress());
 var3.put("app_version", versionName);
 var3.put("imei",getDeviceId());
}
\end{lstlisting}


\begin{figure}[t!]
    \centering
    \subfigure[Categories of Trigger Conditions in HSO.]{\label{fig:top10trigger}\includegraphics[width=0.45\linewidth]{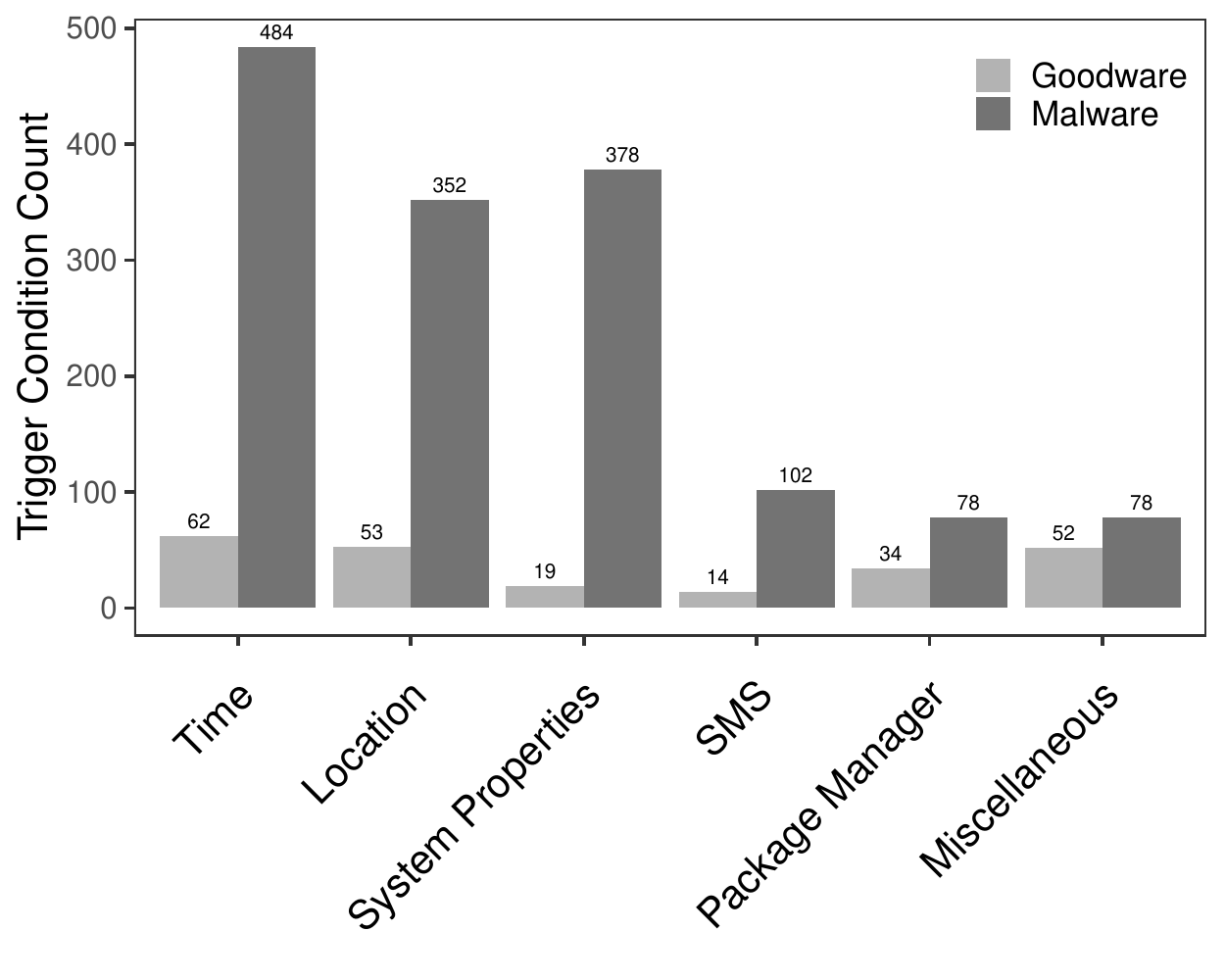}}
    \subfigure[Top 10 Categories of Sensitive APIs in Malware.]{\label{fig:malware_goodware_sensitive_api}\includegraphics[width=0.45\linewidth]{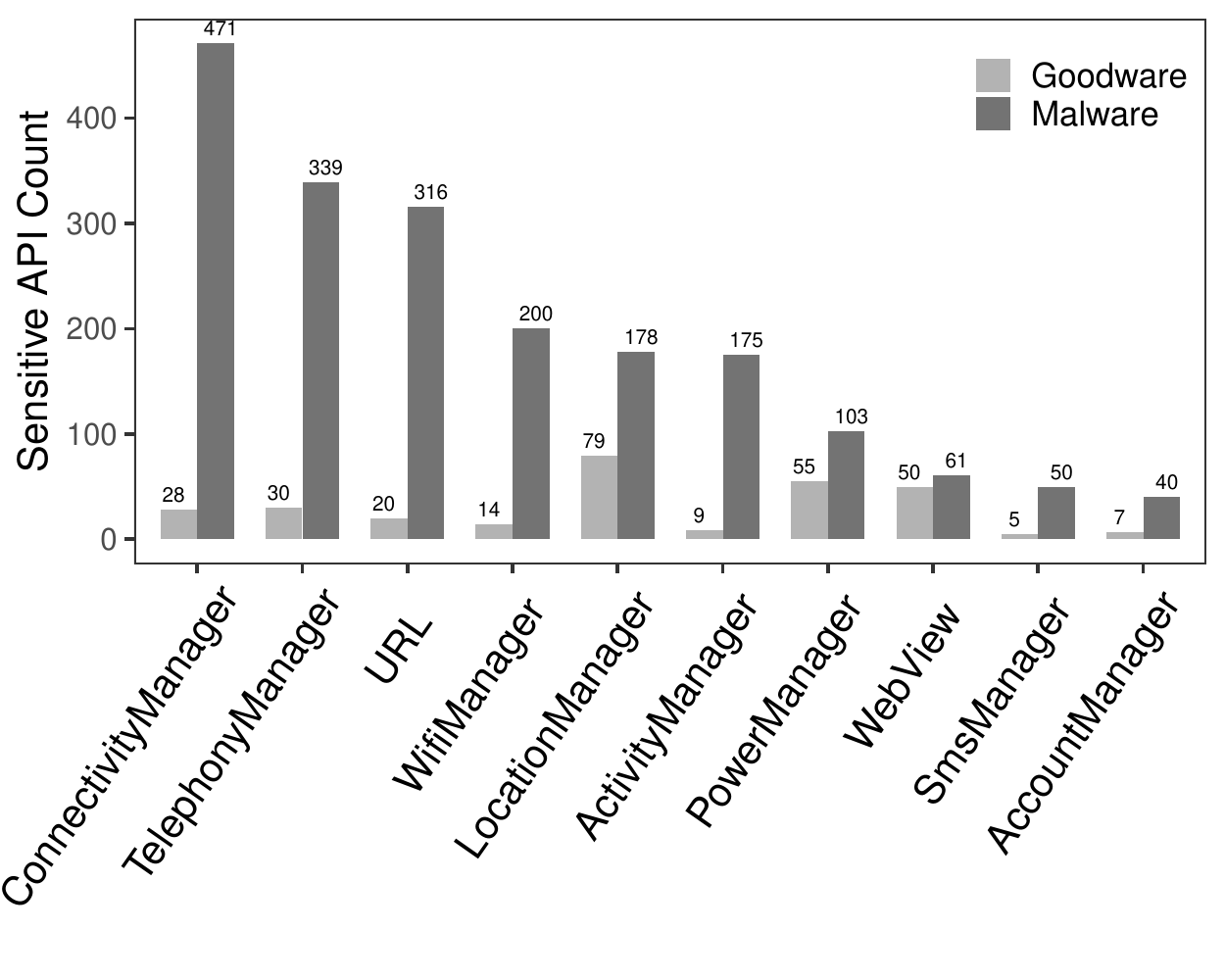}}
	\caption{Categories of Trigger Conditions in HSO and Sensitive APIs in Malware.}
    \label{fig:top10trigger_malware_goodware_sensitive_api}
\end{figure}

\subsection{Sensitive APIs involved in suspicious HSOs}

While the invocation of sensitive APIs does not necessarily mean it is malicious, sensitive APIs deliberately hidden in an HSB 
do raise its suspicion. 
HiSenDroid has identified 134 unique hidden sensitive APIs that appeared 3,195 times in our dataset. 
Figure \ref{fig:malware_goodware_sensitive_api} presents the top ten classes of the most frequently invoked sensitive APIs in \textit{malware set} and \textit{benign set}. The most involved APIs are network related, including the ones in \textit{URL}, \textit{ConnectivityManager}, and \textit{WifiManager} classes. The\textit{WebView} ((displays web pages))
and \textit{SmsManager} (manages SMS operations such as sending text messages) are also prevalently used in HSOs. Other commonly involved API classes include \textit{PowerManager} (controls the power state of the device such as keeping the screen stay on), \textit{LocationManager} (provides access to the system location services such as getting last known location), \textit{ActivityManager} (gives information about activities and services such as getting running tasks on the phone), \textit{TelephonyManager} (provides access to information of the telephony services such as phone number), and \textit{AccountManager} (manages user's online accounts). The detailed most common APIs in HSOs can be found in Table \ref{tab:top10apis}.

Interestingly, while most of the API classes have significantly more instances in malware samples than benign apps, \textit{WebView} is an exception. We therefore took an in-depth look into benign apps with WebView APIs in their HSOs and observed that 34 out of 50 cases are free apps that display advertisement web pages for revenue.

\begin{table}[!h]
\setlength{\belowcaptionskip}{5pt}
\centering
\caption{Details of The Top 10 Classes of Hidden Sensitive APIs in HSO} 
\label{tab:top10apis}
\resizebox{0.65\linewidth}{!}{
\begin{tabular}{| r | l | } 
\hline
Class &  Most Frequent Sensitive API Examples \\
\hline
\multirow{3}{*}{ConnectivityManager}     &  
 net.ConnectivityManager\#getActiveNetworkInfo \\
 &   net.ConnectivityManager\#getNetworkInfo  \\
 &   net.ConnectivityManager\#getAllNetworkInfo  \\
\hline
\multirow{3}{*}{TelephonyManager}     & 
 telephony.TelephonyManager\#getDeviceId  \\
 &  telephony.TelephonyManager\#getSubscriberId \\
  &  telephony.TelephonyManager\#getCellLocation \\
\hline
\multirow{3}{*}{URL}       & 
 net.URL\#openConnection    \\
 &  net.URL\#getContent     \\
  &  net.URL\#openStream     \\
\hline
\multirow{3}{*}{WifiManager}     &  
  net.wifi.WifiManager\#getScanResults\\
 &    net.wifi.WifiManager\#getConnectionInfo  \\
  &    net.wifi.WifiManager\#getWifiState  \\
 \hline
 \multirow{3}{*}{LocationManager}       & 
 location.LocationManager\#getLastKnownLocation    \\
 &  location.LocationManager\#requestLocationUpdates \\
 &  location.LocationManager\#getBestProvider \\
\hline
\multirow{3}{*}{ActivityManager}     & 
 app.ActivityManager\#getRunningTasks  \\
 & app.ActivityManager\#getRecentTasks  \\
  & app.ActivityManager\#moveTaskToFront  \\
\hline
\multirow{3}{*}{PowerManager}     & 
 os.PowerManager.WakeLock\#release \\
 &  os.PowerManager.WakeLock\#acquire() \\
  &  os.PowerManager.WakeLock\#acquire(long) \\
\hline
\multirow{3}{*}{WebView}     & 
 webkit.WebView\#setBackgroundColor  \\
 &  webkit.WebView\#addJavascriptInterface   \\
  &  webkit.WebView\#loadDataWithBaseURL   \\
\hline
\multirow{3}{*}{SmsManager}       & 
 telephony.SmsManager\#sendTextMessage    \\
 &  telephony.SmsManager\#sendMultipartTextMessage  \\
 &  telephony.SmsManager\#sendDataMessage  \\
\hline
\multirow{3}{*}{AccountManager}     & 
accounts.AccountManager\#getAccountsByType \\
& accounts.AccountManager\#getAccounts \\
& accounts.AccountManager\#getUserData \\
\hline

\end{tabular} }
\end{table}

\subsection{Trigger Condition to Hidden Sensitive API Pairs}
\begin{figure}[!t]
    \centering 
    \includegraphics[width=0.8\linewidth]{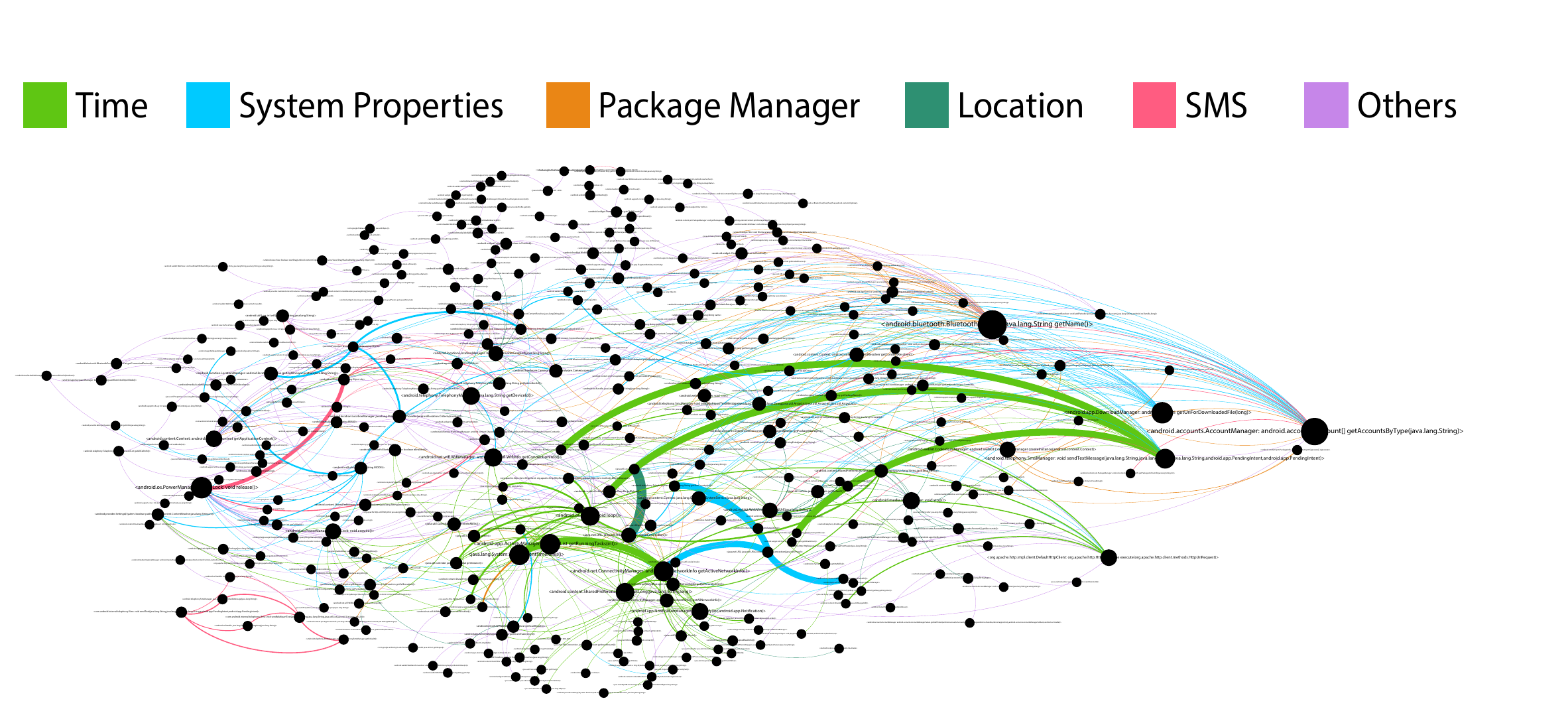}
    \caption{The HSO Trigger-Sensitive API Pairs Graph. }
    \label{fig:Trigger-Sensitive_Pair_graph}
\end{figure}

We now investigate the relationships between trigger conditions and the hidden sensitive APIs accessed in their corresponding HSOs so as to identify common patterns leveraged by attackers to achieve malicious purposes.
Figure~\ref{fig:Trigger-Sensitive_Pair_graph} graphically summarizes such relationships, i.e., trigger-to-hidden-sensitive-API pairs, where each node represents an API in either the trigger conditions or the hidden sensitive branches, while each edge denotes the connections between them. HiSenDroid has identified 404 nodes within which 346 are APIs in trigger conditions, 134 are APIs in hidden sensitive branches, and 15 APIs exist in both triggers and hidden sensitive branches. There are 2,847 edges found between them, which are illustrated in different colors according to their trigger conditions' categories. 

Table \ref{tab:top10pairs} further details the top 10 pairs found in HSOs with their categories and counts. The most frequent HSO patterns are to hide network-related activities behind retrieving the phone's location. For instance, the top one pattern that appeared 135 times in our dataset requests the SIM provider's country code. Based on the user's location, it then determines whether or not to open a web page, and what web pages (e.g., advertisement pages) to display to the user. 
Time-related HSO patterns are also widely found in the detected HSOs. They firstly compare the current system time with preset values. If the condition fulfills (e.g., the app is running for more than ten minutes), they try to initialize a network connection and send out the user's private information such as IMEI, phone number, etc. More than 250 instances in our dataset leverage this pattern to steal users' private information stealthily. Other frequent HSO patterns on the top list are involved in anti-emulator tricks include checking the phone's model name and checking if specific apps are installed (which could indicate if it is an emulator) before acquiring sensitive information.

\begin{table*}[!h]
\setlength{\belowcaptionskip}{5pt}
\centering
\caption{Top 10 Trigger Condition to Hidden Sensitive API Pairs.}
\label{tab:top10pairs}
\resizebox{0.9\linewidth}{!}{
\begin{tabular}{ | l | l | l | c | } 
\hline
Category & Trigger Condition APIs & Hidden Sensitive APIs & Counts\\
\hline
Location &
TelephonyManager\#getSimCountryIso
& java.net.URL\#openConnection
& 135 \\
\hline
Location &
TelephonyManager\#getCellLocation 
& ConnectivityManager\#getActiveNetworkInfo
& 131 \\
\hline
Time &
java.lang.System\#currentTimeMillis 
& ConnectivityManager\#getNetworkInfo
& 80 \\
\hline
Time &
java.lang.System\#currentTimeMillis 
& ConnectivityManager\#getAllNetworkInfo
& 66 \\
\hline
Time &
java.lang.System\#currentTimeMillis 
& ConnectivityManager\#getActiveNetworkInfo
& 57 \\
\hline
System Properties &
android.os.Build\#MODEL
& TelephonyManager\#getSubscriberId
& 54 \\
\hline
System Properties &
android.os.Build\#MODEL
& Settings\$System\#putInt
& 54 \\
\hline
SMS &
android.os.Message\#obj
& PowerManager\$WakeLock\#release
& 52 \\
\hline
Package Manager & 
PackageManager\#getInstalledPackages
& ActivityManager\#getRunningTasks
& 51 \\
\hline 
Time &
java.lang.System\#currentTimeMillis 
& DefaultHttpClient\#execute
& 49 \\
\hline
\end{tabular} }
\end{table*}

\subsection{Suspicious HSOs in Third-party Code}
\label{subsec:library}

Finally, we further look into the identified HSOs to check if they are introduced by app developers or reused by third-party libraries.
For the sake of simplicity, we consider the code only located in the unique app package (also known as the app id) as developers newly implemented code while all the other code (i.e., in packages not connected with the app's domain name) as third-party code (e.g., third-party libraries).
Among the 2,201 HSOs, surprisingly, over half of them (i.e., 1,342) is contributed by third-party code (i.e., 1,173 HSOs in malware and 169 in benign apps), among which malware tends to be more favored to introduce HSOs through third-party code than benign apps.
This experimental evidence suggests that attackers have more incentives to achieve malicious behaviors through third-party code as it allows easy code reuse that makes it much easier to implement new malware.

\subsection{Comparison with state-of-the-art}
\label{subsec:comparison}

We now compare our approach with state-of-the-art works targeting the problem of detecting hidden sensitive operations. To the best of our knowledge, there are two closely related approaches: HSOMiner\cite{pan2017dark} and TriggerScope\cite{fratantonio2016triggerscope}. Unfortunately, the source code of HSOMiner is not publicly available, and it is infeasible to compare against it because they trained data on the authors' labelled dataset, which has also not been publicly released. We have contacted the authors about launching their approach to analyze Android apps. Unfortunately, we have not yet received any response from them. Similarly, the authors of TriggerScope have also not made it publicly available. As a result, we cannot compare with TriggerScope as well. Fortunately, Jordan Samhi has provided a re-implemented version\footnote{https://github.com/JordanSamhi/TSOpen} of TriggerScope based on the details given in its research paper. The re-implemented version is named as TSOpen (referring to as the open implementation of TriggerScope) and has already been leveraged by previous studies~\cite{samhi2021effectiveness}. In this work, we resort to comparing our approach with TriggerScope by actually comparing it with TSOpen.

To set up the experiments for a fair comparison, we run TSOpen on the same 10,000 malware and 10,000 benign apps selected in evaluating HiSenDroid in section 4 (as indicated in the second and third columns in Table~\ref{tab:comparison_TS}). The experiments are executed under the same environment, i.e., the same server and the same timeout threshold (i.e., 20 minutes). 

The experimental results are summarized in Table~\ref{tab:comparison_TS}.
Overall, the number of HSOs found by HiSenDroid in goodware and malware (i.e., 441 and 1,790, respectively) is larger than those found by the TSOpen (i.e., 110 and 237, respectively). 
Recall that when evaluating the performance of HiSenDroid at the beginning of Section~\ref{HiSenDroid:SuspiciousHSO}, we have manually validated the 2,231 suspicious HSOs yielded by HiSenDroid, for which 1,938 are confirmed to be true positives, giving a precision of 86.8\%.
In this work, we further conduct the same manual validation for the results of TSOpen. 
Our manual validation confirms that TSOpen has at least correctly detected 90.2\% of logic bombs.
This result is expected as TSOpen only detects three types of HSOs (i.e., time, location, and SMS) while HiSenDroid aims at detecting a broader scope of HSOs.
To enable a fair comparison\footnote{We consider the original outputs of HiSenDroid and TSOpen for comparison since only a small number of their results could be false positive.}, in this work, we will only consider HiSenDroid's results falling in these three categories.

As highlighted in Table~\ref{tab:comparison_TS} (cf. Columns 6-8), HiSenDroid detects more HSOs in all of the three categories. Among the detected HSOs, we find that 265 HSOs (186, 48, and 31 in time, location, and SMS, respectively) were detected by both tools (as summarized in the fourth row in Table~\ref{tab:comparison_TS}). Besides that, there are 802 HSOs (360, 357 and 85 in time, location, and SMS, respectively) exclusively detected by HiSenDroid, while still 82 HSOs (38, 1, and 43 in time, location, and SMS, respectively) identified by TSOpen are not flagged by HiSenDroid.

On a further investigation, we found the reason why HiSenDroid failed in detecting the 82 HSOs is that HiSenDroid's definition of potentially-sensitive APIs is different from the definition in TSOpen. 
In this work, we consider all the APIs that are protected by Android permissions as potentially sensitive, while TSOpen takes a different approach to pre-select such a set of sensitive APIs\footnote{The sensitive APIs are a part of internal implementation of TSOpen, which is not configurable.}.
Their set of sensitive APIs includes both permission-protected and permission-free APIs.
For example, TSOpen treats the following two APIs, <android.content.BroadcastReceiver: void abortBroadcast()> and <android.os.Handler: boolean sendEmptyMessage(int)>, as sensitive APIs. However, HiSenDroid does not consider them as sensitive because they are not protected by permissions. 
Furthermore, considering that TriggerScope was published in 2016 and the Android API rapidly evolves, it is understandable that certain APIs (especially the latest ones) are not included, resulting in possibly less suspicious HSOs. 
Moreover, as claimed in their paper, TriggerScope only focused on characterizing logic bombs on some given behaviors, while HiSenDroid treated each sensitive API in state-of-the-art Android API-permission mappings~\cite{au2012pscout,backes2016demystifying,aafer2018precise,chaoran2022cross} as a target API, leading to better performance in terms of both quantity and variety in detected HSOs, compared with TriggerScope.

\begin{table}[t!]
\setlength{\belowcaptionskip}{5pt}
\centering
\caption{The comparison results between HiSenDroid and TSOpen.} 
\label{tab:comparison_TS}
\resizebox{\linewidth}{!}{
\begin{tabular}{r | c c c c | c c c } 
\hline
Tool & \# Analyzed  & \# Analyzed & \# HSOs in & \# HSOs in & \# Time & \# Location & \# SMS  \\
Name & Goodware   & Malware & Goodware &  Malware & HSOs   & HSOs   & HSOs  \\
\hline
HiSenDroid & 10,000   & 10,000 & 441(in 322 apps) & 1,790(in 982 apps) & 546 & 405  & 116 \\
TSOpen & 10,000  & 10,000  & 110(in 51 apps)  & 237(in 123 apps)  & 229 & 49 & 69 \\
\hline
Common & 10,000  & 10,000  & 71  &  194 & 186 &  48 & 31 \\
\hline
\end{tabular} }
\end{table}

\subsection{Impact of Code Obfuscation}
\label{subsec:obfuscation}

As experimentally revealed by Zeng~{\cite{zeng2018resilient}} and Moser et al.~{\cite{moser2007limits}}, trigger conditions of HSOs could be obfuscated in order to evade the detection of advanced semantics-based malware analyzers.
Therefore, we are interested in checking to what extent our approach is impacted by obfuscation, especially when applied to pinpoint HSOs in real-world Android apps. 
Since there is no existing dataset that is suitable for our experiment, we resort to preparing such a dataset from scratch, i.e., to form a set of obfuscated app pairs for which each pair contains a non-obfuscated app and its obfuscated counterpart.
We start by randomly selecting 1,000 malware from our dataset and then apply Obfuscapk\cite{aonzo2020obfuscapk} on them to generate their obfuscated counterparts.
Obfuscapk is a modular Python tool designed to directly obfuscate closed-source Android apps.
Obfuscapk supports six types of obfuscation operations, which could be configured to achieve different granularities when obfuscating Android apps.

The six types of operations are summarized as follows.

\begin{enumerate}
\item Nop: Insert junk code. Nop, short for no-operation, is a dedicated instruction that does nothing. This technique just inserts random nop instructions within every method implementation.

\item Rename: operations that change the names of the used identifiers (classes, fields, methods).

\item Reorder: This technique consists of changing the order of basic blocks in the code. When a branch instruction is found, the condition is inverted (e.g., branch if lower than, becomes branch if greater or equal than) and the target basic blocks are reordered accordingly. Furthermore, it also randomly rearranges the code abusing goto instructions.

\item Reflection: This technique analyzes the existing code looking for method invocations of the app, ignoring the calls to the Android framework (see AdvancedReflection). If it finds an instruction with a suitable method invocation (i.e., no constructor methods, public visibility, enough free registers etc.) such invocation is redirected to a custom method that will invoke the original method using the Reflection APIs.

\item Advanced Reflection: Uses reflection to invoke dangerous APIs of the Android Framework. To find out if a method belongs to the Android Framework, Obfuscapk refers to the mapping discovered by Backes et al.~\cite{backes2016demystifying}

\item Encryption: packaging encrypted code/resources and decrypting them during the app execution. When Obfuscapk starts, it automatically generates a random secret key (32 characters long, using ASCII letters and digits) that will be used for encryption.
\end{enumerate}

In this work, we are interested in checking the impact of all of these six types of operations on our approach. Hence, for each of the selected apps and each obfuscation type, we launch Obfuscapk to generate an obfuscated app.
For the 1,000 selected apps, we expect to generate 6,000 obfuscated apps and eventually form 6,000 obfuscated app pairs.
We then launch HiSenDroid to analyze those apps and compare the number of detected HSOs obtained on apps with and without obfuscation.
Table~\ref{tab:comparison_obfuscation_malawre} summarizes our experimental results.

Expectedly, except for reflection, our approach is resilient to all the other four obfuscation types.
Our deep analysis reveals that the reason why HiSenDroid is unaffected by Nop obfuscator is that Nop obfuscator will only insert junk code, which is a dedicated instruction that does nothing. In terms of Rename and Reorder obfuscator, their code transformations will retain the functionality as the original APK thus will not impact our approach. Also, the reason why the Encryption obfuscator has no effect on HiSenDroid is that it will only encrypt constant strings in code, which will not impact the data flow analysis of our approach. In terms of reflection obfuscator and advanced Reflection obfuscator, both trigger conditions and sensitive API invocations can be redirected to 
other code entities by reflection calls, while those entities cannot be always resolved statically since the reflection call targets may not be statically resolved, which would lead to false negatives of HiSenDroid. 
The remaining two types that have an impact on our approach are all related to reflection, which performs complicated code changes that will likely break the data flow processes. 
Nevertheless, even for reflection, our approach can still detect around one-third of HSOs.

To better mitigate the impact of reflection-based obfuscation on our approach, we further propose to strengthen the capability of HiSenDroid by integrating the state-of-the-art reflection analysis tool DroidRA to handle reflection usages~\cite{sun2021taming}. After statically locating the reflective calls, DroidRA can transform a reflection-included Android app to a reflection-free version, where the located reflective calls will be represented by standard java calls. The newly generated reflection-free app would allow HiSenDroid to yield reflection-aware analysis results. Specifically, considering the 345 apps and 821 apps that are obfuscated by reflection calls and advanced reflection calls, respectively, we first apply DroidRA to convert them into 1,166 reflection-free apps. After that, we execute HiSenDroid to perform HSO analysis on these new apps and compare the number of detected HSOs obtained based on the original apps. As a result, HiSenDroid is able to detect all 13 reflection-relevant HSOs which are obfuscated with reflection obfuscation, while detecting 124 (with a success rate of 85.5\%) reflection-relevant HSOs  that are obfuscated with advanced reflection obfuscation. The reason why HiSenDroid fails on detecting a small portion of reflection-relevant HSOs is that DroidRA may not resolve all the advanced reflective calls. For example, DroidRA relies on COAL~\cite{octeau2015composite} solver to infer reflective calls, which might introduce false negatives, leading to reflection calls unresolved and thus can not be successfully detected by HiSenDroid. Nevertheless, our experimental result shows the capability of HiSenDroid in achieving most of the reflection-aware hidden sensitive operation detections.

\begin{table}[t!]
\setlength{\belowcaptionskip}{5pt}
\centering
\caption{The comparison results of HiSenDroid before and after obfuscation techniques in malware.} 
\label{tab:comparison_obfuscation_malawre}
\resizebox{\linewidth}{!}{
\begin{tabular}{r | c c c c c c c c } 
\hline
\textbf{Obfuscator} & \textbf{Nop}  & \textbf{Rename} & \textbf{Reorder} & \textbf{Reflection} & \textbf{Advanced Reflection}  & \textbf{Encryption}\\
\hline
\# HSOs Before Obfuscation   & 144(in 821 apps) & 18(in 332 apps) & 146(in 792 apps)  & 13(in 345 apps)  & 145(in 821 apps)  & 141(in 811 apps)\\
\# HSOs After Obfuscation  & 144(in 821 apps)  & 18(in 332 apps)  & 146(in 792 apps) &  4(in 345 apps) &  64(in 821 apps)  & 141(in 811 apps)\\
\hline
Common & 144  & 18 & 146  &  4 & 64 &  141  \\
\hline
\end{tabular} }
\end{table}

\section{Implication}
\label{HiSenDroid:implication}
After being able to automatically detect suspicious HSOs, we now go one step further to investigate how such HSOs can bring security harms to users. There might be different security implications, in this work, we only focus on sensitive data leaks, which is also part of our initial attempts towards demonstrating the usefulness of identifying suspicious HSOs. Specifically, we are interested in detecting hidden sensitive data flows (HSDFs), i.e., leaking sensitive data collected through HSOs.
To the best of our knowledge, hidden sensitive data flow has not yet been explored by our community.
Unfortunately, it has not even been clearly defined.
To this end, we first define HSDF following the previous rules leveraged to define HSOs (cf. Section~\ref{HiSenDroid:motivation}).
Let $S$ denote a sensitive data flow (also known as a private data leak as mentioned in the FlowDroid work~\cite{arzt2014flowdroid}), we consider that a sensitive data flow  happens when a sensitive ``tainted'' information goes from a source (e.g. the API method \emph{getDeviceId}) to a given sink (e.g. the API method \emph{sendTextMessage}).

\textbf{Definition 3 [Hidden Sensitive Data Flow (HSDF)]:}
A sensitive data flow $S$ is an HSDF if the source of $S$ appears in the hidden sensitive branch of a HSO.

Although HSDFs have not yet been specifically exploited by the state-of-the-art, our community has proposed various approaches to detect general sensitive data-flows.
One of the most famous approaches is FlowDroid~\cite{arzt2014flowdroid}, a state-of-the-art static analyzer that performs taint analysis to pinpoint sensitive data leaks flowing from a pre-defined set of \emph{source} methods to \emph{sink} methods.
These \emph{source} and \emph{sink} methods can be easily customized.
In this work, we leverage FlowDroid to detect sensitive data flows related to HSOs.
If a sensitive data flow reported by FlowDroid has its source method invoked in an HSO, we regard it as an HSDF.

By applying FlowDroid\footnote{In this work, the latest \emph{development} branch of FlowDroid{\cite{FlowDroid_develop_branch}} is leveraged for the experiments. It should be roughly equivalent to the FlowDroid 2.8 release.} to 1,304 apps (982 malware and 322 goodware) involving suspicious HSOs, we find that 67 apps further involve HSDFs, accounting to in total 401 HSDFs.
While manually checking the experimental results of FlowDroid and HiSenDroid, we find that 16 sensitive APIs, which are frequently invoked within HSOs to collect system information, are not taken into account by the source set of FlowDroid by default. These APIs (listed in Table~{\ref{tab:Hidden_source_methods}}), after manual confirmation, should still be considered as source methods by FlowDroid as they are responsible for retrieving sensitive data that should not be exposed to other parties. Here, to clarify, when doing the experiment, we include both of the default source and sink methods of FlowDroid and the additional sensitive APIs involved in HSOs in the SourceAndSink.txt file of FlowDroid. During the manual process, we have not found any sensitive API (i.e., involving dangerous operations) that should be additionally considered as a \emph{sink} method by FlowDroid.
Hence, we add the 16 APIs to the source set of Flowdroid and keep its sink set unchanged (hereinafter referred to this version as FlowDroid + HiSenDroid) and relaunch it on the same set of apps.
This time, we are able to disclose 1,110 HSDFs from 1,304 apps.
This result shows that suspicious HSOs could be leveraged to leak users' sensitive information outside of their devices.
As an example shown in Listing~\ref{code:SMS_Message}, the sensitive data \emph{device id} and \emph{subscriber id}, which are unique to the device and hence can be leveraged to uniquely track the phone, are eventually sent outside the device through a text message.

\begin{figure}[!h]
    \centering
    \includegraphics[width=0.65\linewidth]{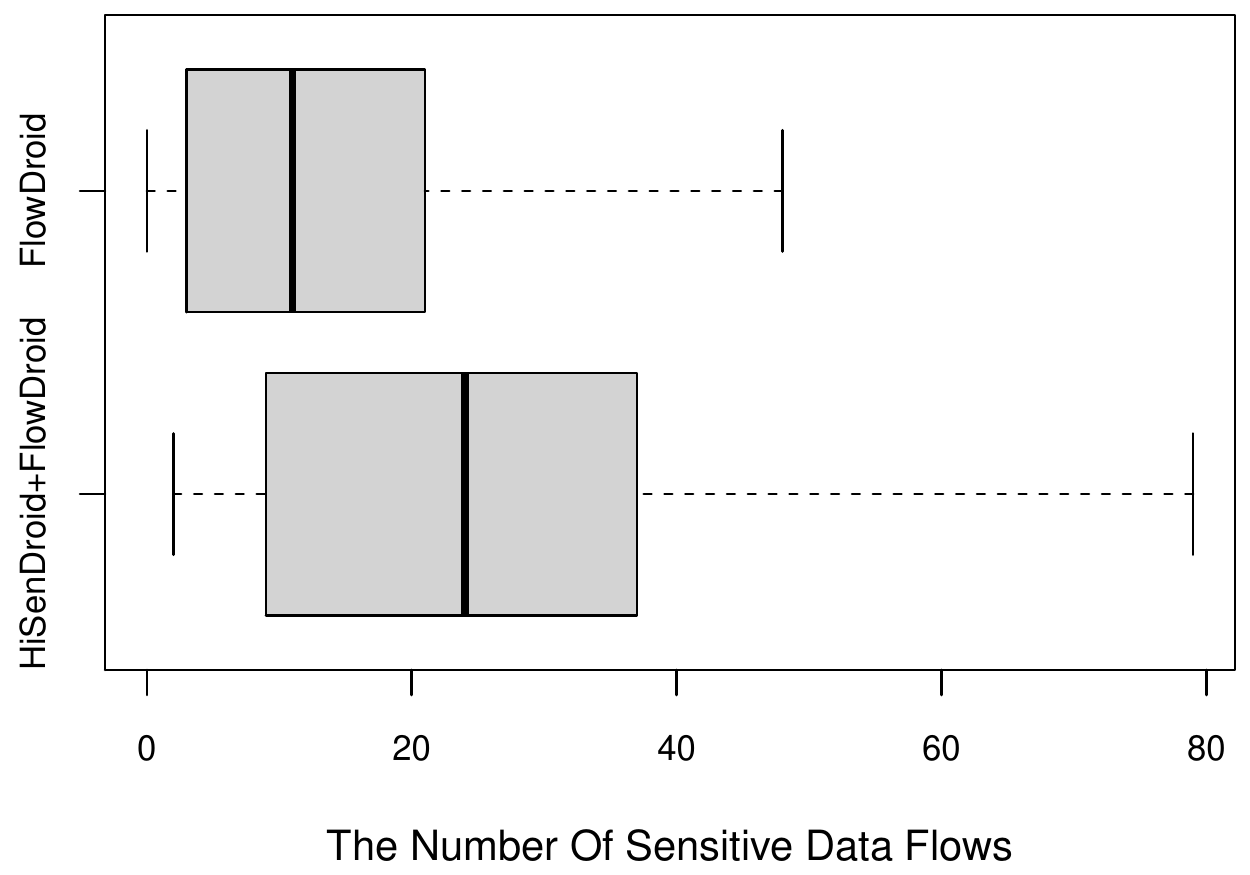}
	\caption{Results of Sensitive data Flows in Android Apps.}
    \label{fig:SDF_FlowDroid}
\end{figure}

Considering general sensitive data-flows (SDF), we compare FlowDroid with HiSenDroid on the same dataset. In general, among the 1,304 apps, HisenDroid+FlowDroid detect 31,215 SDF, which is significantly larger than that of the original FlowDroid (which is 16,946). This result, as expected\footnote{We remind the readers that, in this work, we did not improve FlowDroid by itself but only enlarged its \emph{source} set as some of the sensitive APIs, which are favored by HSOs, are overlooked by FlowDroid.}, does experimentally demonstrate the effectiveness of our approach towards revealing more data flows in Android Apps.
Our experimental results are illustrated in Figure~\ref{fig:SDF_FlowDroid}, which indicates the distribution of the number of sensitive data flows in each app yielded by HisenDroid+FlowDroid and HiSenDroid. This result shows that FlowDroid + HiSenDroid  has significantly improved the original results of FlowDroid, which shows the usefulness of our identified HSOs and suggests that there is a strong need to characterize hidden sensitive operations.

\begin{table}[!h]
\centering
\caption{The list of selected source methods that, by default, are not included by FlowDroid.}
\label{tab:Hidden_source_methods}
\resizebox{0.6\linewidth}{!}{
{\begin{tabular}{ l } 
\hline
API Signature \\
\hline
android.net.wifi.WifiManager\#getConnectionInfo()\\
android.app.ActivityManager\#getRunningTasks  \\
android.app.ActivityManager\#getRecentTasks   \\
android.accounts.AccountManager\#getUserData \\
android.net.ConnectivityManager\#getNetworkInfo  \\
android.provider.Settings\$System\#getUriFor  \\
android.telephony.TelephonyManager\#getNeighboringCellInfo \\
android.telephony.TelephonyManager\#getCellLocation  \\
android.accounts.AccountManager\#getAccountsByType  \\
android.net.wifi.WifiManager\#getScanResults  \\
android.net.wifi.WifiManager\#getConfiguredNetworks  \\
java.net.URL\#openConnection\\
android.net.ConnectivityManager\#getAllNetworkInfo \\
android.net.VpnService\#prepare  \\
android.hardware.Camera\#open   \\
android.net.ConnectivityManager\#getActiveNetworkInfo \\
\hline
\end{tabular} }}
\end{table}

\section{Limitations}
\label{HiSenDroid:limitations}
The main limitation of our approach lies in the backward data-flow analysis, which applies only context-insensitive analysis and thereby may lead to imprecise results. Furthermore, at the moment, our approach is not aware of dynamically loaded code, reflectively accessed methods, and native code.
Subsequently, HiSenDroid may overlook certain app features and hence result in false-negative results.

Second, HiSenDroid data-flow analysis may be susceptible to obfuscation techniques. According to some former research works~\cite{glanz2020hidden, rasthofer2016harvesting, samhi2022jucify}, obfuscation (especially those involving complicated changes of the program code) may cause false negatives of the static analysis approach. Indeed, as demonstrated by Moser~\cite{moser2007limits}, obfuscation is actually a challenge for almost all static program analyzers. Just like all prior efforts on static analysis of HSOs ~\cite{fratantonio2016triggerscope}, ~\cite{pan2017dark}, we do not consider the apps whose branch conditions have been deeply obfuscated. 
Fortunately, the majority of obfuscations applied to Android apps only involve basic transformations (such as renaming~\cite{dong2018understanding}) that do not involve complicated code changes (e.g., structural or logic changes, or invoke sensitive code through reflections, etc.), which will not impact the analysis of our approach. This has also been confirmed by our exploratory study towards understanding the impact of obfuscation on our approach, as discussed in Section~\ref{subsec:obfuscation}. Considering reflection obfuscation, integrating DroidRA with HiSenDroid as a pipeline is demonstrated to be effective in eliminating the impact of reflection calls. Therefore, we believe that the technical capabilities and our results would not be significantly impacted by code obfuscation. Nevertheless, as part of our future work, we plan to integrate other approaches developed by our fellow researchers to mitigate these long-standing challenges, e.g., by applying DroidRA~\cite{sun2021taming,li2016droidra} to mitigate the impact of reflection-enhanced code obfuscations.

Although summarized from many sensitive operations, the definition of HSO rules may not be perfect. Indeed, on the one hand, the set of sensitive operations considered for summarization may not be representative, and the set of apps leveraged to obtain such sensitive operations may not be represented as well. On the other hand, the manual analysis leveraged to summarize the rules may contain errors since it is known that human efforts are prone to errors. Apart from that, the definition of HSO is based on empirical evidence that might not be perfect. There might be complicated cases that do not follow the definition but still manifest themselves as hidden sensitive behaviors in practice, leading to false negatives. This limitation can also apply to the conventional usage analyses since the list of conventional usages is manually summarized based on a given set of apps. The subsequent outputs (i.e., whitelist) may not be representative. Nonetheless, our follow-up study using a set of 20,000 new apps has shown that this impact is negligible. Furthermore, in this work, we have attempted to provide detailed insights to explain why HSOs are reported as such. This knowledge is expected to be useful for practitioners and researchers to characterize conventional usages and for security analysts to understand suspicious HSOs.

Moreover, since the original implementation of TriggerScope is not publicly available, we have resorted to an open re-implementation version of TriggerScope to compare our approach against it. This alternative decision may result in possible biases as the re-implementation may not really represent the original version. Unfortunately, the re-implemented version is the only source we can publicly locate to fulfill the comparison.  As of our future work, we plan to also evaluate the reliability of the re-implementation of TriggerScope so as to mitigate potential biases, if any.

Last but not the least, the performance of the hidden sensitive data flow analysis may be impacted by the collection of sensitive APIs (i.e., sources). On one hand, some sensitive APIs, especially the latest ones, might be overlooked by FlowDroid and hence cannot be considered for pinpointing potential leaks, leading to false-negative results. In this work, our experimental results have confirmed this. On the other hand, some historical sensitive APIs included in FlowDroid's source list might be deprecated and subsequently removed from a certain Android API version~\cite{li2018cid}. There is hence no need to include them when analyzing apps targeting higher API versions, as these APIs will not be used anymore, not even mentioning causing sensitive data leaks. To overcome these impacts, we believe there is a need to keep updating FlowDroid's list of sensitive APIs, in order to achieve a more effective and sound sensitive data flow analysis for Android apps. Furthermore, ideally, FlowDroid should also not be expected to include APIs that are released after itself.

\section{Related Work}
\label{HiSenDroid:related_work}
Hidden sensitive operations have long existed in Android malware as evasive technologies have widely been used by attackers to hide their malicious behaviors.
Our research community has hence proposed various approaches to tackle these issues. We now discuss some of the representative works from two angles, including the evasive techniques that have been proposed to hide malicious code from being identified, and the detection methods proposed to pinpoint such evasive techniques.

\textbf{Evasive Techniques.} There has been a number of research works on hiding malicious behavior from detection, most of which focus on evading the dynamic test platforms such as virtual machines and emulators. Early works target the Windows platform \cite{chen2008towards}, while recently the trend has been moved to Android \cite{petsas2014rage,vidas2014evading,jing2014morpheus,diao2016evading,costamagna2018identifying,liu2021deep,liu2022explainable}. These evasive techniques detect the presence of a simulated environment by either looking into the system properties of the testing platform (e.g., system fingerprints, hardware capabilities, etc.) \cite{petsas2014rage,vidas2014evading,costamagna2018identifying}, or leveraging a reverse Turing test that examines if the app interacts with a human user \cite{diao2016evading}. For instance, Diao et al. \cite{diao2016evading} observed that programmed interaction has specific patterns of input and interaction frequency, which is different from real users. Overall, the evasive techniques usually hide malicious activities in an \emph{if-then-else statement}. The hidden malicious behavior will only be set off when certain conditions are fulfilled (e.g., not in an emulator); otherwise dummy benign operations are triggered. The prevalence of such evasive techniques motivated us to investigate the HSOs in Android apps and propose HiSenDroid to detect them.

\textbf{Detection of Evasive Techniques.} The pervasive evasive techniques (e.g., anti-emulator techniques) have motivated the research community to take countermeasures. Great effort has been spent on detecting known types of hidden behaviors that hampers the dynamic analysis process. These works include detecting anti-emulator techniques \cite{balzarotti2010efficient, lindorfer2011detecting, kirat2014barecloud,kirat2015malgene} and generic logic-bombs \cite{crandall2006temporal,brumley2008automatically,zheng2012smartdroid,fratantonio2016triggerscope,papp2019towards}. The approaches of detecting anti-emulator techniques compare the behavioral deviation of the tested apps on the various environments when feeding them the same input. The fundamental idea is that if the app behaves differently in different environments, it is likely trying to evade one or more analysis platforms (usually referred to as bare-metal analysis in the literature) \cite{balzarotti2010efficient, lindorfer2011detecting, kirat2014barecloud,kirat2015malgene}. While these early works investigate a critical category of hidden operations (i.e., anti-emulator), the proposed methods lack generalization that cannot be applied to detect other types of hidden operations emerging recently.

Besides the detection of anti-emulator techniques, several works are focusing on uncovering other trigger-based behaviors. These approaches leverage symbolic execution or static code analysis and instrumentation to expose the hidden branches in an \emph{if-then-else statement} \cite{crandall2006temporal,brumley2008automatically,zheng2012smartdroid,fratantonio2016triggerscope,papp2019towards}. As examples, Zheng et al. \cite{zheng2012smartdroid} proposed to leverage a static analysis approach to retrieve all UI related events, and use dynamic testing to trigger them and log the invocation of sensitive APIs. Unlike HiSenDroid that leverages static analysis, the dynamic analysis based approach introduces significant system- and time-overhead. The coverage of the dynamic analysis is also in question.
Fratantonio et al. \cite{fratantonio2016triggerscope} proposed TriggerScope to detect hidden triggered behaviors based on the observation that certain triggers (i.e., time, location, and SMS related triggers) always involve the comparison of specific types of input (i.e., system time, system location, and received SMS). Symbolic execution is then leveraged to detect such narrow conditions. While TriggerScope is effective in detecting the above-mentioned three types of logic bombs, it cannot be generalized to detect hidden operations triggered by other types of conditions, such as system property, which has been found pervasive in Android apps.

Similar to HiSenDroid, another line of work attempts to detect unknown types of trigger-based behaviors~\cite{pan2017dark},~\cite{wang2017droid}.
A prominent example is HSOMiner~\cite{pan2017dark}, which extracts static characteristics of hidden behaviors as features and trains a machine learning model to identify the code blocks that observe similar patterns. The major differences between our work and HSOMiner are twofold. First, HSOMiner requires a large number of manually labelled training samples, which involves extensive human experts' effort. Its performance also heavily relies on the manually labelled training data, which is prone to errors. Our method, on the other hand, is an automatic process without human intervention. Second, HSOMiner, as a machine learning based approach, lacks explanations of the decisions. In contrast, our static code analysis based approach outputs the full call traces of detected HSOs, and provides more detailed information for further analysis.

\section{Summary}
\label{HiSenDroid:summary}
In this work, we present to the community a prototype tool called HiSenDroid, which performs a static code analysis to uncover hidden sensitive operations that will only be triggered under special circumstances such as at a specific location or in a certain time period.
Additionally, HiSenDroid goes one step deeper to provide details aiming at helping security analysts understand why a given hidden sensitive operation is flagged as such.
Experimental results over 20,000 apps, including both malicious and benign apps, show that hidden sensitive operations are indeed quite frequently presented in Android apps and HiSenDroid is effective in automatically discovering them.
Moreover, with the help of FlowDroid, a state-of-the-art static taint analyzer, we further experimentally find that hidden sensitive operations could eventually lead to privacy leaks.

%% file: chapter5.tex
\chapter{Mining Android API Usage to Generate Unit Test Cases for Pinpointing Compatibility Issues}

\begin{tcolorbox}
[width=6.2in]
\small{
\textbf{Xiaoyu Sun}, Xiao Chen, Yanjie Zhao, Pei Liu, John Grundy and Li Li. 2022. Mining Android API Usage to Generate Unit Test Cases for Pinpointing Compatibility Issues.  The 37th IEEE/ACM International Conference on Automated Software Engineering (ASE) 2022.
\url{https://arxiv.org/abs/2208.13417}
}
\end{tcolorbox}

\section{Introduction}
\label{JUnitTestGen:introduction}
Unit testing is a type of software testing aiming at testing the effectiveness of a software's units, such as functions or methods.
It is often the first level of testing conducted by software developers themselves (hence a white box testing technique) to ensure that their code is correctly implemented.
Unit testing has many advantages.
First, it helps developers fix bugs early in the development cycle which subsequently saves costs in the end. Indeed, it is known that the cost of a bug increases exponentially with time in the software development workflow.
Second, it makes it possible to achieve regression testing. For example, if developers refactor their code later, it allows them to make sure that the refactored code still works correctly.
Third, unit tests provide an effective means for helping developers understand the unit under testing, i.e., what functions are provided by the unit and how to use them?
Because of the aforementioned advantages, it is recommended to always write unit tests when developing software, and the unit tests should cover as many units as possible.

The Android framework, as one of the largest software projects (with over 500,000 commits), is no exception. 
The Android framework provides thousands of public APIs that are heavily leveraged by app developers to facilitate their development of Android apps. Ideally, each such public API should be provided with a set of unit tests to ensure that the API is correctly implemented and the continuous evolution of the framework will not change its semantics.
Unfortunately, based on our preliminary investigation, less than 30\% of APIs, are provided with unit test cases, leaving the majority of APIs uncovered.
This is unacceptable considering that the Android framework nowadays has become one of the most popular projects (with millions of devices running it).

This poor test coverage of Android APIs has led to serious compatibility issues in the Android ecosystem, as recently shown~\cite{li2018cid,wei2016taming,mutchler2016target,zhang2015compatibility,ham2011mobile,huang2018understanding}.
For example, Li et al. ~\cite{li2018cid} demonstrate that various Android APIs suffer from compatibility issues as the evolution of the Android framework will regularly remove APIs from or add APIs into the framework. Such API removal or addition can result in no such class or method runtime exceptions when the corresponding app is running on certain framework versions.
Liu et al.~\cite{liu2021identifying} further present an approach to detect silently-evolved Android APIs, which could cause another type of compatibility issue as their semantics are altered (while not explicitly documented) due to the evolution of the Android framework.
Moreover, Wei et al.~\cite{wei2019pivot} experimentally show that some Android APIs could even be customized by smartphone manufacturers, leading to another type of compatibility issue that causes Android apps to crash on certain devices while behaving normally on others.
The authors further propose a prototype tool called PIVOT to automatically learn device-specific compatibility issues from existing Android apps. Their experiments on a set of top-ranked Google Play apps have discovered 17 device-specific compatibility issues.

To the best of our knowledge, the state-of-the-art works targeting compatibility issue detection leverage static analysis techniques to achieve their purpose.
However, as known to the community, the static analysis will likely yield false positive results as it has to make some trade-offs when handling complicated cases (e.g., object-sensitive vs. object-insensitive).
In addition, the static analysis will also likely suffer from soundness issues because some complicated features (e.g., reflection, obfuscation, and hardening) are difficult to be handled~\cite{sun2021taming,samhi2022jucify}.
Furthermore, except for syntactic changes, compatibility issues could also be triggered by semantic changes, which are non-trivial to be handled statically.
Indeed, as demonstrated by Liu et al.~\cite{liu2021identifying}, there are various semantic change-induced compatibility issues in the Android ecosystem that remain undetected after various static compatibility issue detection approaches are proposed to the community.

Moreover, static app analysis can only be leveraged to perform post-momentum analysis (i.e., after the compatibility issues are introduced to the community). 
They cannot stop the problems from being distributed into the community -- many Android apps, including very popular ones, still suffer from compatibility issues.
To mitigate this problem, we argue that incompatible Android APIs should be addressed as early as possible, i.e., ideally, at the time when they are introduced to the framework.
This could be achieved by providing unit tests for every API introduced to the framework and regressively testing the APIs against Android devices with different manufacturers and different framework versions.
However, it is time-consuming to manually write and maintain unit tests for each Android API (which probably explains why there is only a small set of APIs covered by unit tests at the moment).
There is hence a need to automatically generate compatibility unit tests for Android APIs.

In this work, we present a prototype tool, JUnitTestGen, that attempts to automatically generate test cases for Android APIs based on their practical usage in real-world apps.
Specifically, after locating existing API usages in real-world Android apps, JUnitTestGen performs field-aware, inter-procedural backward data-flow analysis to infer the API caller instance and its parameter values.
JUnitTestGen then leverages the inferred information to reconstruct a minimal executable code snippet for the API under testing.
Experimental results on thousands of Android apps show that JUnitTestGen is effective in generating test cases for Android APIs. It achieves an 80.4\% of success rate in generating valid test cases. 
These test cases subsequently allow our approach to pinpoint various types of compatibility issues, outperforming a state-of-the-art generic test generation tool named EvoSuite, which can only generate test cases to reveal a small subset of compatibility issues.
Furthermore, we demonstrate the usefulness of JUnitTestGen by comparing it against a state-of-the-art static analysis-based compatibility issue detector called CiD. JUnitTestGen is able to mitigate CiD's false-positive results and go beyond CiD's capability (i.e., detecting compatibility issues induced by APIs' signature changes) to detect compatibility issues induced by APIs' semantic changes.

Overall, we make the following main contributions in this work:
\begin{itemize}
\item We have designed and implemented a prototype tool JUnitTestGen that leverages a novel approach to automatically generate unit test cases for APIs based on their existing usages.

\item We have set up a reusable testing framework for pinpointing API-induced compatibility issues by automatically executing a large set of unit test cases on multiple Android devices. 

\item We have demonstrated the effectiveness of JUnitTestGen by i) generating valid test cases for Android APIs and pinpointing problematic APIs that could induce compatibility issues if accessed by Android apps, ii) outperforming state-of-the-art tools on real-world apps in detecting a wider range of compatibility issues.
\end{itemize}

The source code\footnote{\url{https://github.com/SMAT-Lab/JUnitTestGen}} and experimental results are all made publicly available in our artifact package.\footnote{\url{https://doi.org/10.5281/zenodo.6507579}}.

\section{Motivation}
\label{JUnitTestGen:motivation}
To overcome the fragmentation problem, our fellow researchers have proposed various approaches to mitigate the usage of compatibility issues in Android apps~\cite{wei2016taming, wei2019pivot, xia2020android}.
These approaches mainly leverage static analysis to achieve their purpose.
Unfortunately, static analysis is known to likely generate false-positive and false-negative results and is yet hard to handle such issues that involve semantics changes in Android APIs~\cite{liu2021identifying}.
Therefore, we argue that there is a need also to invent dynamic testing approaches to complement existing works in handling app compatibility issues.

We hence start by conducting a preliminary study investigating the test case coverage in the Android framework. Specifically, we downloaded the source code of AOSP from API level 21 to 30 and then calculated the number of public APIs\footnote{The APIs in \texttt{platform/frameworks/base} path.} and their corresponding unit test cases provided by Google. Our result reveals that on average \textbf{less than 30\% of Android framework APIs have provided test cases in each API level}, indicating the Android framework has a poor test case coverage.
When more APIs are provided with unit test cases, more compatibility issues of APIs will likely be identified during regression testing. This will enable them to be fixed at an earlier stage to avoid the introduction of compatibility issues in the first place. 
To this end, we propose to effectively and efficiently detect compatibility issues through a dynamic testing approach that fulfills its objective by automatically generating valid test cases by mining API usages from real-world Android apps.

\begin{figure*}[!h]
    \centering
    \includegraphics[width=\textwidth]{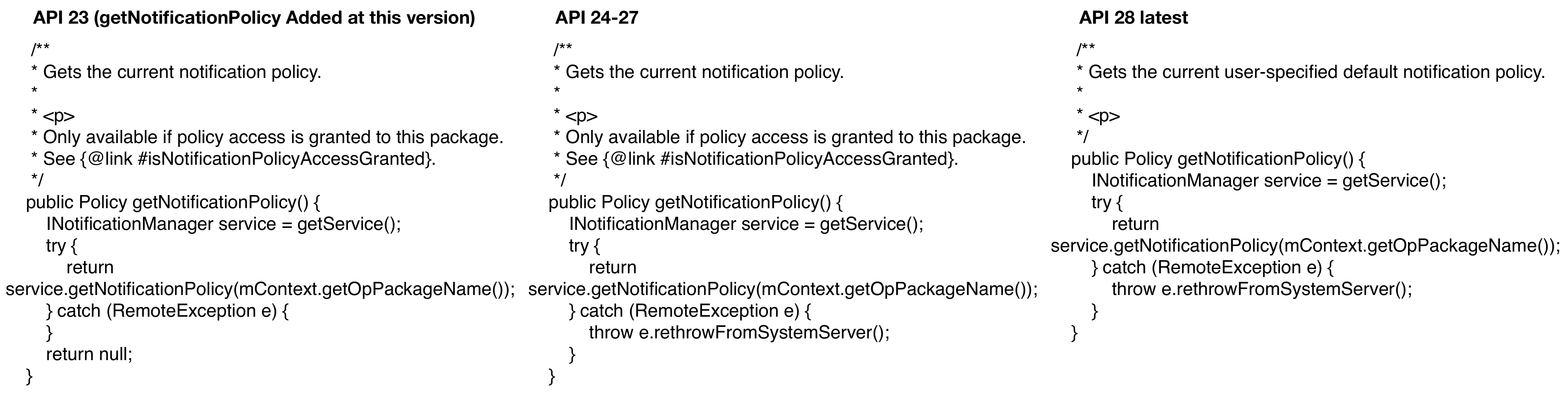}
	\caption{The source code of Android API \emph{getNotificationPolicy()}.}
    \label{fig:getNotificationPolicy}
\end{figure*}

\textbf{Why Dynamic Testing.} We now present a concrete example to motivate why there is a need to generate more and better unit test cases for Android APIs to pinpoint compatibility issues.
There is an Android API called \emph{getNotificationPolicy()}, located in class \emph{NotificationManager}.
At the moment, there are no unit tests provided for this API.
The lack of solid testing for this API has unfortunately led to various problems, as demonstrated by the various discussions on StackOverflow~\cite{stackoverflowPermissions}, one of the most widely used question and answer websites.

Figure~\ref{fig:getNotificationPolicy} presents the brief evolution of the source code of \emph{getNotificationPolicy()}.
This API was introduced to the Android framework at API level 23.
The apps that accessed this API would crash on devices powered by Android with API level 22 or earlier, resulting in forwarding compatibility issues.
During the evolution of the Android framework, the implementation of \emph{getNotificationPolicy()} is quickly changed at API level 24 (i.e., no longer returns null).
Nevertheless, at this point, the comments are not changed, suggesting potential compatibility issues because of changed semantics.
At API level 28, this API is further changed.
This time, only the comments are changed, i.e., the implementation of the API is kept the same.
This change further suggests that this API may be involved with compatibility issues as the source code of the API and its comments are inconsistent at a certain period (over six API levels).

\begin{lstlisting}[
caption={Code example of invoking API \emph{getNotificationPolicy()}.},
label=code:getNotificationPolicy_app, firstnumber=1, abovecaptionskip=0pt, belowcaptionskip=0pt, aboveskip=0.5pt, belowskip=0.8pt]
@Override
protected void onCreate(Bundle savedInstanceState) {
 super.onCreate(savedInstanceState);
 setContentView(R.layout.activity_main);
 
 NotificationManager mng = (NotificationManager) this.getSystemService("notification");
 NotificationManager.Policy policy = mng.getNotificationPolicy();
 int priorityCallSenders = policy.priorityCallSenders;
}\end{lstlisting}

The actual implementation of \emph{getNotificationPolicy()} is through the complicated inter-process communication mechanism (e.g., the API is defined via Android Interface Definition Language (AIDL). Thus, it is non-trivial to confirm if there is a compatibility issue after API level 23 by only (statically) looking at the Java code of the framework. It is still a known challenge to statically analyze Java and C code at the same time.

We resort to a dynamic approach to check if \emph{getNotificationPolicy()} suffers from compatibility issues.
Specifically, we implement a simple Android app with minimal lines of code to invoke the API (i.e., lines 6-7 in Listing~\ref{code:getNotificationPolicy_app}) and also the usage of the return value of this API(i.e., lines 8 in Listing~\ref{code:getNotificationPolicy_app}).
The minimal and targeted SDKs of this app are set to be 21 and 30, respectively.
We then launch this app on ten emulators covering Android API levels from 21 to 30.
As expected, the app throws \emph{NoSuchMethodError} as the API is not yet introduced at API level 21 and 22.
From API level 23 to 27, the app throws \emph{SecurityException} with the message ``Notification policy access denied'', due to a lack of declaration of Android permissions. Even though the permission was granted, the API can still introduce compatibility issues at API level 23 because the return value can be null, causing \emph{NullPointerException} later on (e.g., lines 8 in Listing 1)). 
Surprisingly, since API level 28, the app does not throw any exception, even though no permissions are declared as well.
We then go one step further to track the detailed implementation of \emph{service.getNotificationPolicy(String)} and found that the enforcement of policy access (i.e., line 3 in Listing~\ref{code:getNotificationPolicy_change}) is removed in the API at API level 28, which explains why the app no longer crashes since API level 28.

\begin{lstlisting}[
caption={Code changes of method \emph{getNotificationPolicy(String)}, which includes the underline implementation of Android API \emph{getNotificationPolicy()}.},
label=code:getNotificationPolicy_change,
firstnumber=1,abovecaptionskip=0pt, belowcaptionskip=0pt, aboveskip=0.5pt, belowskip=0.5pt]
 @Override
 public Policy getNotificationPolicy(String pkg) {
-  enforcePolicyAccess(pkg, "getNotificationPolicy");
  final long identity = Binder.clearCallingIdentity();
  try {
   return mZenModeHelper.getNotificationPolicy();
  } finally {
   Binder.restoreCallingIdentity(identity);
 }}
\end{lstlisting}

These observed runtime behaviours strongly indicate that the direct invocation of \emph{getNotificationPolicy()} will very likely result in two compatibility issues: (1) a method is not yet defined and (2) method semantics have been altered.
Ideally, such issues -- especially the latter case -- should not be introduced into the Android framework.
However, due to the lack of API unit tests, such issues are non-trivial to identify and avoid.
It is hence essential to provide more and better unit test cases for all Android APIs.
Since it is time-consuming to achieve this manually, we argue that there is a strong need to provide automated approaches to automatically generate unit test cases for Android APIs to identify compatibility issues as early as possible.
In this work, we propose to generate unit test cases for Android APIs by learning from existing Android API usages.

\textbf{Why mining API usage.} In addition, generic test generation tools~\cite{fraser2011evolutionary, pacheco2007randoop,Kex} mainly targeting on satisfying coverage criteria for classes, while compatibility issues are mainly caused by the fast-evolving of APIs~\cite{li2018cid}, which makes them insufficient in detecting compatibility issues. Specifically, such generic test case generation approaches (such as EvoSuite) are tailored to generate tests based on the source code of classes, lacking API usage knowledge~\cite{kechagia2019effective}, including both API calling context and API dependency knowledge. This information is crucial to setting up the environment for successfully calling Android APIs properly to detect compatibility issues. In addition, all of these tools completely ignore semantic-level behaviours at the API level, leading to many compatibility issues undetected. Especially in Android, APIs often come with usage
caveats, such as constraints on call order~\cite{ren2020api}. Thus, it is essential to capture API dependencies~\cite{zhang2014semantics} involved in the calling context before invoking the target API. However, it is challenging for traditional test generation tools to achieve this since they generate test suites only based on source code, which lacks API dependency knowledge. Thus, the insufficiency of the generic coverage-based test case generation approach motivates us to mine API usage to generate much more effective test cases in detecting compatibility issues.

\section{Approach}
\label{JUnitTestGen:approach}
\begin{figure*}[!h]
    \centering
    \includegraphics[width=\linewidth]{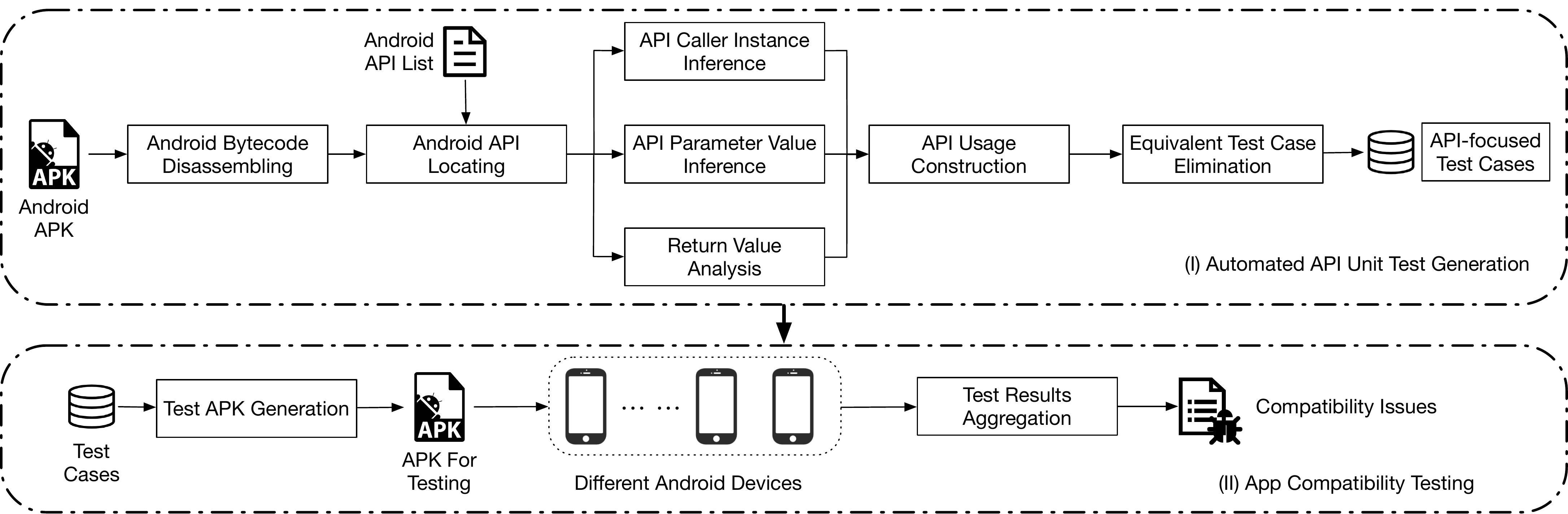}
	\caption{The working process of our approach.}
    \label{fig:JUnitTestGen_methodology}
\end{figure*}

The main goal of this work is to automatically generate unit tests for the Android framework to provide better coverage at the unit test stage of as many Android APIs as possible. Fig.~\ref{fig:JUnitTestGen_methodology} outlines the process of JUnitTestGen, which is made up of two modules involving a total of nine steps.
We first locate target API invocations after disassembling the APK
bytecode. We then apply inter-procedural data-flow analysis to identify the API usage, including API caller instance inference and API parameter value inference. We then execute these generated test cases on Android devices with different Android versions (i.e., API levels). The following elaborates on the detailed process of each module.

The first module, \emph{Automated API Unit Test Generation}, includes five sub-processes to automatically generate unit test cases for Android APIs.
This module takes as input an Android API (or an API list) for which we want to generate unit tests and an Android app that invokes the API and outputs a minimum executable code snippet involving the given API.
This code snippet is the API's unit test case.

The second module, \emph{App Compatibility Testing}, includes two sub-processes.
This module takes as input the previously generated unit test cases to build an Android app, allowing direct executions of the test cases on Android devices running different framework versions (i.e., API levels). 
The output of this second module is the execution results of the test cases concerning different execution environments. Using this, JUnitTestGen can then determine all the Android APIs suffering from potential compatibility issues.
We now detail these two modules of JUnitTestGen below.

\subsection{Automated API Unit Test Generation}
\label{sub:unit_test_generation}

As shown in Figure~\ref{fig:JUnitTestGen_methodology}, the first module of JUnitTestGen, namely Automated API Unit Test Generation, is made up of five steps.

\textbf{I.1 Android Bytecode Disassembling.} JUnitTestGen takes Android APKs as input. 
Given an Android app, the first step of JUnitTestGen is to disassemble the Dalvik bytecode to an intermediate representation. In this work, we leverage Soot~\cite{vallee2010soot}, which provides precise features for code analysis. In particular, it supports 3-address code intermediate  representation Jimple and accurate call-graph analysis framework Spark\cite{lhotak2003scaling}. On top of Soot, JUnitTestGen is able to convert Android APK's bytecode into Jimple precisely.

\textbf{I.2 Android API Locating.}
The second step towards locating the practical usages of Android APIs is quite straightforward.
JUnitTestGen visits every method of each application in its Jimple representation, i.e., statement by statement, to check if any of the APIs in the input list is invoked. If so, we record the location and mark the usage of the API as located.

\textbf{I.3 API Caller Instance Inference.}
Android APIs are invoked as either \emph{static methods} (also known as \emph{class methods}) or \emph{instance methods}. A static method can be called without needing an object of the class, while an instance method requires an object of its class to be created before it can be called.
Listing~\ref{code:CALL_static_instance_APIs} shows an example for each method type in the Jimple representation. The static method (\emph{boolean equals()}) can be called directly as long as the API signature (\emph{<android.text.TextUtils: boolean equals(java.lang.CharSequence, java.lang.CharSequence)>}) is acquired, which has already been done in step I.2. However, to invoke the instance method (\emph{void setLayoutParams()}), its calling object \emph{\$r5} needs to be identified. Specifically, a backward data-flow analysis is needed to locate both the definition statement of \emph{\$r5} and any intermediate APIs (or methods) on the call trace that may change the behaviour of the caller instance.

\begin{lstlisting}[
caption={Examples of calling static and instance APIs.},
label=code:CALL_static_instance_APIs,
firstnumber=1]
//Calling a Static API
<android.text.TextUtils: boolean equals(java.lang.CharSequence,java.lang.CharSequence)>

//Calling an Instance API
$r5.<android.widget.ListView: void setLayoutParams(android.view.ViewGroup$LayoutParams)>
\end{lstlisting}

Identifying the calling object and constructing the call trace is a non-trivial task. As the calling object's instantiation could depend on multiple method calls (e.g., can be defined in other methods and passed on via callee's returned values), the caller instance inference process needs to be inter-procedural. Furthermore, each of the involved methods in the instantiation process may further require specific parameter values, which need to be properly prepared in order to successfully create the calling object.
Hence, the caller instance inference process requires backtracing not only the direct caller instance but also many other variables leveraged by the app to instantiate the calling object.

\setlength{\intextsep}{3pt} 
\begin{algorithm}
 \scriptsize
  \caption{API Caller Instance Inference.}\label{API_Caller_Instance_Inference}
  \begin{algorithmic}[1]
    \Require $api_t$: the target API
    \Ensure $M$: a list of method invocations towards $api_t$
    \Function {calculateMinimumExecutableContext}{$api_t$}
        \State $M = \emptyset$
        \State $stmt_t\gets $ statement contains $api_t$
        \If {$api_t$ is a static API}
            \State $M\gets$ The API signature of $api_t$
        \EndIf
        \If {$api_t$ is an instance API}
            \State $assignStmt\gets $ \Call{getDefinitionStmt}{$stmt_t$}
            \State $M\gets $ \Call{constructCallTrace}{$assignStmt, \emptyset$}
        \EndIf
        \State \Return{$M$}
    \EndFunction
  \end{algorithmic}
\end{algorithm}

\begin{algorithm}
\scriptsize
  \caption{Construct Call Trace.}\label{Construct_Call_Trace}
  \begin{algorithmic}[1]
    \Require $stmt_{co}$: A definition statement of the calling object
    \Statex $visitedStmts$ : The set of statements visited by the analysis.
    \Ensure $callTrace$: a list of method invocations towards $stmt_{co}$
    \Function{constructCallTrace}{$stmt_{co}, visitedStmts$}
        \State $callTrace \gets \emptyset$
        \If{($stmt_{co}$ in $visitedStmts$) \textsf{OR} ($stmt_{co}$ is Constant) \textsf{OR} ($stmt_{co}$ is Syetem API)}
            \State \Return{$callTrace$}
        \EndIf
        \State $visitedStmts \gets visitedStmts \cup stmt_{co}$
        \If{$stmt_{co}$ is ParameterRef}
            \State $stmt_{inv}\gets getInvolvingStmtsfromCallGraph(stmt_{co})$
            \For{$each\ s \in stmt_{inv}$}
                \State $definitionStmt_{param}\gets $ \Call{getDefinitionStmt}{$s$}
                \State $callTrace\gets $ \Call{constructCallTrace}{$definitionStmt_{param}, visitedStmts$}
            \EndFor
        \EndIf
        \If{$stmt_{co}$ is InvokeExpr}
            \State $stmt_{inv}\gets getInvolvingStmtsfromCallGraph(stmt_{co})$
            \For{$each\ s \in stmt_{inv}$}
                \State $method_{host} \gets $ the host method of $s$
                \State $returnStmt \gets getReturnStmtFromMethod(method_{host})$
                \State $definitionStmt_{returnStmt}\gets $ \Call{getDefinitionStmt}{$returnStmt$}
                \State $callTrace\gets $ \Call{constructCallTrace}{$definitionStmt_{returnStmt}, visitedStmts$}
            \EndFor
        \EndIf
    \EndFunction
  \end{algorithmic}
\end{algorithm}

Algorithm~\ref{API_Caller_Instance_Inference} gives the details of the approach. Given a target API as input, we apply a backward data flow analysis to identify the minimum executable context of the target API (lines 1\textasciitilde12). As shown in Algorithm~\ref{API_Caller_Instance_Inference}, we describe the API caller inference process for both static methods (lines 4\textasciitilde6) and instance methods (lines 7\textasciitilde10). For static methods, we return its corresponding API signature, which has been obtained in step I.2. For instance methods, we first locate the definition statement of the calling object through invoking method \emph{getDefinitionStmt} (line 8), which returns the definition statement for a local variable. This method walks the inter-procedural control flow graph from the target API statement in reverse order, aiming to look for the nearest assignment statement defining the API's calling object.
After that, with the help of the function \emph{constructCallTrace}(i.e., defined in Algorithm ~\ref{Construct_Call_Trace}) (line 9), we can extract the call trace corresponding to the calling object. As shown in Algorithm~\ref{Construct_Call_Trace}, we handle parameter callers (lines 7\textasciitilde13) and method callers (lines 14\textasciitilde22), respectively. If the definition statement of the calling object comes from a parameter reference, we first retrieve all the statements at which the invocation occurs, and then for each of the statements, we recursively construct its call trace by calling the method \emph{constructCallTrace}. A similar process has been applied to handle method callers (lines 14\textasciitilde22), which recursively construct the call trace involving statements of the calling object. The recursive process will not terminate until any of the conditions have been satisfied in line 3, i.e., either the statement is a constant, or it has been visited before, or it is an Android system API.

We elaborate on this process with a Jimple code example presented in Listing~\ref{code:Code_Example_API_Caller_Instance_Return_Value}.
In this example, the target Android API to test (i.e., \emph{queryDetailsForUid(int,String,long,long,int)}) is invoked in line 26 by the calling object \emph{\$r3}, where \emph{\$r3} is a returned value of a self-defined method \emph{getNetworkStatsManager(Context)} (line 22). We then step into the definition of the method \emph{getNetworkStatsManager(Context)} (lines 1\textasciitilde7), and further backtrace the variables \emph{\$r2} and \emph{\$r1} along the call chain. \emph{\$r1} retrieves the network stats service from the application context \emph{\$r0} (line 4), and finally, the backtrace terminates at \emph{\$r0} (line 3), where all unknown variables are resolved. 

\begin{lstlisting}[
caption={Code example demonstrating the usage of API \emph{queryDetailsForUid}. The code snippet is extracted from app \emph{com.eyoung.myutils}.},
label=code:Code_Example_API_Caller_Instance_Return_Value,
firstnumber=1]
public static android.app.usage.NetworkStatsManager getNetworkStatsManager(android.content.Context)
{
 $r0 := @parameter0: android.content.Context;
 $r1 = $r0. getSystemService("netstats");
 $r2 = (android.app.usage.NetworkStatsManager) $r1;
 return $r2;
}

public static int getUid(android.content.Context) throws android.content.pm.PackageManager$NameNotFoundException
{
 $r0 := @parameter0: android.content.Context;
 $r1 = $r0.getPackageManager();
 $r2 = $r0.getPackageName();
 $r3 = $r1.getApplicationInfo($r2, 1);
 i0 = $r3.<android.content.pm.ApplicationInfo: int uid>;
 return i0;
}

public static float getCurAppFlow(android.content.Context) throws android.content.pm.PackageManager$NameNotFoundException
{
 $r0 := @parameter0: android.content.Context;
 $r3 = DeviceInfoUtil.getNetworkStatsManager($r0);
 $l1 = System.currentTimeMillis();
 $i0 = DeviceInfoUtil.getUid($r0);
 //Target API
 $r5 = virtualinvoke $r3.<android.app.usage.NetworkStatsManager: android.app.usage.NetworkStats queryDetailsForUid(int,java.lang.String,long,long,int)>(0, "", 0L, $l1, $i0);
}
\end{lstlisting}

In addition to inter-procedural data-flow analysis, JUnitTestGen also needs to be field-aware.
When performing backward data-flow analysis, the access of fields may break the original flow and hence may lead to unexpected results if not properly handled.

To mitigate this, we transform a field involved in the call trace of the API under testing (in the analyzed app) into a local variable in the generated test case.
The local variable will be initiated following the same method as it is assigned in the original app.
We then search the whole class to check how the field's value is assigned and subsequently apply the same method to initialize the local variable.
If we cannot find the field's assignment or the assigned value is complicated to be reconstructed, we will use a dummy object to mock the required value.

In addition, JUnitTestGen needs to handle branches in the backward dataflow analysis. Specifically, we leverage the Inter-procedural Control Flow Graph ( ICFG~\cite{bodden2012inter}), which is a combination of call-graph(CG) and control flow graph(CFG), to identify the minimum executable context of the target API. Here, CG is a graph representing the calls between methods over the entire program, while CFG is a graph that represents the control flows in a single method. ICFG treats each statement (or a set of sequential statements) as a node, including branch statements that enable path-sensitive analyses, i.e., the propagation of different information along different branches. With ICFG, we are able to implement branch analysis by analyzing the structure of the graph. To this end, for those methods involving multiple branches, JUnitTestGen will separate each branch to form a different test case.

\textbf{I.4 API Parameter Value Inference.}
To support API compatibility testing, e.g., to ensure that the API will, in any case, be reached once the test case is executed, we propose to directly assign possible values to the API's parameters inside the test case, i.e., the test case per se will not contain any parameter.
To achieve this, JUnitTestGen also infers possible values for each parameter of the API to be tested.
This step follows the same strategy i.e., the approach adopted to infer API caller instances, to infer the API's parameter values.  This is done by performing inter-procedural backward data-flow analysis. We apply the same algorithms described in Algorithm~\ref{API_Caller_Instance_Inference} and Algorithm~\ref{Construct_Call_Trace} on each parameter object to figure out the exact value.

Unfortunately, some Android APIs' parameter values may involve sophisticated operations when building their run-time values that are non-trivial to correctly retrieve statically. Specifically, if the analysis process does not end up at a constant/Android system API statement, the value of a parameter is regarded as being undiscovered.
To mitigate this, we introduce a set of pre-defined rules to generate dummy values for such APIs that have their parameter values hard to retrieve practically.
Some of the representative rules for generating dummy values for such hard-to-retrieve parameters are:

\begin{itemize}
\item For the eight primitive data types in Java (such as int, double, etc.) or their wrapper data types (such as Integer, Double, etc.) -- we provide random values for each of them that conform to their types. 
\item For the String  data type in Java -- we generate a random alphanumeric string.
\item For the Array parameter whose basic type is the eight primitive data types (or their wrapper data types) in Java (such as int, Integer, etc.) -- we generate an Array variable with random primitive values.
\item For  Android system-related objects (or the intermediate objects in the calling object's instantiation process), we use a heuristic approach to obtain the corresponding constructors to create their instances. If an object has multiple constructors, we select the simplest one (with the least number of parameters) to achieve the highest possibility of constructing a valid object.
\end{itemize}

\textbf{I.5 Return Value Analysis.}
To support detecting compatibility issues caused by return values (e.g., a given API may return A at API level X and B at API level Y), we propose to output the return value of the target API at the end of the test case. To achieve this, JUnitTestGen adds a statement at the end of the test case to further record the API's return value.

\textbf{I.6 API Usage Construction.}
After obtaining the caller instance, the invocation statements along the call trace and the return object, we can now recover the call sequence from program entry to the target Android API by reversing the retrieved statements step by step.
Based on the results of API caller instance inference along, JUnitTestGen will generate a test case containing the same number and type of parameters as the API to be tested.
For example, as shown in Listing~\ref{code:generated_test_case} at line 3, the generated test case contains five parameters, in the same type and order of the API under testing.

\begin{lstlisting}[
caption={Examples of the generated test cases for API \emph{queryDetailsForUid (int networkType, String subscriberId, long startTime, long endTime, int uid)}.},
label=code:generated_test_case,
firstnumber=1]
//for supporting generic testing
@Test
public void testQueryDetailsForUid(int var1, String var2,long var3, long var4, int var5) throws Exception {
 Context var6 = InstrumentationRegistry.getTargetContext();
 Object var7 = var6.getSystemService("netstats");
 NetworkStatsManager var8 = (NetworkStatsManager) var7;
 var8.queryDetailsForUid(var1, var2, var3, var4, var5);
}

//for supporting compatibility testing
@Test 
public void testQueryDetailsForUid() throws Exception {
 long var1 = System.currentTimeMillis();
 
 Context var2 = InstrumentationRegistry.getTargetContext();
 PackageManager var3 = var2.getPackageManager();
 String var4 = var2.getPackageName();
 ApplicationInfo var5 = var3.getApplicationInfo(var4, 1);
 int var6 = var5.uid;
 
 NetworkStats var7 = testQueryDetailsForUid(0, "", 0L, var1, var6);
// Output return value
 out(var7);
}
\end{lstlisting}

Taking the results of the API parameter value inference step, JUnitTestGen will generate another test case containing no parameters (line 10 in Listing~\ref{code:generated_test_case}).
This test case will directly call the former test case with prepared parameter values.
This test case is specifically designed to support API compatibility testing.
The former test case, on the contrary, is designed to serve a more general purpose.
With the help of fuzzing testing approaches (to generate possible parameter values for the test case), we expect the former test case could be leveraged to discover not only compatibility issues but also design defects such as bugs and security issues.
This trade-off allows JUnitTestGen to generate test cases that are at least suitable for identifying signature-based compatibility issues (e.g., a given API is no longer available in a certain framework), although it may not be effective enough to help identify semantic change involved compatibility issues.

Please note that there are several special classes, such as \emph{InstrumentationRegistry}, involved in the generated unit test cases.
These classes are part of the Android Testing Support Library provided by Google for supporting instrumented unit tests.
Compared to traditional unit tests, also known as local unit tests which can run on the JVM, instrumented unit tests require the Android system to run (e.g., through physical Android devices or emulators).
Since this requires us to generate actual Android apps to run on Android devices or emulators, instrumented tests are much slower than local unit tests.

Nevertheless, we still choose to use instrumented tests to examine the compatibility of Android APIs. This is because instrumented tests provide more fidelity than local unit tests, which we have found is essential to reveal potential compatibility issues, especially those that involve device-specific issues.

\textbf{I.7 Equivalent Test Case Elimination.}
Considering that the constructed test cases could be equivalent (i.e., duplicated), it is necessary to filter them out to save subsequent testing time and resources. Here, based on the concept of semantic equivalence defined in operational semantics~\cite{jiang2009automatic}, we consider that two test cases are equivalent if they share the same API invocation sequence. To this end, we first obtain the API invocation sequence for each test case and then examine the discrepancy between any two of them to check if $O_a$ = $O_b$, where $O_a$ and $O_b$ are the lists of API invocations in two different test cases. Based on this rule, we are able to eliminate equivalent tests (i.e., the first test case is retained). After this step, for the sake of simplicity, if there are still multiple test cases retained for a given API, we will select the small-scale one (with the least number of method invocations) for supporting follow-up analyses.

\subsection{App Compatibility Testing}
Using the unit test cases generated by the first module, the second module of JUnitTestGen leverages them to check if the corresponding Android APIs will likely induce app compatibility issues.
It first assembles all the test cases into an Android app and then aggregates their execution results against different devices running different Android frameworks.
We now briefly detail these two steps below.

\textbf{II.1 Test APK Generation.}
As discussed earlier, we have to resort to instrumented unit tests to examine Android APIs' incompatibilities.
This process essentially requires us to generate an Android app (or APK) to be installed and executed on Android systems.
Fortunately, Google has provided such a mechanism to achieve this purpose, i.e., supporting instrumented tests for a limited number of Android APIs. In this work, we directly reuse this mechanism to generate the test APK for all the unit tests automatically generated by JUnitTestGen.

\textbf{II.2 Test Results Aggregation.}
After the test APK is generated, we can distribute it for execution on multiple devices.
Since the test APK contains only known test cases, it is quite straightforward to execute it fully.
The only challenge that lies in this step is to select the right set of devices on which to execute the test cases to reveal as many incompatible APIs as possible.
Crowdsourced app testing could be an approach to achieve this purpose.

After installing and executing the generated test APK on multiple devices, the last step is in aggregating the test results to highlight potential compatibility issues in the app.
Inspired by the experimental setup of the work proposed by Cai et al.~\cite{cai2019large}, we consider an API as a potential incompatible case if 1) its corresponding test case can successfully run on a nonempty set of devices while failing on others; 2) its corresponding test case returns different values when running on different SDKs.

\section{Evaluation}
\label{JUnitTestGen:evaluation}
Our JUnitTestGen aims to generate unit test cases covering as many Android APIs as possible, so as to allow the discovery of more API-induced compatibility issues in apps.
To evaluate if this goal has been fulfilled, we propose to answer the following three research questions.
\begin{description}
\setlength\itemsep{0.05em}

\item[RQ1] To what extent can JUnitTestGen generate executable unit test cases for Android APIs?

\item[RQ2] How effective is JUnitTestGen in discovering API-induced Compatibility Issues? 

\item[RQ3] How does JUnitTestGen compare with existing tools in detecting compatibility issues?

\end{description}

\subsection{Experimental Setup}
To investigate the success rate of JUnitTestGen in producing valid test cases, we randomly select 1,000 Android apps for each target SDK version between 21 (i.e., Android 5.0) and 30 (i.e., Android 11.0\footnote{The latest version at the time when we conducted this study.}) from AndroZoo to prepare the experimental dataset. 
Here, we select 1,000 apps for each target SDK version because compatibility issues mainly lie in the evolution of APIs on different Android SDK versions~\cite{li2018cid}. Here, the criteria for app selection are based on the targetSdkVersion, which is the most appropriate API level on which the app is designed to run. Hence, the overall dataset for the experiment contains 10,000 Android apps whose target API versions are distributed equally across ten API levels. 
Our experiment runs on a Linux server with Intel(R) Core(TM)
i9-9920X CPU @ 3.50GHz and 128GB RAM. The timeout setting for analyzing each app with JUnitTestGen is 20 minutes. 
In this experiment, we generate the test cases for all Android APIs that have been invoked in the apps. However, users of JUnitTestGen could provide a customized list of APIs to only generate test cases based on their interests.

\subsection{RQ1-Effectiveness}

Our first research question concerns the effectiveness of JUnitTestGen in mining API usages from existing Android apps to generate valid unit tests for Android APIs. In this work, we consider a test case to be valid if (1) the generated code snippet can be successfully compiled on all API levels and (2) the test case does not throw an exception before the execution point of the API on all API levels.
The first condition ensures that the test case is syntactically correct. The second condition makes sure that the API's execution environment is properly set up. In other words, the second condition ensures that the exceptions we collected from valid test cases are exceptions thrown by the API under testing, which is essential for examining if the API will induce compatibility issues.

\begin{figure}[t!]
    \centering
    \setlength{\belowcaptionskip}{1pt}
    \includegraphics[width=0.85\textwidth]{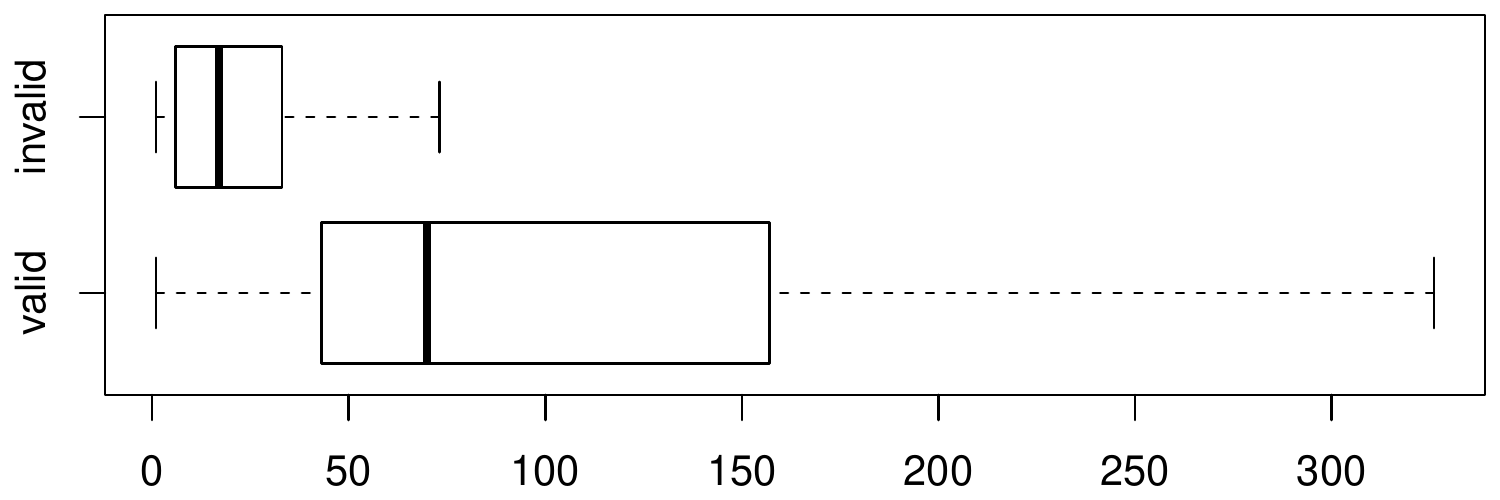}
	\caption{Distribution of the number of valid and invalid test cases per APK.}
    \label{fig:valid_per_apk}
\end{figure}

\textbf{Result.}
Among the 10,000 randomly selected apps, JUnitTestGen generates in total 1,032,182 test cases.
After eliminating the equivalent test cases, 66,499 of them are retained as distinct test cases, w.r.t. 28,367 distinct Android APIs. For the sake of simplicity, we select the small-scale test case (with the least number of invocation sequences) for each API (i.e., 28,367 test cases) for further study. By compiling and executing these 28,367 test cases, we further confirm that 5,562 of them are invalid (i.e., 22,805 of them are valid), giving a success rate of 80.4\% in generating valid test cases. In addition, our manual analysis on 100\footnote{The sample size is determined based on a confidence level at 95\% and a confidence interval at 10(\url{https://www.surveysystem.com/sscalc.htm}).} randomly selected test cases confirm that these test cases generated by JUnitTestGen are indeed valid.

Figure~\ref{fig:valid_per_apk} summarizes the distribution of the number of valid and invalid test cases generated per app.
The median number of valid and invalid test cases generated per app are 70 and 17, respectively, while their average are 106.29 and 25.37, respectively. Like most other state-of-the-art approaches, even though our static analysis approach has limitations so that it cannot generate valid test cases for every API, especially the complex ones, \textbf{our approach is still capable of generating more valid test cases than invalid ones.}
This demonstrates the effectiveness of our approach in mining Android API usages to generate API unit test cases. 

We note that the success rate of generating valid test cases  -- ~80.4\% at the moment -- is important but not crucial to our work. Theoretically, as long as we increase the number of Android apps considered for learning, we would likely be able to generate valid test cases for the given API under testing. For the test cases that are regarded as invalid, we further manually look into their root causes.
Our in-depth analysis reveals that the invalid cases are mainly caused by the lack of prerequisites (e.g., resource files), especially in UI-related APIs. For example, UI objects can hardly be programmatically initialized without certain resource files.

\begin{figure}[t!]
    \centering
    \includegraphics[width=0.85\textwidth]{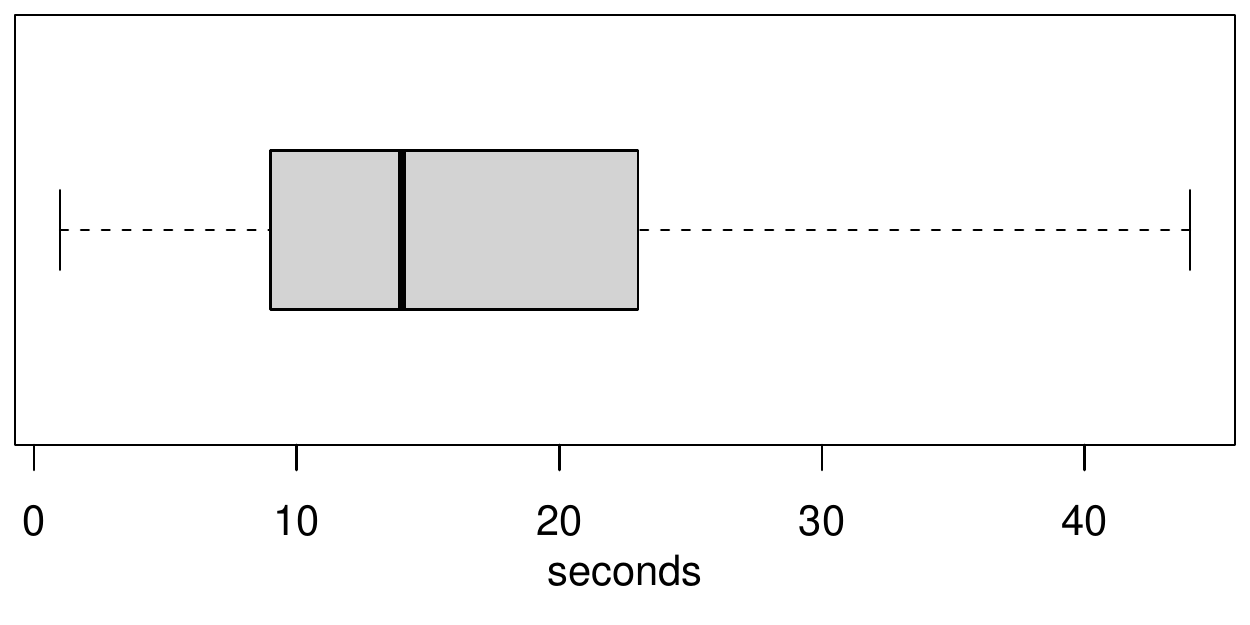}
	\caption{Distribution of time spent by JUnitTestGen to generate test cases per APK.}
    \label{fig:time_cost}
\end{figure}

We further look at the efficiency of JUnitTestGen by reporting its time cost when applied to generate test cases. Figure~\ref{fig:time_cost} summarizes the distribution of time consumed by JUnitTestGen per app. On average, it takes 17.45 seconds to process an app. The median time cost is 14 seconds. The fact that the time spent by JUnitTestGen to process an Android app is quite small suggests that it is practical to apply JUnitTestGen to analyze large-scale Android apps.

\begin{boxedtest}
By learning from existing API usages, our approach can automatically generate API unit test cases. Despite various challenges posed by advanced Android programming features, our static analysis approach can still achieve over 80\% of the success rate in generating valid test cases.
\end{boxedtest}

\subsection{RQ2-Performance on Real-World Applications}
The ultimate goal of JUnitTestGen is to help better identify API compatibility issues that occur during Android system evolution. To this end, in this research question, we evaluate, based on the generated valid unit test cases, to what extent can JUnitTestGen help in identifying API-induced compatibility issues.

We consider an Android API has a compatibility issue if the execution results on different Android SDK versions are inconsistent. More specifically, an API is considered to have a compatibility issue if any of the following happens: (1) the corresponding test case runs successfully on some Android SDK versions but fails (e.g., throws errors or exceptions) on others; or (2) the corresponding test case throws different errors or exceptions on different Android SDK versions (e.g., throws \emph{NoSuchMethodError} on some versions, while throws \emph{IllegalArgumentException} on others); or (3) the return values of a target API are non-identical on different SDK versions.

\textbf{Result.}
Recall that JUnitTestGen successfully generates 22,805 distinct valid test cases covering 22,805 unique Android APIs, based on the randomly selected 10,000 apps. In this work, these 22,805 test cases are respectively executed on ten Android emulators running API levels from 21 to 30.

By comparing the experimental results against the aforementioned three rules, \textbf{we are able to locate 3,488 compatibility issues covering 2,695 Android APIs.} Note that some APIs may involve more than one compatibility issue.
To confirm whether the APIs identified by JUnitTestGen indeed have compatibility issues, we manually examined 100 randomly selected APIs reported to have compatibility issues, 100 of them are confirmed to have compatibility issues (i.e., true positive results). Here, we remind the readers that all of the compatibility issues are actually identified through dynamic analysis, which is expected to be highly accurate. In the manual validation process, we manually examine the APIs' implementation in Android framework source code across different SDK versions and compare it with the release notes in the official API documentation.

According to the root causes of the compatibility issues, we further categorize them into the following types. 
\begin{itemize}
\item \textbf{Type 1: Signature-based compatibility issue.} This type refers to the incompatibility caused by adding new APIs, deprecating existing APIs, or changing existing APIs' signatures, such as changing parameters or return types.

\item \textbf{Type 2: Semantics-based compatibility issue.} This type refers to the incompatibility caused by the same API (i.e., the signature is not changed) behaving differently on different Android API levels. Based on the APIs' behaviours, semantic-based compatibility issues are further breakdown into the following two sub-types.

\textbf{1) Type 2.1 Semantics-based compatibility issue: Different Errors}: The APIs categorized into this type will throw exceptions or errors (including app crashes) on devices running some SDK versions but will not (either running successfully or throw different exceptions/errors) when running on devices with other SDK versions. \par
\textbf{2) Type 2.2 Semantics-based compatibility issue: Different Return Values}: APIs of this type will not directly cause compatibility issues. Under the same experimental setting, these APIs will return different values when running on different SDKs.
However, if the different return values are not specifically distinguished, the subsequent code that uses these return values may behave differently on different devices, leading to also compatibility issues.
\end{itemize}

Among the 3,488 identified compatibility issues, 438 of them suffer from signature-based issues, while 3,050 are caused by semantic issues. Within the semantic-based compatibility issues, 946 of them are caused by different errors (Type 2.1), while the remaining 2,104 cases observe different return values (Type 2.2). Table~\ref{tab:catgory_exception} summarizes the possible errors/exceptions that cause the signature/semantic compatibility issues.

As expected, the most common error is no such method error, which can be caused by (1) the API being deprecated and removed from the framework, or (2) the API not yet introduced.
Both reasons are introduced by the evolution of the framework.
As also revealed by Li et al.~\cite{li2018cid}, the fast evolution of the Android framework has indeed introduced a lot of APIs that will likely induce compatibility issues.
Except for signature-based issues, which are relatively easy to be statically identified (for example, by comparing the framework codebase of different versions), our approach has also found 2,974 issues (over five times the number of Type 1 issues) involving semantic changes of APIs, which are non-trivial to be identified statically~\cite{liu2021identifying}.

\begin{table}[!h]
\scriptsize
\centering
\setlength{\belowcaptionskip}{-1pt}
\caption{Categories and statistics of the observed error/exception types associated with compatibility issues.} 
\label{tab:catgory_exception}
\resizebox{0.85\linewidth}{!}{
\begin{tabular}{l l c} 
\hline
Issue Type &  Errors/Exceptions & Count\\
 \hline
\multirow{3}{*}{Signature}       & 
 NoSuchMethodError   & 270 \\
 &  NoClassDefFoundError & 163  \\
& NoSuchFieldError & 5\\

\\
\multirow{27}{*}{Semantic}  & 
 SecurityException  & 196\\
 &  NullPointerException & 189\\
 &  RuntimeException & 139\\
 & Resources\$NotFoundException & 113 \\
 &  IllegalArgumentException & 67 \\ 
 &  NoSuchElementException & 42\\
 & Crash & 41 \\
 &  IllegalStateException & 24\\
 & AndroidRuntimeException & 23\\
 & IOException & 15 \\
 &  ArrayIndexOutOfBoundsException & 15\\
 & FileNotFoundException & 12\\
 & PackageManager\$NameNotFoundException & 10 \\
 & IndexOutOfBoundsException & 10 \\
 & ClassCastException & 9 \\
 & IllegalAccessError & 9 \\
 & ActivityNotFoundException & 8 \\
 & StringIndexOutOfBoundsException & 5 \\
 & UnsupportedOperationException & 5 \\
 & BadTokenException & 3 \\
 & CanceledException & 3 \\
 & ErrnoException & 2 \\
 & KeyChainException & 2 \\
 & SQLiteCantOpenDatabaseException & 1 \\
 & ParseException & 1 \\
 & StackOverflowError & 1 \\
 & NumberFormatException & 1 \\
\hline
\end{tabular} 
}
\end{table}

Below, we elaborate on real-world compatibility issues for each type of case study.

\textbf{Case Study 1: Signature-based Compatibility Issue.}
The API \emph{android.content.pm.LauncherApps\#hasShortcutHostPermission} has been reported to contain a signature compatibility issue. The corresponding test case\emph{(whose API usage is extracted from app ch.deletescape.\\lawnchair.plah\footnote{SHA-256:bf4e6e7fb594cd9db4b168a68f70157ad9c1fea0192e0bd5d9a39d1c38802639})} throws \emph{NoSuchMethodError} on Android SDK version 21 to 23 but can be successfully executed on Android SDK version 24 to 30. This result suggests that the API was introduced to the Android system since API level 24; therefore, it would cause an error if the containing app runs on devices with Android SDK version earlier than 24. 
However, according to the official Android API documentation, this API was added in API level 25 \cite{hasShortcutHostPermission}, which is imprecise according to our result. We further checked the source code of Android SDK 24 and confirmed its existence.

\textbf{Case Study 2: Semantics-based compatibility issue caused by different Exceptions.}
The API \emph{android.app.NotificationManager
\#notify} has been reported to contain a semantic compatibility issue. The corresponding test case\emph{(whose API usage is extracted from app com.ag.dropit\footnote{SHA-256:30f7f72cebeffd7c6e26489198ee5ad244bd44b076dd9cb59865d8b0e82a86af})} can be successfully executed on Android SDK version 21 to 22 but throws \emph{IllegalArgumentException} on Android SDK version 23 to 30. We manually looked into its source code in the Android codebase and found that the actual implementation of this API has been changed since API level 23, which added a sanity check of object \emph{mSmallIcon}. This explains why it throws \emph{IllegalArgumentException} when there is no valid small icon from the API level after 23. Unfortunately, the official Android API documentation does not reflect this change, which is misleading.

\textbf{Case Study 3: Semantics-based compatibility issue caused by different return values.}
The API \emph{(extracted from app cleaner.\\clean.booster\footnote{SHA-256:def5db37b3a68de62a0472e872700092060bdec3e875d4f476fcda52795bceb2})} \emph{<android.text.format.Formatter: String formatShortFileSize(Context, long)>} has a return value-induced compatibility issue. The format of the return values vary on different API levels: given the 1L(The long data type of value 1) as the second parameter, the return value on API level 21 to 22 is \emph{``1.0B''}, the return value on API level 23 is \emph{``1.0 B''} (with additional whitespace between 1.0 and B), while the return value on API level 24 to 30 is \emph{``1 B''}. The difference in return values can introduce potential issues if app developers rely on the return value to implement future functions without checking the running API level.
For example, if app developers cast the return value from String to Byte afterwards, it may throw an exception if users ignore the value discrepancy on different API levels.

\begin{boxedtest}
Our approach is useful in automatically pinpointing API-induced compatibility issues. It also goes beyond the state-of-the-art to be capable of detecting not only signature-based compatibility issues but also more significant semantics-based compatibility issues, i.e., the corresponding APIs' signatures are kept the same, while their semantics are altered.
\end{boxedtest}

\subsection{RQ3-Comparison with State-of-the-art}
Considering the main purpose of our work is generating test cases for detecting compatibility issues, both generic test case generation approaches, such as EvoSuite~\cite{fraser2011evosuite} and compatibility issues detection tools, such as CiD~\cite{li2018cid}, are selected as the baselines to evaluate our approach. We evaluate the performance of JUnitTestGen, EvoSuite~\cite{fraser2011evosuite} and CiD~\cite{li2018cid} in detecting compatibility issues. Overall, table~\ref{tab:comparison_RQ3} lists the number of compatibility issues found by JUnitTestGen, EvoSuite and CiD. We then break down the comparative results as follows:

\begin{table}[t!]
\centering
\caption{The comparison results between {JUnitTestGen} and Evosuite, CiD.}
\vspace{2mm}
\label{tab:comparison_RQ3}
\resizebox{0.7\linewidth}{!}{
\begin{tabular}{r | c c c} 
\hline
Tool & \# Type 1 & \# Type 2.1 & \# Type 2.2 \\
\hline
{JUnitTestGen} & 438  & 946 & 2,104  \\
Evosuite & 36 & 0  & 0  \\
CiD & 864  & -  & -  \\
\hline
\end{tabular} }
\end{table}

\textbf{Comparison with EvoSuite.}
To compare JUnitTestGen with generic test case generation tools, we choose EvoSuite as the baseline because EvoSuite has been considered the state-of-the-practice test generation tool, which achieved the highest score at the SBST 2021 Tool Competition~\cite{vogl2021evosuite}. EvoSuite uses an evolutionary search approach to generate and optimize test suites toward satisfying an entire coverage criterion for Java classes. We remind the readers that the objective of EvoSuite is to generate tests for classes, not directly aiming at generating tests for APIs, which are the main target when concerning compatibility issues happening in Android apps.
Since Evosuite can only generate tests based on source code, we resort to AOSP from SDK 21 to 30 for Evosuite to generate test cases.
In total, Evosuite successfully generates 5,335 test cases. 
We then execute all of them on SDK versions from 21 to 30 and apply the same rules for determining compatibility issues as used in JUnitTestGen. 

As shown in Table~\ref{tab:comparison_RQ3}, EvoSuite only finds 36 signature-based compatibility issues, and no semantic compatibility issues are identified. We further manually check the test cases generated by EvoSuite and observe that the false negatives (compared with JUnitTestGen) are mainly caused by overlooking API dependency information. For example, some APIs can only be invoked by system services, which makes EvoSuite fail to generate valid tests without knowing the usage of such APIs. Missing API dependency information makes EvoSuite insufficient in pinpointing compatibility issues. In other words, our comparison result reveals that mining API usage from apps is beneficial for finding compatibility issues.

\textbf{Comparison with CiD.}
To the best of our knowledge, no work has been devoted to detecting compatibility issues in Android apps dynamically. CiD~\cite{li2018cid} is the closest work to ours on detecting compatibility issues. CiD model and compare API signatures on different SDK versions to detect compatibility issues. We thus evaluate the performance of JUnitTestGen and CiD on the same dataset in RQ1, which contains 10,000 Android apps.

In total, CiD detects 864 compatibility issues, as highlighted in Table~\ref{tab:comparison_RQ3}.
Out of the 864 compatibility issues detected by CiD, JUnitTestGen successfully identified 3,050 cases that have been overlooked by CiD. We further randomly analyze 50 false negatives of CiD and find that these issues are caused by the lack of semantics analysis of the API implementation. Specifically, when analyzing the evolution of APIs, CiD only examines the change of API signatures (including name, type, and parameters); hence it is not capable of detecting APIs that modify the implementation details but retain the same signature. Also, we find other 51 false positives of CiD are caused by imprecisely extracting the usage of the APIs, due to the context-insensitive approach of CiD when building the conditional call graph (CCG).
On the other hand, we find that JUnitTestGen miss 375 cases reported in CiD (i.e., false negatives in JUnitTestGen). This is mainly caused by the sophisticated usage of some APIs, which cannot easily be initialized programmatically (e.g. UI-related APIs). For example, some APIs may involve the initialization of UI objects that cannot be initialized programmatically. This makes it very challenging to automatically generate unit test cases in some circumstances.
However, we argue that this limitation can be alleviated by manually adding prerequisite resources to create more valid tests.
Overall, the comparison results reveal the weakness of static analysis approaches in detecting semantic compatibility issues, i.e., yields false-positive results and is hard to detect issues involving semantic changes in methods. 
It also demonstrates that our approach can indeed find more diverse compatibility issues and hence is promising to complement existing static approaches.

\begin{boxedtest}
JUnitTestGen outperforms the state-of-the-practice test generation tool, EvoSuite, and the state-of-the-art static compatibility detection tool, CiD, in pinpointing compatibility issues caused by the fast evolution of Android APIs. 
This experimental evidence shows the necessity to perform dynamic testing to pinpoint compatibility issues in Android apps, and it should take API usage dependencies into consideration when generating test cases to fulfill the dynamic testing approach.
\end{boxedtest}

\section{Discussion}
\label{JUnitTestGen:Discussion}
We now discuss the potential implications and limitations of this work.

\subsection{Implications}
\textbf{Better Supplementing Compatibility Analysis: } 
JUnitTestGen is able to automatically generate tests for Android APIs for detecting both signature-based and semantics-based compatibility issues. Previous static analysis works~\cite{li2018cid} overlooked the semantics-based ones (i.e., APIs have the same signature but different implementation), which are more challenging to detect statically. Together with the state-of-the-art static analysis-based methods, our proposed method can provide a more comprehensive overview of compatibility issues in Android APIs. 
Therefore, we argue that there is a need to invent hybrid approaches to take advantage of both static analysis and dynamic analysis to conquer compatibility issues.

\textbf{Beyond Compatibility Testing: }
JUnitTestGen is not only useful in pinpointing compatibility issues in Android APIs but also could be easily adopted to automatically generate test cases for other purposes.
For instance, in the cases that an API takes various parameters as input, it can work with other testing approaches such as fuzz testing to explore the API's implementation dynamically.
It hence goes beyond compatibility testing and provides a more general-purpose form of API testing.

\textbf{Go Beyond Android.}
Our approach performs static program analysis to learn and generate test cases from existing API usages, which are not strongly attached to Android apps. We believe it could be easily adapted to analyze other Java projects, e.g., to automatically generate test cases for popular Java libraries. Although our approach cannot be directly applied to analyze projects written in other programming languages than Java, we believe the idea and methodology proposed in this paper could still work.
We plan to explore these research directions in our future work. We also encourage our fellow researchers to explore this direction further.

\section{Limitations}
\label{JUnitTestGen:limitations}
The main limitation of JUnitTestGen lies in its backward data-flow analysis when inferring API caller instance and API parameter values.
Indeed, as already known in the community, it is non-trivial to implement a sound data-flow analysis.
Other researchers often accept trade-offs to obtain relatively good results, and this is the same in our case.
When the variables to be backwardly retrieved are complex, e.g., involving constructing various intermediate objects, JUnitTestGen rely on their simplest constructor to initialize the objects to generate a valid test case. Unfortunately, some of the constructors are too complex to initialize, leading to invalid test cases. Some other APIs may involve the initialization of UI objects that cannot be initialized programmatically (hence random values are leveraged to handle such cases). This makes it very challenging to automatically generate unit test cases in all circumstances.

Second, currently, our JUnitTestGen data-flow analysis is agnostic to some advanced programming features, such as reflective and native calls. This may further impact the soundness of our approach.
As part of our future work, we plan to integrate approaches developed by our fellow researchers to mitigate those long-standing challenges (e.g., applying DroidRA~\cite{li2016droidra} to mitigate the impact of reflective calls on our static analysis approach.).

Third, the capability of our approach is limited by the number of Android APIs leveraged by real-world apps. Our approach cannot generate unit test cases for such APIs that have never been accessed by real-world Android apps.
Nevertheless, we argue that this impact is not significant as the APIs that have never been used by app developers should have a low priority to be tested than the others that are frequently accessed.
Compared to the latter case, the former ones will not cause problems such as crashes to Android apps.
Subsequently, it will not impact the user experience and the reputation of the app developers.

Fourth, considering the generated parameter values are inferred from real-world apps, which might not reveal all possible semantic-related compatibility issues. In other words, the capability of our approach is limited by the values of parameters leveraged by real-world apps. As for our future work, we plan to integrate fuzzing techniques~\cite{zhang2020bigfuzz,patra2016learning, sui2011effective} into our approach so as to trigger as many compatibility issues as possible.

Fifth, our definition of compatibility issue is based on the observation that the same test case throws different errors or exceptions on different Android SDK versions. However, this might introduce false negatives because the tests that throw the same exception across all SDK versions are ignored. Nevertheless, we argue that this type of false-negative requires further human validation and thus cannot be determined automatically. In addition, related works~\cite{cai2019large} also have not considered this situation.

Sixth, the tests generated by our approach may suffer from flaky tests. Indeed, non-deterministic return values may appear under different execution environments, leading to false positives. However, it is a non-trivial task to tackle flaky tests ~\cite{zolfaghari2021root} because the root causes of flaky tests can be quite sophisticated. Nevertheless, as part of our future work, we plan to integrate other approaches developed by our fellow researchers to mitigate this long-standing challenge, e.g., by applying FlakeScanner~\cite{dong2021flaky} to mitigate the impact of flaky tests on our approach.

Seventh, the types of compatibility issues detected in our approach are the ones that are related to exceptions or return values. However, this will certainly not be complete to cover all possible cases. For example, the value of variables in the same API may evolve on different SDK versions. Indeed, as summarised by Liu et al.~\cite{liu2022automatically}, apart from compatibility issues raised by API signature/semantic changes and return value differences, there exist other types of compatibility issues, such as those introduced by field evolution, callback method changes, etc. Nevertheless, as also highlighted by Liu et al.~\cite{liu2022automatically}, the number of such compatibility issues is quite limited, suggesting that the impact of such cases on our approach may not be significant.

Eighth, the main objective of the generated test cases in this work is to pinpoint compatibility issues.
The quality of these test cases (e.g., readability, overlaps, and maintenance) has hence not been considered. We therefore commit to further investigating the quality of the generated test cases in our future work.

Last but not least, our approach relies on existing code examples to generate test cases. However, the selected code examples may contain sub-optimal or erroneous API usages. Nevertheless, we argue that this impact is not significant as we extract code examples from real-world Android apps from Google Play, for which thousands of users might have used (hence tested) them in practice.

\section{Related Work}
\label{JUnitTestGen:related_work}
Android API evolution is a critical issue in software maintenance ~\cite{nagappan2016future,li2018moonlightbox,martin2016survey,li2016accessing,oliveira2018android,lamothe2020a3,li2018characterising,zhou2016api,dig2005role,kapur2010refactoring,liu2022deep,sun2021characterizing}. McDonnell et al. ~\cite{mcdonnell2013empirical} have shown that the Android
system updates 115 APIs per month on average, while app developers usually adopt the new APIs much more slower. The slow adoption of API updates may raise various issues, such as security and compatibility. An empirical study on StackOverflow conducted by Linares-Vasquez et al. ~\cite{linares2014api} suggests that API updates would trigger more discussions, especially if APIs are removed from the Android system. They also revealed that users are in more favour of apps that use less fault and change-prone APIs \cite{linares2013api, bavota2014impact}, as these apps would likely introduce fewer failures, crashes and other bugs.

Android developers have long been suffered from compatibility issues due to the fast-evolving and fragmented nature of the Android ecosystem \cite{xia2020android,kamran2016android,li2017static,nayebi2012state,zhao2022towards}. 
Researchers have proposed several solutions for detecting compatibility issues of Android APIs. Wei et al. \cite{wei2016taming, wei2018understanding} conducted an empirical study to investigate fragmentation-induced API compatibility issues and proposed a tool named FicFinder to detect such APIs. FicFinder identifies APIs with compatibility issues based on heuristic rules manually derived from a limited number of Android apps, which is expected to introduce high false negatives (i.e., missing undiscovered compatibility issues). Thereafter, several works have been proposed to leverage data-driven techniques that automatically mine compatibility issues from various sources such as Android code base and real-world apps. Li et al. proposed a tool named CiD to detect potential compatibility issues by mining the history of the Android framework source code. CiD identifies Android APIs' lifetimes and finds if an app's declared supported versions conflict with its used APIs. 

Comparable to our method, several works have been proposed to mine API usage from real-world apps. Scalabrino et al. \cite{scalabrino2019data, scalabrino2020api} considered the APIs wrapped in a version check condition (e.g., \emph{if (Build.VERSION.SDK\_INT >= 21)}) to potentially have compatibility issues and developed a tool named ACRYL to extract such APIs from real-world apps. However, ACRYL can only detect APIs whose compatibility issue is already known by the developers (i.e., they are enclosed in the version check conditions by the developers), while our method is capable of detecting zero-day compatibility issues that the app developers are not yet aware of, or even Google itself. 
Other generic test generation tools, such as EvoSuite and Randoop\cite{pacheco2007randoop}, are able to generate tests for Java classes. However, these tools do not directly aim at generating tests for APIs and they have been demonstrated as insufficient in pinpointing compatibility issues because of the lack of API usage knowledge.

\section{Summary}
\label{JUnitTestGen:summary}
In this work, we presented a novel prototype tool, JUnitTestGen, that mines existing Android API usages to generate API-focused unit test cases automatically for pinpointing potential compatibility issues caused by the fast evolution of the Android framework.
Experimental results on thousands of real-world Android apps show that (1) JUnitTestGen is capable of automatically generating valid unit test cases for Android APIs with an 80.4\% success rate;
(2) the automatically generated test cases are useful for pinpointing API-induced compatibility issues, including not only signature-based but also semantics-based compatibility issues; and (3) JUnitTestGen outperforms the state-of-the-practice test generation tool, EvoSuite, and the state-of-the-art static compatibility detection tool, CiD, in pinpointing compatibility issues.

%% file: chapter6.tex
\chapter{Conclusions and Future Research}

\section{Key Contributions}
This PhD project aimed to develop automatic tools tailor for detecting security and compatibility issues in Android applications. 
Particularly, I proposed three new approaches and prototype tools to detect these issues from three different aspects.

First, I proposed \textbf{a new static analysis-based approach, {\sc \textbf{SEEKER}}}, which can detect privacy leaks originated from Android sensors in Android applications. 
Despite being needed to support the implementation of many diverse Android apps, mobile phone sensors can also be abused to achieve malicious behaviors. There have been many reports of  apps that exploit sensors in Android devices to conduct malicious activities. For example, sensor data are known to be leaked out because they are not protected by any permissions in Android. There is hence a strong need to regulate the usage of mobile sensors so as to keep them from being exploited by malicious attackers. However, despite the fact that various efforts have been put in achieving this, i.e., detecting privacy leaks in Android apps, we have not yet found approaches to automatically detect sensor leaks in Android apps.

To address these issues, I proposed a novel tool, SEEKER~\cite{sun2021characterizing}, for characterizing sensor leaks in Android apps. Specifically, I extended the famous FlowDroid~\cite{arzt2014flowdroid} tool to detect sensor-based data leaks in Android apps. SEEKER conducts sensor-focused static taint analyses directly on the Android apps’ bytecode and reports not only sensor-triggered privacy leaks but also the sensor types involved in the leaks. 

Our experimental results on a large scale of real-world Android apps indicate that our new SEEKER tool is effective in identifying all types of potential sensor leaks in Android apps. Our tool is not only capable of detecting sensor leaks, but also pinpointing general privacy leaks that are triggered by class fields. 
Although there are related works on sensor usage analysis, to the best of our knowledge, there is no other work that thoroughly analyses Android sensor leakage. Unlike previous works, our tool is the first one to characterize all kinds of sensor leaks in Android apps. I extended FlowDroid for supporting field sources detection (i.e., merged to FlowDroid via pull \#385 on Github\cite{FlowDroidMerge}), which I believed could be adapted to analyze other sensitive field-triggered leaks.
To benefit our fellow researchers and practitioners towards achieving this, I have made our approach open source at the following Github site.

The key contributions of this work are as follows: 

(i) I have designed and implemented a prototype tool, SEEKER (\underline{Se}nsor l\underline{e}a\underline{k} find\underline{er}), that leverages static analysis to automatically detect privacy leaks originated from Android sensors.

(ii) I applied SEEKER to analyze both malware and benign apps at a large scale. Our results show many sensor leaks that are overlooked by the state-of-the-art static analysis tool.

(iii) I have demonstrated the effectiveness of our tool by evaluating the sensor leaks it highlights.

Second, I proposed \textbf{a prototype program analysis tool, {\sc \textbf{HiSenDroid}}}, which performs a static code analysis to uncover hidden sensitive operations that will only be triggered under special circumstances such as at a specific location or in a certain time period.
Additionally, HiSenDroid goes one step deeper to provide details aiming at helping security analysts understand why a given hidden sensitive operation is flagged as such.
Experimental results over 20,000 apps, including both malicious and benign apps, show that hidden sensitive operations are indeed quite frequently presented in Android apps and HiSenDroid is effective in automatically discovering them.
Moreover, with the help of FlowDroid, a state-of-the-art static taint analyzer, I further experimentally find that hidden sensitive operations could eventually lead to privacy leaks.

The key contributions of this work are as follows:

(i) I designed and implemented a prototype tool HiSenDroid for analyzing hidden sensitive operations. I released HiSenDroid as an open source project \cite{HiSenDroid} for supporting security analysts in their analysis needs and fostering further researches in this direction.

(ii) I evaluated HiSenDroid on a large-scale dataset that contains 10,000 benign and 10,000 malware samples, and discovered emerging anti-analysis techniques employed by malware samples, such as fulfilling certain restrictions related to \emph{time}, \emph{location}, \emph{SMS message}, \emph{system properties}, \emph{package manager}, and other logics. 

(iii) With the help of FlowDroid~\cite{arzt2014flowdroid}, a static taint analyzer, I further experimentally showed that HSOs have been recurrently leveraged by attackers to leak sensitive user information.

Third, \textbf{I proposed a field-aware, inter-procedural backward data-flow analysis tool, {\sc \textbf{JUnitTestGen}}}, which mines Android API usage to generate unit test cases for pinpointing compatibility issues. Despite being one of the largest and most popular projects, the official Android framework has only provided test cases for less than 30\% of its APIs.
Such a poor test case coverage rate has led to many compatibility issues that can cause apps to crash at runtime on specific Android devices, resulting in poor user experiences for both apps and the Android ecosystem.
To mitigate this impact, various approaches have been proposed to automatically detect such compatibility issues.
Unfortunately, these approaches have only focused on detecting signature-induced compatibility issues (i.e., a certain API does not exist in certain Android versions), leaving other equally important types of compatibility issues unresolved. One unresolved type is related to semantic changes of APIs, which are non-trivial to pinpoint as it is yet not possible to comprehend automatically all the semantics of code.
In this work, I proposed a novel prototype tool, JUnitTestGen, to fill this gap by mining existing Android API usage to generate unit test cases.
After locating Android API usage in given real-world Android apps, JUnitTestGen performs inter-procedural backward data-flow analysis to generate a minimal executable code snippet (i.e., test case).
Experimental results on thousands of real-world Android apps show that JUnitTestGen is effective in generating valid unit test cases for Android APIs. I showed that these generated test cases are indeed helpful for pinpointing compatibility issues, including ones involving semantic code changes.

In this work, I have made the following key contributions:

(i) I have designed and implemented a prototype tool JUnitTestGen that leverages a novel approach to automatically generate unit test cases for APIs based on their existing usages.

(ii) I have set up a reusable testing framework for pinpointing API-induced compatibility issues by automatically executing a large set of unit test cases on multiple Android devices. 

(iii) I have demonstrated the effectiveness of JUnitTestGen by i) generating valid test cases for Android APIs and pinpointing problematic APIs that could induce compatibility issues if accessed by Android apps, ii) outperforming state-of-the-art tools on real-world apps in detecting a wider range of compatibility issues.

\section{Future Work}

My long-term research goal is to conduct comprehensive program analysis research in order to ensure better security and reliability of software systems. I am committed to solving difficult software problems using practical and elegant solutions. 

The state-of-the-art static analysis techniques are not sufficient to find various defects in real-world applications. This  leaves many software defects undetected, causing serious problems for both users and developers. For example, attackers are reported to use complicated language features (e.g., reflection, obfuscation, and encryption) to hide malicious operations. So I have the ambition to propose new approaches to do code unification to perform much more sound and comprehensive results. Specifically, I plan to unify native code, and javascript with the bytecode in Android applications to achieve this. After enhancing the static analysis techniques, I plan to further apply such techniques to help detect vulnerabilities or software defects which are overlooked before. This line of thinking provides for a number of opportunities for future research, incluidng but not limited to:

\begin{enumerate}
       \item \textbf{React Native Code Unification: } Android code unification is needed to perform comprehensive static analysis for hybrid Android apps. In Android apps, dex bytecode cohabits with JavaScript code which can be used through the React Native Interface. Due to the challenge presented to analyze JavaScript code, it is most of the time overlooked by existing approaches. This limitation is a severe threat to validity since malicious behaviour can be implemented in JavaScript code. Therefore, I have the ambition to propose a model unifying both the bytecode and the JavaScript code in Android apps. I plan to propose a first step toward this direction at the call-graph level and with more granularity at the statement level relying on heuristic-based defined statements.

       \item \textbf{Typestate Misuse Detection in Android: } Android frameworks often provide complex functionality in order to support rich state transitions, which is a non-trivial task for programmers to understand and adopt correctly~\cite{aldrich2009typestate}. One essential aspect of correct API use is \emph{typestate}~\cite{mishra2016asynchrony}, which defines legitimate sequences of operations that can be performed upon a given state type. A \emph{typestate property} is a finite state machine~\cite{Finite_state_machine}, representing a sequence of behavioural type refinements such as ``method A must be called after method B returns successfully''. However, programmers are reported frequently ignore such disciplines~\cite{kruger2019crysl}, causing Android defects and lost productivity when typestate APIs are misused in practice~\cite{aldrich2009typestate}. Even worse, applications that fail to follow correct state transitions (i.e., typestate misuses) may also be infected with vulnerabilities~\cite{li2022path,liu2016understanding,pan2020static,yan2017machine,wang2020typestate}, including data corruption~\cite{xu2015collision,conti2015losing}, privacy leaks~\cite{lee2015preventing,serna2012info} and hijacking attacks~\cite{caballero2012undangle,fan2017boosting}. \\
       
       To address this problem, I plan to design and implement a prototype tool to automatically detect typestate API misuse in real-world Android apps. Specifically, I am going to statically model the correct typestate usage based on the API comments in Android official documents, indicating the explicit disciplines of how typestate APIs should be performed. After that, I plan to perform intra- and inter-procedural data-flow analysis to extract the typestate API usage in real-world applications. After that, I will compare the typestate API usage with the correct typestate rules for pinpointing the violations. I believe that my tool can help app developers identify typestate misuse cases at the development stage to avoid these problems from being distributed into the community.
       
        \item \textbf{Dynamic Software Testing: } The current state-of-the-art works targeting software defects detection leverage static analysis techniques to achieve their purpose. However, all software detects cannot be statically detected under every circumstance, introducing false positives and false negatives. To mitigate this problem, I plan to leverage dynamic program testing techniques to perform defects analysis as dynamic malware analysis methods generally provide better precision than purely static methods. \\
        
        To trigger malicious behaviours, I plan to develop a tool for automatically generating inputs that will trigger certain sensitive APIs (e.g., the APIs are protected by permission) in the source code. By dynamically triggering the target APIs, I am able to determine whether there exist malicious operations. In addition, I will then integrate my tool with state-of-the-art static taint analysis tools (e.g., FlowDroid~\cite{arzt2014flowdroid}) to offer better precision. 
\end{enumerate}

\section{Summary}

In this Ph.D. thesis I focused on detecting security and compatibility issues in Android applications. 
I developed three tools, i.e., {\sc SEEKER}, {\sc HiSenDroid}, {\sc JUnitTestGen} for improving the traditional program analysis techniques in detecting wider range of security and compatibility issues. Specifically, SEEKER helps extend FlowDroid for supporting field sources detection, HiSenDroid reveals hidden sensitive operations which have been overlooked by state-of-the-art tools and JUnitTestGen is an effective tool to detect compatibility issues through a dynamic testing approach that fulfills its objective by automatically generating valid test cases. I plan to enhance the soundness and comprehensiveness of existing  program analysis techniques for identifying wider range of security and compatibility issues in our future work.